\documentclass[aos]{imsart}

\RequirePackage{amsthm,amsmath,amsfonts,amssymb}
\RequirePackage[numbers,sort&compress]{natbib}
\RequirePackage[colorlinks,citecolor=blue,urlcolor=blue]{hyperref}
\RequirePackage{graphicx}
\usepackage{booktabs}
\usepackage{comment}
\startlocaldefs
\theoremstyle{plain}

\newtheorem{theorem}{Theorem}
\newtheorem{corollary}[theorem]{Corollary}
\newtheorem{lemma}{Lemma}

\theoremstyle{remark}

\endlocaldefs

\usepackage{ulem} 
 \usepackage{mathtools} 

\usepackage{enumerate}
\usepackage{multicol}
\usepackage{multirow}
\usepackage{fancyhdr}
\usepackage{epsfig}

\usepackage{xparse}

\usepackage[english]{babel}															

\usepackage{multirow}
\usepackage{caption}
\usepackage{subcaption} 

\newcommand{\var}{\operatorname{Var}}

\newcommand{\col}{\operatorname{col}}

\newcommand{\tr}{\operatorname{tr}}
\newcommand{\diag}{\operatorname{diag}}
\newcommand{\bA}{\mathbf{A}}
\newcommand{\bB}{\mathbf{B}}
\newcommand{\bC}{\mathbf{C}}
\newcommand{\bD}{\mathbf{D}}
\newcommand{\bE}{\mathbf{E}}
\newcommand{\bF}{\mathbf{F}}
\newcommand{\bG}{\mathbf{G}}

\newcommand{\bI}{\mathbf{I}}
\newcommand{\bJ}{\mathbf{J}}

\newcommand{\bM}{\mathbf{M}}

\newcommand{\bO}{\mathbf{O}}
\newcommand{\bP}{\mathbf{P}}
\newcommand{\bQ}{\mathbf{Q}}
\newcommand{\bR}{\mathbf{R}}
\newcommand{\bS}{\mathbf{S}}

\newcommand{\bU}{\mathbf{U}}
\newcommand{\bV}{\mathbf{V}}
\newcommand{\bW}{\mathbf{W}}
\newcommand{\bX}{\mathbf{X}}

\newcommand{\bZ}{\mathbf{Z}}
\newcommand{\ba}{\mathbf{a}}
\newcommand{\bb}{\mathbf{b}} 
\newcommand{\bc}{\mathbf{c}} 
\newcommand{\bd}{\mathbf{d}}
\newcommand{\be}{\mathbf{e}}
\newcommand{\bbf}{\mathbf{f}}

\newcommand{\bii}{\mathbf{i}}

\newcommand{\bm}{\mathbf{m}}

\newcommand{\bu}{\mathbf{u}}

\newcommand{\bx}{\mathbf{x}}
\newcommand{\by}{\mathbf{y}} 

\newcommand{\balpha}{\boldsymbol{\alpha}}
\newcommand{\btheta}{\boldsymbol{\theta}}

\newcommand{\bgamma}{\boldsymbol{\gamma}}

\newcommand{\bbeta}{\boldsymbol{\beta}}

\newcommand{\bLambda}{\boldsymbol{\Lambda}}

\newcommand{\btau}{\boldsymbol{\tau}}

\newcommand{\bzero}{\boldsymbol{0}}
\newcommand{\bone}{\boldsymbol{1}}

\newcommand{\siga}{{\sigma}_{\alpha}}  
\newcommand{\sigb}{{\sigma}_{\beta}}  
\newcommand{\siggama}{{\sigma}_{\gamma}}  
\newcommand{\sige}{{\sigma}_{e}}

\newcommand{\sumig}{\sum_{i=1}^g}
\newcommand{\sumjh}{\sum_{j=1}^h}

\NewDocumentCommand{\barRia}{t^}{%
\IfBooleanTF{#1}
{\barRiaAux}
{\bar{R}_{i.}^{(a)}}%
}
\NewDocumentCommand{\barRiaAux}{m}{%
\bar{R}_{i.}^{(a)#1}%
}

\NewDocumentCommand{\barRjb}{t^}{%
\IfBooleanTF{#1}
{\barRjbAux}
{\bar{R}_{.j}^{(b)}}%
}
\NewDocumentCommand{\barRjbAux}{m}{%
\bar{R}_{.j}^{(b)#1}%
}

\NewDocumentCommand{\barRijab}{t^}{%
\IfBooleanTF{#1}
{\barRijabAux}
{\bar{R}_{ij}^{(ab)}}%
}
\NewDocumentCommand{\barRijabAux}{m}{%
\bar{R}_{ij}^{(ab)#1}%
}

\NewDocumentCommand{\barRijkw}{t^}{%
\IfBooleanTF{#1}
{\barRijkwAux}
{{R}_{ijk}^{(w)}}%
}
\NewDocumentCommand{\barRijkwAux}{m}{%
{R}_{ijk}^{(w)#1}%
}

\newcommand{\btaugamma}[3]{\btau_{\gamma,(#1,#2,#3)}}
\newcommand{\btaualpha}[1]{\btau_{\alpha_{#1}}}
\newcommand{\btaubeta}[1]{\btau_{\beta_{#1}}}
\newcommand{\taualpha}[2]{\tau_{\alpha_{#1,#2}}}
\newcommand{\taubeta}[2]{\tau_{\beta_{#1,#2}}}
\newcommand{\btaugammagh}[5]{\btau_{#1,#2,#3}^{(#4,#5)}}
\newcommand{\btaugammaght}[5]{\btau_{#1,#2,#3}^{(#4,#5)\top}}
\newcommand{\taugamma}[6]{\tau_{(#1,#2,#3),#6}^{(#4,#5)}}

\newcommand{\barbJ}[1]{\Tilde{\bJ}_{#1}}
\newcommand{\tildebJ}{\bar{\bJ}}
\newcommand{\barbone}{\tilde{\bone}}
\newcommand{\tildebone}{\bar{\bone}}
\newcommand{\barbI}{\tilde{\bI}}
\newcommand{\tildebI}{\bar{\bI}}

\begin{document}

\begin{frontmatter}
\title{On the inverse of covariance matrices for unbalanced crossed designs}
\runtitle{On the inverse of covariance matrices for unbalanced crossed designs}

\begin{aug}
\author[A]{\fnms{Ziyang}~\snm{Lyu}\ead[label=e1]{Ziyang.Lyu@unsw.edu.au}\orcid{0000-0003-3307-4148}},
\author[A]{\fnms{S.A.}~\snm{Sisson}\ead[label=e2]{Scott.Sisson@unsw.edu.au}\orcid{0000-0001-8943-067X}},
\and
\author[B]{\fnms{A.H.}~\snm{Welsh}\ead[label=e3]{Alan.Welsh@anu.edu.au}\orcid{0000-0002-3165-9559}}
\address[A]{UNSW Data Science Hub, and School of Mathematics and Statistics,
University of New South Wales\printead[presep={,\ }]{e1,e2}}

\address[B]{Research School of Finance, Actuarial Studies and Statistics,
Australian National University\printead[presep={,\ }]{e3}}
\end{aug}

\begin{abstract}
This paper addresses a long-standing open problem in the analysis of linear mixed models with crossed random effects under unbalanced designs: how to find an analytic expression for the inverse of $\bV$, the covariance matrix of the observed response.  The inverse matrix $\bV^{-1}$ is required for likelihood-based estimation and inference.  However, for unbalanced crossed designs, $\bV$ is dense and the lack of a closed-form representation for $\bV^{-1}$, until now, has made using likelihood-based methods computationally challenging and difficult to analyse mathematically.
We use the Khatri--Rao product to represent $\bV$ and then to construct a modified covariance matrix whose inverse admits an exact spectral decomposition. 
Building on this construction, we obtain an elegant and simple approximation to $\bV^{-1}$ for asymptotic unbalanced designs. 
For non-asymptotic settings, we derive an accurate and interpretable approximation under mildly unbalanced data and establish an exact inverse representation as a low-rank correction to this approximation, applicable to arbitrary degrees of unbalance.  
Simulation studies demonstrate the accuracy, stability, and computational tractability of the proposed framework.
\end{abstract}



\end{frontmatter}

\section{Introduction}
Linear mixed models with crossed random effects are widely used in education, psychology, and recommender systems, where observations are affected by multiple categorical factors in a fully crossed structure. Unlike nested designs, crossed structures involve non-hierarchical interactions among factors, resulting in complex correlation structures. As a result, the marginal covariance matrix of the observed response, denoted by $\bV$, is typically dense. Computing or approximating $\bV^{-1}$ is essential for likelihood-based estimation and inference, including maximum likelihood (ML) estimation, restricted maximum likelihood (REML) estimation, and empirical best linear unbiased prediction (EBLUP), because $\bV^{-1}$ directly determines the form and precision of these procedures.

However, deriving the analytical form of $\bV^{-1}$ for unbalanced crossed designs, where the number of observations varies across cells defined by the factor levels, has been an open problem for more than fifty years. \cite{searle1973wanted} first recognized the challenge of expressing $\bV^{-1}$ in closed form under unbalanced data, and \cite{wansbeek1982another} later proposed a recursive approximation. However, the explicit dependence of its elements on unknown parameters called variance components remains unresolved, and a general analytical or tractable representation of $\bV^{-1}$ is not available.


The highly complex structure of the variance matrix $\bV$ under unbalanced crossed designs has meant that very few studies focus on crossed random models, and those that do, avoid using likelihood-based methods. 
For example, the asymptotic analysis of \cite{jiang2025asymptotic} for generalized linear crossed random models employed a second-order Laplace approximation of a conditional expectation to achieve likelihood-free inference that does not require evaluating $\bV^{-1}$. 
Other recent studies have, to varying degrees, explored approximations related to $\bV^{-1}$ in simplified crossed random-effects settings. 
For example, \cite{gao2017efficient} proposed a moment-based method for estimating variance components, and \cite{ghosh2022backfitting} introduced a backfitting algorithm for computing generalized least squares estimators. 
However, these methods are limited to special cases where each cell contains only one observation and no interaction terms are present, and thus do not address the general structure of $\bV^{-1}$ under arbitrary cell sizes or with interaction effects. 
Therefore, obtaining an explicit and tractable expression for $\bV^{-1}$ remains one of the central unresolved problems in the theory of crossed mixed models.

We consider data arranged in a two-way table in which the cells correspond to the crossing of two categorical factors, referred to as factor A (rows) and factor B (columns), respectively.  
Let \( [y_{ijk}, \bx_{ijk}^\top]^\top \) represent the \( k \)th observed vector in cell \( (i,j) \), where \( y_{ijk} \) is a scalar response variable, and \( \bx_{ijk} \) is a $p$-dimensional vector of covariates or explanatory variables.
Indices \( i \) and \( j \) identify the rows and columns of the table, respectively, and \( k \) identifies the specific observation within a cell.  When the number of observations in cell $(i,j)$, $m_{ij} \ge 1$, the data can be modelled using
the two-way crossed random effect with interaction model
\begin{equation}\label{two-way cross model}
	y_{ijk}=\mu(\bx_{ijk})+\alpha_i+\beta_j+\gamma_{ij}+e_{ijk},
\end{equation}
for $i=1,\ldots,g$, $j=1,\ldots,h$, $k=1,\ldots,m_{ij}$,
where $\mu(\bx_{ijk})$ is the conditional mean or regression function of the response given the covariate $\bx_{ijk}$, $\alpha_i$ is the random effect due to the $i$th row, $\beta_j$ is the random effect  due to the $j$th column, $\gamma_{ij}$ is the interaction random effect  for row $i$ and column $j$, and $e_{ijk}$ is the error term.  The random variables $\alpha_i$, $\beta_j$, $\gamma_{ij}$ and $e_{ijk}$ are assumed mutually independent with zero means and variances $\sigma_\alpha^2$, $\sigma_\beta^2$,  $\sigma_{\gamma}^2$, and $\sigma_e^2$, respectively.  Because our focus is on the random effects, we leave the regression function $\mu(\bx_{ijk})$ in general form; see \cite{lyu2024increasing,lyu2025asymptotics} for a useful specific form.
%
The no interaction model assumes $\siggama^2=0$ (and hence $\gamma_{ij}=0$), which is required for identifiability when $m_{ij}=1$. Our results in this case follow as a special case of the general model (\ref{two-way cross model}).  

Under the model (\ref{two-way cross model}),  the covariance matrix of the response $\bV$ is dense. For balanced designs, where $m_{ij}=m$, $\bV$ can be written using Kronecker products \cite{searle1979dispersion}, which facilitates theoretical analysis and explicit matrix inversion. Building on this structure, \cite{lyu2024increasing,lyu2025asymptotics} studied the asymptotic properties of ML and REML estimators and EBLUPs under balanced settings. However, this representation does not hold in unbalanced designs, where the structure of $\bV$ becomes substantially more complicated. Consequently, obtaining accurate and tractable approximations to $\bV^{-1}$ in general unbalanced settings remains a central challenge, both for computational efficiency and in terms of enabling rigorous asymptotic analysis.

This paper develops a unified framework for approximating and analyzing $\bV^{-1}$ under general unbalanced crossed random effects designs. Section~\ref{Sec2} introduces notation and the Khatri--Rao product, a generalization of the Kronecker product. Section~\ref{Sec3} defines a modified covariance matrix with a closed-form inverse 
that yields a tractable approximation to $\bV^{-1}$ under the asymptotic setting and under the non-asymptotic setting with mild unbalance. We also derive an exact expression for a low-rank correction to the approximation under the non-asymptotic setting with severe unbalance. Section~\ref{Sec4} presents simulation results that confirm the accuracy of the approximation, and Section~\ref{Sec5} concludes with a discussion of the broader implications and possible extensions of this work.

\section{Preliminaries}\label{Sec2}
\subsection{Khatri--Rao product}

To overcome the analytical and computational challenges posed by unbalanced crossed designs, we introduce the Khatri--Rao product  \citep{khatri1968solutions, liu2008hadamard}, which generalizes the Kronecker product traditionally used with balanced designs. Unlike the Kronecker product, which requires identical replication across cells, the Khatri--Rao product accommodates variation in cell sizes and enables an exact, structured expression for the covariance matrix even when the data are unbalanced. 
To our knowledge, this is the first use of the Khatri--Rao product to represent unbalanced crossed random effects structures.

Consider matrices $\bA=(a_{ij})$ of order $m\times n$ and $\bB=(b_{kl})$ of order $p\times q$. Let $\bA=(\bA_{ij})$ be partitioned with $\bA_{ij}$ denoting the $(i,j)$th block submatrix of order $m_i\times n_j$, and $\bB=(\bB_{kl})$ be partitioned with $\bB_{kl}$ denoting the $(k,l)$th block submatrix of order $p_k\times q_l$ ($\sum_i m_i=m, \sum_j n_j=n, \sum_k p_k=p $ and $\sum_l q_l=q$). Assume that $\bA$ and $\bB$ have the same number of row and column partitions. 
The  Khatri--Rao product of two suitably partitioned matrices is then
\begin{align*}
\bA \circledast \bB = \left( \bA_{ij} \otimes \bB_{ij} \right)_{ij},
\end{align*}
where the $ (i,j)$-th block of $\bA \circledast \bB$ is the Kronecker product of the corresponding blocks $ \bA_{ij} $ and $\bB_{ij}$. Specifically, each block $\bA_{ij} \otimes \bB_{ij}$ has size $ m_i p_i \times n_j q_j$. The overall size of $\bA \circledast \bB$ is therefore $(\sum_i m_i p_i ) \times ( \sum_j n_j q_j )$.
Throughout, $\otimes$ denotes the Kronecker product, and $\circledast$ denotes the Khatri--Rao product.
For example, if $\bA$ and $\bB$ are both  partitioned matrices with
\begin{align*}
\bA = \begin{bmatrix}
	\bA_{11} & \bA_{12} \\
	\bA_{21} & \bA_{22}
\end{bmatrix}
=
\left[
\begin{array}{cc|c}
	1 & 2 & 3 \\
	4 & 5 & 6 \\
	\hline
	7 & 8 & 9
\end{array}
\right]
, \quad
\bB = \begin{bmatrix}
	\bB_{11} & \bB_{12} \\
	\bB_{21} & \bB_{22}
\end{bmatrix}
=
\left[
\begin{array}{c|cc}
	1 & 4 & 7 \\\hline
	2 & 5 & 8 \\
	3 & 6 & 9
\end{array}
\right],
\end{align*}
then
\begin{align*}
\bA \circledast \bB =
\begin{bmatrix}
	\bA_{11} \otimes \bB_{11} & \bA_{12} \otimes \bB_{12} \\
	\bA_{21} \otimes \bB_{21} & \bA_{22} \otimes \bB_{22}
\end{bmatrix}
=
\left[
\begin{array}{cc|cc}
	1 & 2 & 12 & 21 \\
	4 & 5 & 24 & 42 \\\hline
	14 & 16 & 45 & 72 \\
	21 & 24 & 54 & 81
\end{array}
\right].
\end{align*}

\subsection{Notation}
We define $\col[\cdot]$ to be the column-stacking operator. When stacking vectors indexed by multiple indices, we adopt the convention that the rightmost index varies fastest: for $\col[\bx_{ijk}]$, the stacking proceeds from $k$, then $j$, and finally $i$ (i.e., $\bx_{111}, \bx_{112}, \ldots, \bx_{ghm_{gh}}$).
We use the notation $(\cdot)_{\mathrm{row}}$ to mean each row of the matrix is treated as a block submatrix; $(\cdot)_{\mathrm{col}}$ to mean each column is treated as a block submatrix; and $(\cdot)_{\mathrm{cell}}$ to mean each scalar entry is treated as a block submatrix.

We let the total number of observations be $n = \sum_{i=1}^{g} \sum_{j=1}^{h} m_{ij}$, and define $m_L = \min_{ij}(m_{ij})$ and $m_U = \max_{ij}(m_{ij})$. Let the vector of cell sizes be $\bm = [m_{11}, \ldots, m_{gh}]^\top$, and define $\bM = \diag(m_{11}, \ldots, m_{gh})$.
Let $\bone_a$ denote the $a$-dimensional vector of ones and $\bI_a$ the $a \times a$ identity matrix.
To use the Khatri-Rao product effectively, define $\bone_{\bm}$ as the $n \times 1$ vector of ones, partitioned into $gh$ subvectors as $\bone_{\bm} = [\bone_{m_{11}}^\top, \bone_{m_{12}}^\top, \ldots, \bone_{m_{gh}}^\top]^\top$, where each $\bone_{m_{ij}}$ is a vector of ones of length $m_{ij}$.
Define $\bI_{\bm}$ as the block-diagonal identity matrix with $gh$ blocks as $\bI_{\bm} = \diag(\bI_{m_{11}}, \ldots, \bI_{m_{gh}})$, where each $\bI_{m_{ij}}$ is the $m_{ij}\times m_{ij}$ identity matrix.
We define $\bI_{\bm}^M = \diag(m_{11}\bI_{m_{11}}, \ldots, m_{gh}\bI_{m_{gh}})$, which can be regarded as the block-diagonal expansion of $\bM$ onto the observation level. 
Define $\bJ_{\bm}= \bone_{\bm}\bone_{\bm}^\top$ as the $n \times n$ all-ones matrix partitioned as $\bJ_{\bm} = (\bJ_{[m_{ij}: m_{i'j'}]})$, where each block $\bJ_{[m_{ij}: m_{i'j'}]}$ is an $m_{ij}\times m_{i'j'}$ matrix of ones. The block indexed by $(i,j)$ in the row direction and $(i',j')$ in the column direction appears at the $((i-1)h+j,\; (i'-1)h+j')$-th position of the overall $gh\times gh$ block structure.

We employ two systematic normalizations. The bar-normalization ($\;\tilde{\cdot}\;$) rescales each block by the square root of its size, producing unit-length blockwise all-one vectors and projection-type matrices. The tilde-normalization ($\;\bar{\cdot}\;$) instead divides by the block size, producing an averaging operator. Together, these definitions provide a unified way to represent block-structured matrices and streamline subsequent derivations.
Specifically, define
\begin{align*}
&\barbone_g=\frac{1}{\sqrt{g}}\bone_g,
\quad \barbone_h=\frac{1}{\sqrt{h}}\bone_h,
&& \tildebone_g=\frac{1}{g}\bone_g,
\quad
\tildebone_h=\frac{1}{h}\bone_h,
\\&\barbone_{\bm}=[\frac{1}{\sqrt{m_{11}}}\bone_{m_{11}}^\top,\ldots,\frac{1}{\sqrt{m_{gh}}}\bone_{m_{gh}}^\top]^\top,
&&\tildebone_{\bm}=[\frac{1}{m_{11}}\bone_{m_{11}}^\top,\ldots,\frac{1}{m_{gh}}\bone_{m_{gh}}^\top]^\top,
\\&\barbI_{\bm}=\diag(\frac{1}{\sqrt{m_{11}}}\bI_{m_{11}},\ldots,\frac{1}{\sqrt{m_{gh}}}\bI_{m_{gh}}),
&&\tildebI_{\bm}=\diag(\frac{1}{m_{11}}\bI_{m_{11}},\ldots,\frac{1}{m_{gh}}\bI_{m_{gh}}).
\end{align*}
It then follows that
$\barbJ{g} = \barbone_{g}\barbone_{g}^\top = \frac{1}{g} \bJ_g$, 
\,$\barbJ{h} = \barbone_{h}\barbone_{h}^\top = \frac{1}{h} \bJ_h$,
\,$\barbJ{\bm} = \barbone_{\bm}\barbone_{\bm}^\top = \barbI_{\bm} \bJ_{\bm} \barbI_{\bm}$, 
\,$\tildebJ_{\bm} =\tildebone_{\bm}\tildebone_{\bm}^\top	= \tildebI_{\bm} \bJ_{\bm} \tildebI_{\bm}$, 
and $\barbJ{\bm} \barbJ{\bm} = gh \barbJ{\bm}$.

The following identities in Lemma \ref{lma1} will be used throughout the paper. They play a key role in simplifying the analysis of crossed random effects models under unbalanced designs, and enable structural insight into likelihood-based inference and prediction. Other related mixed product properties are presented in Lemma~\ref{lma2}, given in the Supplementary Material with illustrative examples.

\begin{lemma}\label{lma1}
Suppose $\ba$ is a $g \times 1$ vector and $\bb$ is an $h \times 1$ vector.  
Let $\bP$ and $\bR$ be $g \times g$ matrices, and $\bQ$ and $\bS$ be $h \times h$ matrices.  
Then the following equalities hold.
		\begin{align*}
			&       \left[ (\ba \otimes \bb)_{\mathrm{row}} \circledast \barbone_{\bm}\right]^\top 
			\left[(\bP\otimes \bQ)_{\mathrm{cell}} \circledast \barbJ{\bm}\right]
			= [(\ba^\top \bP) \otimes (\bb^\top \bQ)]_{\mathrm{col}} \circledast \barbone_{\bm}^\top,  
			\\&
			\left[(\bP\otimes \bQ)_{\mathrm{cell}} \circledast \barbJ{\bm}\right]
			\left[(\bR\otimes \bS)_{\mathrm{cell}} \circledast \barbJ{\bm}\right]
			=[(\bP\bR)\otimes(\bQ\bS)]_{\mathrm{cell}}\circledast\barbJ{\bm},
			\\&
			\left[(\bP\otimes \bQ)_{\mathrm{cell}} \circledast \tildebJ_{\bm}\right]
			\left[(\bR\otimes \bS)_{\mathrm{cell}} \circledast \tildebJ_{\bm}\right]
			=[(\bP\otimes \bQ)\bM^{-1}
			(\bR\otimes \bS)]_{\mathrm{cell}}\circledast \tildebJ_{\bm},
            \\&
			\left[(\bP\otimes \bQ)_{\mathrm{cell}} \circledast \tildebJ_{\bm}\bI_{\bm}^f\right]
			\left[(\bR\otimes \bS)_{\mathrm{cell}} \circledast \tildebJ_{\bm}\right]
			=[(\bP\otimes \bQ)f(\bm)\bM^{-1}
			(\bR\otimes \bS)]_{\mathrm{cell}}\circledast \tildebJ_{\bm},
            \\&
			\left[(\bP\otimes \bQ)_{\mathrm{cell}} \circledast \bI_{\bm}^f\tildebJ_{\bm}\right]
			\left[(\bR\otimes \bS)_{\mathrm{cell}} \circledast \tildebJ_{\bm}\right]
			=[(\bP\otimes \bQ)\bM^{-1}
			(\bR\otimes \bS)]_{\mathrm{cell}}\circledast (\bI_{\bm}^f\tildebJ_{\bm}),        
        \end{align*}
where  $\bI_{\bm}^f=\diag(f(m_{11})\bI_{m_{11}},\ldots,f(m_{gh})\bI_{m_{gh}})$, $f(\bm)=\diag(f(m_{11}),\ldots,f(m_{gh}))$, and $f(m_{ij})$ is a scalar function.
\end{lemma}

\subsection{Matrix form}
Let $\by=[y_{111},\ldots,y_{ghm_{gh}}]^\top$, $\bX=[\bx_{111},\ldots,\bx_{ghm_{gh}}]^\top$,
$\balpha=[\alpha_1,\ldots,\alpha_g]^\top$, $\bbeta=[\beta_1,$\ldots$,\beta_h]^\top$, $\bgamma=[\gamma_{11},\ldots,\gamma_{gh}]^\top$, and $\be=[e_{111},\ldots,e_{ghm_{gh}}]^\top$. 
As before, the indices cycle in lexicographic order, with the rightmost index \( k \) varying fastest.
The  matrix formation of (\ref{two-way cross model}) is
\begin{align}\label{matrix two-way cross model}
	\by=\mu(\bX)+\bZ\bu+\be,
\end{align}
where $\bu=[\balpha^\top,\bbeta^\top,\bgamma^\top]^\top$ and $\bZ=[\bZ_1,\bZ_2,\bZ_3]$ with
$\bZ_1=(\bI_g\otimes\bone_h)_{\text{row}}\circledast\bone_{\bm}$,
$\bZ_2=(\bone_g\otimes\bI_h)_{\text{row}}\circledast\bone_{\bm}$ and
$\bZ_3=(\bI_g\otimes\bI_h)_{\text{row}}\circledast\bone_{\bm}$.
We let $\btheta=[\siga^2,\sigb^2,\siggama^2,\sige^2]^\top$, $\bG(\btheta)=\var(\bu)=\diag(\siga^2\bI_g,\sigb^2\bI_h,\siggama^2\bI_{gh})$ and $\var(\be)=\sige^2\bI_n$. 
The covariance matrix of the random effects, $\bD(\btheta) = \var(\bZ\bu)$, can be written as $\bZ\bG(\btheta)\bZ^\top$. Using the inner and matrix product relations in Lemma~\ref{lma1}, 
the variance-covariance matrix of $\by$, can be expressed using the Khatri-Rao product as  
\begin{align}
    \bV(\btheta)=\sige^2\bI_n+\bD(\btheta), \label{Matrix V}
\end{align}
where
\begin{align}
\bD(\btheta)=\sigma_\alpha^2(\bI_g\otimes\bJ_h)_{\text{cell}}\circledast\bJ_{\bm}+\sigma_\beta^2(\bJ_g\otimes\bI_h)_{\text{cell}}\circledast\bJ_{\bm}+\sigma_\gamma^2(\bI_g\otimes\bI_h)_{\text{cell}}\circledast\bJ_{\bm}. \label{Matrix D}
\end{align}
The variance-covariance matrix for the model~(\ref{two-way cross model}) without interaction can be obtained by omitting $\gamma$, $\siggama^2$, and $\bZ_3$ from the general form. In this paper, we primarily focus on the model with interaction. Results for the model~(\ref{two-way cross model}) without interaction arise as a special case 
and are provided in the Supplementary Material. 
Throughout, we occasionally omit the dependence on $\btheta$ in the notation for various matrices (such as $\bD$ and $\bV$) when this causes no ambiguity. 

\section{Inverse Covariance Matrix}\label{Sec3}
\subsection{Spectral decomposition of a modified covariance matrix}

In the covariance matrix (\ref{Matrix V}), $\bD$ captures the structured dependence induced by the row and column random effects, while $\sigma_e^2 \bI_n$ accounts for independent observation-level noise. Specifically, $\bD$ is a dense, low-rank matrix whose off-diagonal entries reflect correlations between observations sharing a common row or column factor level. In contrast, $\sigma_e^2 \bI_n$ is a diagonal, full-rank matrix that contributes homogeneous variability across all observations.  
In balanced designs, where each cell $(i,j)$ contains the same number of observations, the structure of $\bV$ remains relatively regular. However, in unbalanced designs with substantial variation in cell sizes ${m_{ij}}$, the interplay between $\sigma_e^2 \bI_n$ and $\bD $ becomes increasingly irregular. The lack of alignment between the uniform error structure and heterogeneous replication across cells, together with the fact that $\bD$ is dense, complicates both theoretical analysis and numerical computation. In particular, the absence of exploitable structure in the combined matrix makes it difficult to derive a closed-form expression for $\bV^{-1}$.

To address these challenges, we first consider a modified covariance matrix for the model with interaction~(\ref{two-way cross model}) of the form
\begin{equation} \label{eq:Vtilde}
\check{\bV} = \frac{\sigma_e^2}{m_U} \bI_{\bm}^M + \bD,
\quad\text{where  $\bI_{\bm}^M = \diag(m_{11}\bI_{m_{11}}, \ldots, m_{gh}\bI_{m_{gh}})$. }
\end{equation}
This construction replaces the observation-level noise component with a block-diagonal form that scales according to each cell size. The matrix $\check{\bV} $ retains the same correlation structure as $\bV$ through $\bD$, while introducing a smoothed error structure that can also reflect the local sampling density.
Importantly, when the design is balanced so that $m_{ij} \equiv m$ for all $i,j$, we have $\sige^2/m_U\bI_{\bm}^M = \sige^2 \bI_n$ and hence $\check{\bV}  = \bV $. Thus, $\check{\bV}$ preserves the key features of the original model while enabling a form that is analytically more tractable in unbalanced designs.

To study $\bV^{-1}$ in the analytically challenging general setting, we begin by deriving an explicit expression for $\check{\bV}^{-1}$. 
This provides a tractable starting point for studying $\bV^{-1}$. Since the only difference between $\check{\bV}$ and $\bV$ lies in the diagonal noise component, this discrepancy plays a key role in the approximation accuracy. All proof details can be found in the Appendix and Supplementary Material.

Let $\btaualpha{i} = [\taualpha{i}{1},\ldots,\taualpha{i}{g}]^\top$, $i = 1, \ldots, g - 1$, be $g-1$ orthonormal vectors, each of dimension $g \times 1$, which are orthogonal to $\bone_g$, so that
\begin{align}\label{orth: alpha}
	\bone_g^\top \btaualpha{i} &= \sum_{s=1}^g \taualpha{i}{s}= 0, 
	&&\text{and }
	\btaualpha{i'}^\top\btaualpha{i} = \sum_{s=1}^g \taualpha{i}{s}\taualpha{i'}{s}=
	\begin{cases}
		1 & \text{if } i = i' \\
		0 & \text{if } i \ne i'
	\end{cases} \quad \text{for } i, i' = 1, \ldots, g - 1.
\end{align}
Let $\btaubeta{j}=[\taubeta{j}{1},\ldots,\taubeta{j}{h}]^\top$, $j = 1, \ldots, h - 1$, be $h - 1$ orthonormal vectors, each of dimension $h \times 1$, which are orthogonal to $\bone_h$.
Similarly, let $\btaugamma{i}{j}{k}=[\btaugammaght{i}{j}{k}{1}{1},\ldots,\btaugammaght{i}{j}{k}{g}{h}]^\top$, $i = 1, \ldots, g;\, j = 1, \ldots, h;\, k = 1, \ldots, m_{ij} - 1$, be $\sum_{i=1}^g \sum_{j=1}^h (m_{ij} - 1)$ vectors, each of dimension $n \times 1$, where each component $\btaugammagh{i}{j}{k}{s}{t}$ is an $m_{st} \times 1$ vector defined as $
\btaugammagh{i}{j}{k}{s}{t}=[\taugamma{i}{j}{k}{s}{t}{1},\ldots,\taugamma{i}{j}{k}{s}{t}{m_{st}}]^\top$,
for $ s = 1, \ldots, g;\, t = 1, \ldots, h$.
Each $\btaugammagh{i}{j}{k}{s}{t}$ is orthonormal and orthogonal to $\bone_{m_{st}}$.
We can always find such sets of vectors.

Let $\bC$ be a $n\times n$ matrix partitioned as $\bC^\top = \col[\bC_1, \bC_2, \ldots, \bC_8]$, with $\bC_1$--$\bC_8$ defined in Table \ref{Matrix C:component}.
\begin{table}[]
	\centering
	\caption{Definition and dimension of blocks in the partition of $\bC$.}\label{Matrix C:component}
   \renewcommand{\arraystretch}{1.3} %
	\begin{tabular}{l p{7cm}l}
		\toprule
		Block & Definition& Dimension \\
		\midrule
		$\bC_1$ & $(\barbone_g\otimes\barbone_h)_{\text{row}}\circledast \barbone_{\bm}$ & $n \times 1$ \\
		\hline
		$\bC_2$ & $\col\left[(\barbone_g\otimes \barbone_h)_{\text{row}} \circledast \btaugamma{g}{h}{k}\right]_{k=1}^{m_{gh}-1}$ & $n \times (m_{gh} - 1)$   \\
		\hline
		$\bC_3$ & $\col\left[(\barbone_g \otimes \btaubeta{j})_{\text{row}} \circledast \barbone_{\bm}\right]_{j=1}^{h-1}$ & $n \times (h - 1)$  \\
		\hline
		$\bC_4$ & $\col\left[(\barbone_g\otimes \btaubeta{j})_{\text{row}} \circledast \btaugamma{g}{j}{k}\right]_{\substack{j=1,\ldots,h-1 \\ k=1,\ldots,m_{gj}-1}}$ & $n \times \sum_{j=1}^{h-1}(m_{gj} - 1)$ \\
		\hline
		$\bC_5$ & $\col\left[(\btaualpha{i} \otimes \barbone_h)_{\text{row}} \circledast \barbone_{\bm}\right]_{i=1}^{g-1}$ & $n \times (g - 1)$ \\
		\hline
		$\bC_6$ & $\col\left[(\btaualpha{i} \otimes \barbone_h)_{\text{row}} \circledast \btaugamma{i}{h}{k}\right]_{\substack{i=1,\ldots,g-1 \\ k=1,\ldots,m_{ih}-1}}$ & $n \times \sum_{i=1}^{g-1}(m_{ih} - 1)$ \\
		\hline
		$\bC_7$ & $\col\left[(\btaualpha{i} \otimes \btaubeta{j})_{\text{row}} \circledast \barbone_{\bm}\right]_{i=1,j=1}^{g-1,h-1}$ & $n \times (g - 1)(h - 1)$ \\
		\hline
		$\bC_8$ & $\col\left[(\btaualpha{i} \otimes \btaubeta{j})_{\text{row}} \circledast \btaugamma{i}{j}{k}\right]_{\substack{i=1,\ldots,g-1 \\ j=1,\ldots,h-1 \\ k=1,\ldots,m_{ij}-1}}$ & $n \times \sum_{i=1}^{g-1} \sum_{j=1}^{h-1} (m_{ij} - 1)$ \\
		\bottomrule
	\end{tabular}
\end{table}
The dimension of $\bC$ is $n \times n$, because
\begin{align*}
	n=\sumig\sumjh m_{ij}=&1+ (m_{gh} - 1)+  (h - 1)+\sum_{j=1}^{h-1}(m_{gj} - 1)+(g - 1)
	\\&+\sum_{i=1}^{g-1}(m_{ih} - 1)+(g - 1)(h - 1)+\sum_{i=1}^{g-1} \sum_{j=1}^{h-1} (m_{ij} - 1).
\end{align*}
We can now prove the following theorem which gives the spectral decomposition of the matrix  $ m_U \barbI_{\bm} \check{\bV} \barbI_{\bm}$.
A detailed proof is provided in Appendix~\ref{proof1}.
\begin{theorem} \label{theom1}
	Let $\btaualpha{i}$ ($i=1,\ldots,g-1$), $\btaubeta{j}$ ($j=1,\ldots,h-1$), and $\btaugamma{i}{j}{k}$ ($i=1,\ldots,g;\, j=1,\ldots,h;\, k=1,\ldots,m_{ij}-1$) be the orthonormal vectors defined above.
    Let $\bC$ be the $n \times n$ orthogonal matrix with $\bC^\top = [\bC_1, \ldots, \bC_8]$, where each block $\bC_i$ is specified in Table~\ref{Matrix C:component}. Then $\bC^\top \bC = \bI_n$. 
	 For the model with interaction in (\ref{two-way cross model}),  we have
	\begin{displaymath}
		\begin{aligned}
			& m_U\bC \barbI_{\bm} \check{\bV} \barbI_{\bm}\bC^\top \\&
            =
			\diag\bigl[
			\lambda_1,\, \lambda_0\bI_{[m_{gh}-1:m_{gh}-1]},\, \lambda_3\bI_{h-1},\, \lambda_0\bI_{[\sum_{j=1}^{h-1}(m_{gj}-1):\sum_{j=1}^{h-1}(m_{gj}-1)]},\, \lambda_5\bI_{g-1}, \\
			&\quad
			\lambda_0\bI_{[\sum_{i=1}^{g-1}(m_{ih}-1):\sum_{i=1}^{g-1}(m_{ih}-1)]},\, \lambda_7\bI_{(g-1)(h-1)},\, \lambda_0\bI_{[\sum_{i=1}^{g-1}\sum_{j=1}^{h-1}(m_{ij}-1):\sum_{i=1}^{g-1}\sum_{j=1}^{h-1}(m_{ij}-1)]}
			\bigr] \\
			&= \bLambda_\lambda,
		\end{aligned}
	\end{displaymath}
	where the five distinct characteristic roots of $ m_U \barbI_{\bm} \check{\bV} \barbI_{\bm}$ are 
    $\lambda_0=\sige^2$,
	$\lambda_1 = \sige^2+m_U\sigma_\gamma^2 + hm_U\sigma_\alpha^2 + gm_U\sigma_\beta^2$, 
	$\lambda_3 = \sige^2+m_U\sigma_\gamma^2 + hm_U\sigma_\alpha^2$,
	$\lambda_5 = \sige^2+m_U\sigma_\gamma^2 + gm_U\sigma_\beta^2$,
	and $\lambda_7 = \sige^2+m_U\sigma_\gamma^2$.
\end{theorem}

The purpose of weighting the modified covariance matrix  $\check{\bV}$ is to eliminate the effect of heterogeneous cell sizes, and enable the use of the Khatri--Rao product properties established in Lemmas~\ref{lma1} and~\ref{lma2}. After obtaining the eigenvalues of $ m_U \barbI_{\bm} \check{\bV} \barbI_{\bm}$ via Theorem \ref{theom1}, its inverse can be computed explicitly. The inverse $\check{\bV}^{-1}$ is then recovered by multiplying by the diagonal matrix $\barbI_{\bm}^{-1}$ as shown in Corollary \ref{corollary1}. Detailed derivations are provided in Appendix~\ref{proof2}.

\begin{corollary}\label{corollary1}
Let $\barbJ{a}^{0} = \bI_a$ and $\barbJ{a}^{1} = \barbJ{a}$. 
Consider the inverse of the modified covariance $\check{\bV}$ in \eqref{eq:Vtilde} for the model with interaction \eqref{two-way cross model}, denoted by $\check{\bV}^{-1}$. Then, from Theorem \ref{theom1}, we have
\begin{align}
\check{\bV}^{-1}
    	=\frac{m_U}{\lambda_0}\tildebI_{\bm}+m_U\left[\sum_{\bii=\bzero_{2}}^{\bone_{2}}\delta_{i_1i_2} (\barbJ{g}^{i_1}\otimes \barbJ{h}^{i_2})\right]_{\text{cell}}\circledast\tildebJ_{\bm},\label{inverse tildeV}
\end{align}
where  $\delta_{00}={1}/{\lambda_7}-1/{\lambda_0}$, $ \delta_{01}={1}/{\lambda_5}-{1}/{\lambda_7}$,$\delta_{10}={1}/{\lambda_3}-{1}/{\lambda_7}$ and $	\delta_{11}={1}/{\lambda_1}-{1}/{\lambda_3}-{1}/{\lambda_5}+{1}/{\lambda_7}$.
Here, the exponents $i_1$ and $i_2$ are also used as subscripts of $\delta$ for notational convenience and readability. They are treated as components of the binary vector $\bii$, with the subscripts written in reverse natural order to allow interpretation of $\bii$ as binary numbers. The summation $\sum_{\bii = \bzero_2}^{\bone_2}$ denotes the sum over all four binary vectors from $00$ to $11$.
\end{corollary}
We now make some remarks on Corollary~\ref{corollary1}.
\begin{enumerate}
    \item 
 In studying the asymptotic properties of estimators (such as ML and REML estimators), we usually require $g, h\to\infty$. In this situation, we can obtain an approximation to (\ref{inverse tildeV}), namely
\begin{equation}\label{vtildeinv approx}
    \check{\bV}^{-1}= \frac{m_U}{\lambda_0}\tildebI_{\bm}+(\frac{m_U}{\lambda_7}-\frac{m_U}{\lambda_0})(\bI_g\otimes\bI_h)_{\text{cell}}\circledast\tildebJ_{\bm}+
 \bO_{[n:n]}\left(\frac{1}{\min(g,h)m_L^2}\right), 
\end{equation}
where $\bO_{[n:n]}(a)$ denotes an $n \times n$ matrix with all entries of order $O(a)$.
Details of the derivation for~\eqref{vtildeinv approx} are provided in the proof of Lemma~\ref{lema: matrix kth product} in Appendix~\ref{appendix:lemma-theorem}. We retain $m_L$ in the remainder term even though it is finite, as this choice facilitates extensions to settings where $m_L \to \infty$. In this paper, however, we focus on the case where the minimum cell size $m_L$ is finite, which is more relevant.

\item 
 Under the balanced design where $m_{ij} = m$ for all $i$ and $j$, we have $\bV = \check{\bV}$ and hence $\bV^{-1} = \check{\bV}^{-1}$. In this case, we can rewrite (\ref{inverse tildeV}) as
\begin{align}
		\bV^{-1}
		=&\frac{1}{\lambda_0}[\bI_g\otimes\bI_h\otimes(\bI_m-\barbJ{m})]+\frac{1}{\lambda_7}[(\bI_g-\barbJ{g})\otimes(\bI_h-\barbJ{h})\otimes\barbJ{m}]\nonumber\\
        &+\frac{1}{\lambda_3}[(\bI_g-\barbJ{g})\otimes\barbJ{h}\otimes\barbJ{m}]
		+\frac{1}{\lambda_5}[\barbJ{g}\otimes(\bI_h-\barbJ{h})\otimes\barbJ{m}]+\frac{1}{\lambda_1}[\barbJ{g}\otimes\barbJ{h}\otimes\barbJ{m}],\label{bV inverse}
\end{align}
which coincides with the expression given in \cite{searle1979dispersion}. Thus, the exact inverse $\bV^{-1}$ under the balanced case can be viewed as a special case of the inverse of the modified covariance matrix $\check{\bV}^{-1}$. The expression (\ref{bV inverse}) has been utilized in recent studies (e.g., \cite{lyu2024increasing,lyu2025asymptotics}) to derive the asymptotic properties of ML and REML estimators as well as EBLUPs. 

\item For more general models, such as $k$-way crossed random-effects models with unbalanced designs ($k \ge 2$), the above approach continues to apply. One only needs to construct $2^{k}$ block matrices following the rule summarized in Table~\ref{Matrix C:component}. This yields a similar spectral decomposition for the modified covariance matrix and its inverse. The subsequent methodology for approximation and analysis of $\bV^{-1}$ can then be extended to $k$-way crossed random-effects models with unbalanced designs.
\end{enumerate}

\subsection{Inverse of covariance matrix $\bV$}

Under unbalanced designs ($m_{ij}\neq m$), $\bV^{-1}$ no longer admits an elegant closed form such as~\eqref{bV inverse}, since $\bV = \check{\bV}+\sige^2\bI_{\bm}^{\Delta}$ with $\bI_{\bm}^{\Delta} = \diag[(1 - m_{ij}/m_U)\bI_{m_{ij}}]$, and the heterogeneity in cell sizes cannot be eliminated from $\bI_{\bm}^{\Delta}$. In this section, we explore the analytic structure of $\bV^{-1}$ from that of $\check{\bV}^{-1}$ under asymptotic and non-asymptotic cases.

We first consider the asymptotic setting where $g,h\to\infty$ and $m_L$ is finite (or diverging to infinity). In this case, $\check{\bV}^{-1}$ admits a block-diagonal approximation as in~\eqref{vtildeinv approx}, which provides insight into the form of $\bV^{-1}$. 
The following approximation is critical for analyzing the property of likelihood-based estimators and inferences.

\begin{theorem}\label{theorem4}
Suppose $g,h\to\infty$ and all variance components are bounded. Then
\begin{align}
\bV^{-1}
=\frac{1}{\sige^2}\bI_n
-\frac{\siggama^2}{\sige^2}
\diag\left[
\frac{1}{\sige^2+m_{ij}\siggama^2}\bJ_{m_{ij}}
\right]
+\bO_{[n:n]}\!\left(
\frac{1}{\min(g^{1-\varepsilon},h^{1-\varepsilon})m_L^2}
\right),\label{Vinv Approximation}
\end{align}
where $0\le\varepsilon<1$. When $\sige^2/(2m_L^2\siggama^2)\le1$ or $m_{ij}=m$, we have $\varepsilon=0$.
\end{theorem}
 Theorem~\ref{theorem4} provides a block-diagonal representation of $\bV^{-1}$ that can substantially simplify the asymptotic analysis of likelihood-based estimation and prediction. This approximation retains only the two leading terms and permits the direct use of the properties of Khatri–Rao products in simplifying the associated estimating equations. Moreover, this exact analytic form greatly improves computational efficiency.  When $m_{ij}=m$, the expression in~\eqref{Vinv Approximation} simplifies to
\begin{align*}
\bV^{-1}
=\frac{1}{\sige^2}\bI_n
-\frac{\siggama^2}{\sige^2(\sige^2+m\siggama^2)}(\bI_g\otimes\bI_h\otimes\bJ_m)
+\bO_{[n:n]}\!\left(
\frac{1}{\min(g,h)m^2}
\right),
\end{align*}
which coincides with~\eqref{bV inverse} as $g,h\to\infty$.

Although the resulting form \eqref{Vinv Approximation} is surprisingly elegant and simple, deriving this approximation involves substantial  algebraic effort. In brief, by the Woodbury matrix identity,
\begin{align}
\bV^{-1}
=(\check{\bV}+\sige^2\bI_{\bm}^{\Delta})^{-1}
=\check{\bV}^{-1}
-\check{\bV}^{-1}\Bigl(\frac{1}{\sige^2}\bI_{\bm}^{\Delta-1}+\check{\bV}^{-1}\Bigr)^{-1}\check{\bV}^{-1}.
\label{Woodbury}
\end{align}
Given the explicit expression for $\check{\bV}^{-1}$ in Corollary~\ref{corollary1}, the key task is to obtain an analytic expression for $(1/{\sige^2}\bI_{\bm}^{\Delta-1}+\check{\bV}^{-1})^{-1}$.  Following equation~\eqref{vtildeinv approx} as $g,h\to\infty$, the term $1/{\sige^2}\bI_{\bm}^{\Delta-1}+\check{\bV}^{-1}$ can be decomposed into a block-diagonal matrix, denoted by $\bU_2$, together with a remainder term $\bO_{[n:n]}(1/[\min(g,h)m_L^2])$.
The classical truncated Neumann series, commonly used for approximate matrix inversion, cannot be applied here, since it requires the operator norm (or spectral radius) of the perturbation to satisfy $\|\bU_2^{-1}\bO_{[n:n]}(1/[\min(g,h)m_L^2])\|<1$. This condition fails in the asymptotic setting because the matrix dimension diverges as $g,h\to\infty$, while each entry of the remainder matrix is of order $1/[\min(g,h)m_L^2]$.  
We overcome this difficulty by exploiting the properties of Khatri–Rao products, which ensure that each entry of $[\bU_2^{-1}\bO_{[n:n]}\!\left(1/[\min(g,h)m_L^2]\right)]^l$ converges to zero for $l = 1, 2, \ldots, \infty$. The detailed proof is provided in Appendix~\ref{Proof Theorem4}.

We now consider the non-asymptotic setting, where $g$ and $h$ are finite.  
Following elementary expansion (see e.g.~\cite[Section 4.5]{jiang2010large}),
we write 
$\bV^{-1} = \check{\bV}^{-1} + \check{\bV}^{-1}(\check{\bV}-\bV)\bV^{-1}$  
and use this expansion to examine the extent of unbalancedness under which $\bV^{-1}$ can be approximated by $\check{\bV}^{-1}$ while preserving the structured form observed in the balanced case~\eqref{bV inverse}.  
The following result establishes an approximation under mildly unbalanced data. The detailed proof is provided in Appendix~\ref{proof3}.
\begin{theorem}\label{theorem2}
Let $\Delta = (m_U - m_L)/m_U$. Suppose that $g$, $h$, and $m_U$ are bounded, and that all variance components are bounded with $0 \le \Delta < 1/2$. Then the $r$th-order approximation to the inverse of the covariance matrix $\bV$ is
\begin{equation}\label{Approxi Vinv mild unbalance}
   \bV^{-1} = \left( \sum_{l=0}^{r} (-\sige^2)^{l} (\check{\bV}^{-1} \bI_{\bm}^{\Delta})^l \right) \check{\bV}^{-1} + \bO_{[n:n]}\!\left( \frac{\Delta^{r+1}}{(1 - \Delta)^{r+1}} \right),
\end{equation}
where $\bI_{\bm}^{\Delta} = \diag[(1 - m_{ij}/m_U)\bI_{m_{ij}}]$, and $\bO_{[n:n]}(a)$ denotes an $n\times n$ matrix with all entries of order $O(a)$.
\end{theorem}
We write the $r$th-order approximation as $\bV_{(r)}^{-1}$.
The approximation error is of order $\Delta^{r+1} / (1 - \Delta)^{r+1}$. 
When $\Delta = 0$ (corresponding to the balanced case), the expression reduces exactly to~\eqref{bV inverse}.  
For small $\Delta$ (i.e., close to zero and strictly less than $0.5$), the error term vanishes rapidly even for small $r$, and the resulting approximation maintains a structure closely resembling that of the balanced case.  
In such settings, truncating at $r=0$, $1$, or $2$ yields very accurate closed-form approximations to $\bV^{-1}$, given in~\eqref{vtildeinv approx}, \eqref{aproxl1}, and~\eqref{aproxl2}, respectively (Appendix \ref{proof3}).  
When $\Delta$ is relatively large (but still less than $0.5$), a higher-order $r$ may be required for satisfactory accuracy.   While the resulting expressions involve more terms, they remain closed form and computationally tractable due to the use of the Khatri--Rao product, which inherits many advantageous properties from the Kronecker product.

The approximation in Theorem \ref{theorem2} is only valid for mildly unbalanced data (i.e., $\Delta < 0.5$). However, in practice, datasets can be severely unbalanced with the minimum cell size much smaller than the maximum and hence $\Delta$ near $1$. In such cases, we can apply the Sherman--Morrison formula \citep{henderson1981deriving} to express $\bV^{-1}$ as a sequence of rank-one updates to $\check{\bV}^{-1}$, applied in lexicographic order over $(i,j,k)$.
A detailed proof is provided in Appendix~\ref{proof4}.

\begin{theorem}\label{theorem3}
Define $\bE_{ijk} = \sigma_e^2 \left(1 - {m_{ij}}/{m_U} \right) \bI_{n(ijk)}$, where $\bI_{n(ijk)}$ denotes the $n \times n$ diagonal matrix with a one at the diagonal position corresponding to the lexicographic index of $(i,j,k)$, and zeros elsewhere.  
For each $(a,b,c)$ with $a \in \{1,\ldots,g\}$, $b \in \{1,\ldots,h\}$, and $c \in \{1,\ldots,m_{ab}\}$, 
define $\mathcal{L}_{ab} = \{(i,j):\ i < a \text{ or } (i = a,\ j < b)\}$ and
$\bW_{abc+1} = \check{\bV} 
+ \sum_{(i,j) \in \mathcal{L}_{ab}} \sum_{k=1}^{m_{ij}} \bE_{ijk}
+ \sum_{k=1}^{c} \bE_{abk},
$
with $\bW_{111} = \check{\bV}$. Assume that each matrix $\bW_{abc}$ is nonsingular. Then
\begin{align*}
\bW_{111}^{-1} = \check{\bV}^{-1} \text{ and } \quad \bW_{abc+1}^{-1} = \bW_{abc}^{-1} - \kappa_{abc} \bW_{abc}^{-1} \bE_{abc} \bW_{abc}^{-1},
\end{align*}
where $\kappa_{abc} =1/[1+\tr(\bW_{abc}^{-1}\bE_{abc})]$, and where $abc+1$ means to increment $abc$ to the next element in the lexicographic order.
In particular, the inverse of $\bV$ is  
\begin{align}
\bV^{-1} = \bW_{gh\,m_{gh}}^{-1} - \kappa_{ghm_{gh}} \bW_{gh\,m_{gh}}^{-1} \bE_{gh\,m_{gh}} \bW_{gh\,m_{gh}}^{-1}.\label{recirsove bV inve}
\end{align}
\end{theorem}

Theorem~\ref{theorem3} shows that $\bV^{-1}$ can be constructed recursively from $\check{\bV}^{-1}$ through a sequence of rank-one Sherman--Morrison updates. The resulting expression involves only $\check{\bV}^{-1}$ and a collection of rank-one matrices $\bE_{abc}$, with typical terms of the form $\check{\bV}^{-1}\bE_{abc}\check{\bV}^{-1}$, possibly appearing in nested products. This representation remains valid regardless of the extent of unbalancedness and whether or not an asymptotic setting ($g,h\to\infty$) is considered. 
Although  the recursive form in~\eqref{recirsove bV inve}  is more efficient than direct inversion—since it relies solely on $\check{\bV}^{-1}$ and rank-one matrices—it still poses computational challenges when $g$ and $h$ are large. For asymptotic analyses, it is therefore preferable to employ the approximation in~\eqref{Vinv Approximation}, which yields a simple analytic form.

In contrast, in non-asymptotic settings where $g$ and $h$ are fixed, the recursive representation does not suffer from such computational limitations. Under mildly unbalanced data, the approximation in~\eqref{Approxi Vinv mild unbalance} provides accurate and tractable closed-form expressions. For severely unbalanced designs, however, the recursive formulation in~\eqref{recirsove bV inve} is particularly useful. Although it lacks a closed analytic expression, it remains well-defined in finite dimensions and can be efficiently implemented. Compared with~\eqref{Approxi Vinv mild unbalance}, which involves multiplication by the diagonal matrix $\bI_{\bm}^{\Delta}$, the recursive approach is often more convenient both analytically and computationally, as it relies only on simple rank-one matrices $\bE_{abc}$.

\section{Numerical Results}\label{Sec4}

This section assesses the accuracy of the approximations in~\eqref{Vinv Approximation} and~\eqref{Approxi Vinv mild unbalance} to $\bV^{-1}$ under crossed random-effects models with both asymptotic and non-asymptotic unbalanced designs, respectively. Throughout, data are generated from the model in~\eqref{two-way cross model}, and the variance components are fixed at $(\sigma_\alpha^2,\sigma_\beta^2,\sigma_{\gamma}^2,\sigma_e^2)=(5,7,3,4)$. For each simulation setting, we generate $N=200$ independent replicates. The two cases below differ only in the grid size $(g,h)$ and in how the cell sizes $\{m_{ij}\}$ (and, in Case~2, the truncation order $r$) are varied.

\noindent
\textit{Case 1 (Asymptotic setting).}
We study increasing grids with $(g,h)\in\{10,20,50,70,100\}\times\{15,25,45,75,95\}$, yielding 25 configurations. For each configuration, the cell sizes are generated as $m_{ij}\sim\mathrm{Uniform}(1,15)$. This produces 25 simulation settings.

\noindent
\textit{Case 2 (Non-asymptotic setting).}
We consider relatively small grids with $(g,h)\in\{10,20\}\times\{15,25\}$. For each configuration, we examine eight levels of unbalance determined by $m_L\in\{10,20\}$ and $\Delta\in\{0.15,0.25,0.35,0.45\}$, generating cell sizes as $m_{ij}\sim\mathrm{Uniform}(m_L,\, m_L/(1-\Delta))$. This yields 32 settings in total. For each replicate in Case~2, the approximation to $\bV^{-1}$ is computed using truncation orders $r\in\{0,1,2,3,4,5\}$.

To assess the accuracy of the approximations, we compute the Average Inversion Residual (AIR) for both the estimated $\hat{\bV}^{-1}$ from~\eqref{Vinv Approximation} and $\hat{\bV}_{(r)}^{-1}$ from~\eqref{Approxi Vinv mild unbalance}, defined as
\begin{align*}
\mathrm{AIR}
=\frac{1}{N}\sum_{b=1}^{N}\frac{1}{n^{(b)}}\bigl\|
\bV^{(b)}\hat{\bV}^{-1(b)}-\bI_{n^{(b)}}\bigr\|_F,
\end{align*}
where $\|\cdot\|_F$ denotes the Frobenius norm, 
a $(b)$ superscript denotes a quantity related to the $b$th independent replicate,
$\hat{\bV}^{-1(b)}$ represents either $\hat{\bV}^{-1(b)}$ or $\hat{\bV}^{-1(b)}_{(r)}$ depending on the approximation used, and $n^{(b)}$ denotes the number of observations in the $b$th replicate.
\begin{figure}[!h]
    \centering
    \begin{subfigure}[b]{0.45\linewidth}
        \centering
        \includegraphics[width=\linewidth]{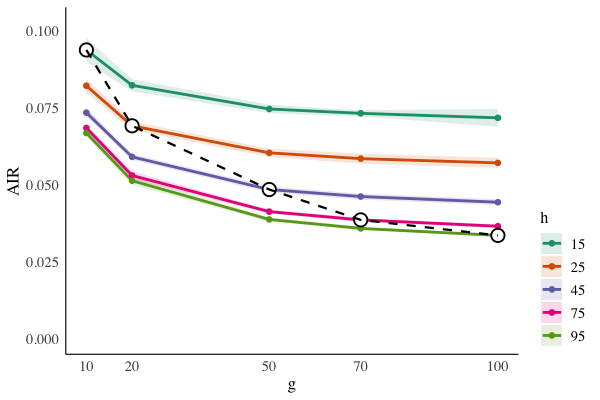}
        \subcaption{Case 1: Asymptotic setting}\label{fig:sub1}
    \end{subfigure}
    \hfill
    \begin{subfigure}[b]{0.45\linewidth}
        \centering
        \includegraphics[width=\linewidth]{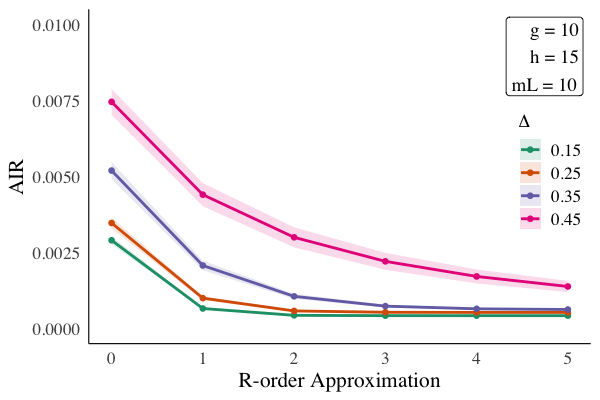}
        \subcaption{Case 2: Non-asymptotic setting}\label{fig:sub2}
    \end{subfigure}
    \caption{Average Inversion Residual (AIR) for both asymptotic and non-asymptotic settings, with shaded areas representing one standard deviation across $N=200$ replications.}
    \label{fig:air}
\end{figure}

Figure~\ref{fig:air} presents the simulation results, where the lines show the mean AIR values and the shaded regions indicate one standard deviation across the replications.

In Figure~\ref{fig:sub1}, for fixed small $h$ (e.g., $h=15$ or $25$), the AIR decreases as $g$ increases from 10 to 50 and then stabilizes as $g$ increases further to 100. In contrast, for fixed large $h$ (e.g., $h=75$ or $95$), the AIR continues to decrease as $g$ increases from 10 to 100, with a less pronounced flattening. This pattern arises because the remainder order of each entry in the approximation depends on $O(1/[\min(g,h)m_L^2])$. Since $\sige^2/(2m_L^2\siggama^2)<1$, Theorem~\ref{theorem4} implies $\varepsilon=0$. When $h$ is fixed and small, the remainder order depends primarily on $h$, causing the AIR to initially decrease and then level off as $g$ grows. Conversely, when $h$ is large, the remainder order depends on $g$, resulting in a continued decrease in the AIR. When both $g$ and $h$ increase simultaneously (the dashed line with circle markers), $\min(g,h)$ increases and the remainder order becomes smaller, leading to a more pronounced reduction in the AIR. These results are consistent with the theoretical guarantees of Theorem~\ref{theorem4}.

In Figure~\ref{fig:sub2}, for fixed $g$, $h$, $m_L$, and $\Delta$, the AIR decreases monotonically as the truncation order $r$ increases from 0 to 5, confirming Theorem~\ref{theorem2}: higher-order approximations yield improved accuracy in estimating the inverse covariance matrix under mildly unbalanced designs. Moreover, holding $g$, $h$, $m_L$, and $r$ fixed, the AIR decreases as $\Delta$ becomes smaller, indicating that the approximation improves as the design becomes more balanced. This trend is again consistent with the theoretical results established in Theorem~\ref{theorem2}.
The full simulation results are in the Supplementary Material.

\section{Discussion}\label{Sec5}

This paper develops a unified and tractable framework for approximating and analyzing the inverse of the covariance matrix in linear mixed models with crossed random effects under unbalanced designs. A modified covariance matrix is introduced whose inverse admits an exact spectral decomposition, providing an elegant and accurate approximation to the true inverse in asymptotic unbalanced settings. For non-asymptotic designs, the inverse of the modified matrix also provides a close analytical approximation under mildly unbalanced data, similar to that in the balanced case. For severely unbalanced data, the exact inverse can be constructed through a sequence of rank-one updates to the modified matrix inverse, offering a computationally feasible alternative to direct inversion.  These results support both theoretical developments and scalable inference in settings where standard methods become intractable.

A key innovation of our approach lies in the introduction of the Khatri--Rao product to model covariance structures under unbalanced designs. While Kronecker products have long served as a foundational tool in balanced crossed designs, their utility diminishes when replication varies across cells. The Khatri--Rao product enables a structured yet flexible formulation that accommodates cell-specific replication, and preserves key algebraic properties that facilitate matrix decomposition, inversion, and asymptotic analysis. To our knowledge, this is the first application of the Khatri--Rao product to such models, and it opens new possibilities for modelling and computation in a wide range of applications involving unbalanced data. A promising direction for future work is to extend the asymptotic theory of likelihood-based estimators and predictions to settings with bounded unbalanced cell sizes, building on the approximations and recursive structures established here. In principle, our approach can also extend to more general models, such as $k$-way crossed random-effects models with unbalanced designs ($k>2$).

\begin{appendix}
\section{Proofs}\label{appendix:proof}
The proofs of Theorems and Corollaries are presented in this section. The supporting lemmas used in these proofs are  proved in Sections~\ref{appendix:lema}.

 \subsection{Proof of Theorem~\ref{theom1}}\label{proof1}
\begin{proof}
Dividing the  modified covariance matrix $\check{\bV}$ by the corresponding cell size over the maximum cell size, we have
	\begin{align}
		m_U\barbI_{\bm}\check{\bV}\barbI_{\bm}
		&=\sige^2 \bI_{n}+m_U\sigma_\alpha^2(\bI_g\otimes\bJ_h)_{\text{cell}}\circledast\barbJ{\bm}+ m_U\sigma_\beta^2(\bJ_g\otimes\bI_h)_{\text{cell}}\circledast\barbJ{\bm}\nonumber\\&\qquad
        + m_U\sigma_\gamma^2(\bI_g\otimes\bI_h)_{\text{cell}}\circledast\barbJ{\bm}. \label{scalar-cov-random}
	\end{align}
	The weighted modified covariance matrix of the random effects $m_U\barbI_{\bm}\check{\bV}\barbI_{\bm}$ has four terms: one in $\bI_n$, one in $(\bI_g\otimes\bJ_h)_{\text{cell}}\circledast\barbJ{\bm}$, one in $(\bJ_g\otimes\bI_h)_{\text{cell}}\circledast\barbJ{\bm}$, and one in $(\bI_g\otimes\bI_h)_{\text{cell}}\circledast\barbJ{\bm}$. It is therefore clear that any vector which eliminates    $(\bI_g\otimes\bJ_h)_{\text{cell}}\circledast\barbJ{\bm}$, $(\bJ_g\otimes\bI_h)_{\text{cell}}\circledast\barbJ{\bm}$ and $(\bI_g\otimes\bI_h)_{\text{cell}}\circledast\barbJ{\bm}$ must be a characteristic vector of $m_U\barbI_{\bm}\check{\bV}\barbI_{\bm}$ with associated root $\sige^2$.
	
	We begin by considering vectors of the form $(\btaualpha{i} \otimes \btaubeta{j})_{\text{row}} \circledast \btaugamma{i}{j}{k}$, where $\btaualpha{i}$, $\btaubeta{j}$ and $\btaugamma{i}{j}{k}$ are orthogonal to $\bone_g$, $\bone_h$, and $\bone_{\bm}$, respectively. Since the last three terms in (\ref{scalar-cov-random}) involve $\barbJ{\bm} = \barbone_{\bm}\barbone_{\bm}^\top$, it follows from Lemma~\ref{lma2} that the last three terms are eliminated by  $(\btaualpha{i} \otimes \btaubeta{j})_{\text{row}} \circledast \btaugamma{i}{j}{k}$. These vectors are eigenvectors with eigenvalue $\sige^2 $ with multiplicity $\sum_{i=1}^{g-1}\sum_{j=1}^{h-1}(m_{ij}-1)$.
    
    Similarly, the last three terms are also eliminated by other combinations, such as $(\barbone_{g} \otimes \barbone_{h})_{\text{row}} \circledast \btaugamma{i}{j}{k}$, $(\barbone_{g} \otimes \btaubeta{j})_{\text{row}} \circledast \btaugamma{i}{j}{k}$ and $(\btaualpha{i} \otimes \barbone_{h})_{\text{row}} \circledast \btaugamma{i}{j}{k}$, because these combinations all include a common component $\btaugamma{i}{j}{k}$. These vectors are likewise eigenvectors of $m_U\barbI_{\bm}\check{\bV}\barbI_{\bm}$  with eigenvalue $\sige^2$, and with multiplicities $m_{gh}-1$,  $\sum_{j=1}^{h-1}(m_{gj}-1)$, and $\sum_{i=1}^{g-1}(m_{ih}-1)$, respectively.

    Next, we examine vectors of the form $(\btaualpha{i} \otimes \btaubeta{j})_{\text{row}} \circledast \barbone_{\bm}$, where $\btaualpha{i}$ and $\btaubeta{j}$ are orthogonal to $\bone_g$ and $\bone_h$, respectively.  It follows from Lemma~\ref{lma2} that the $\sigma_\alpha^2$ and $\sigma_\beta^2$ terms (involving $\barbJ{g}=\barbone_g\barbone_g^\top$ and $\barbJ{h}=\barbone_h\barbone_h^\top$) vanish, while the $\siggama^2$ term remains active due to its alignment with $\bone_{\bm}$, as indicated by Lemma~\ref{lma1}.  Therefore, these vectors are eigenvectors with eigenvalue $\sige^2 + m_U\sigma_\gamma^2$, with total multiplicity $(g - 1)(h - 1)$.
    
	Similarly, consider the vectors $(\btaualpha{i} \otimes \barbone_{h})_{\text{row}} \circledast \barbone_{\bm}$, where  $\btaualpha{i}$ is orthogonal to $\bone_g$, eliminating the $\sigma_\beta^2$ term (involving $\barbJ{g}=\barbone_g\barbone_g^\top$). The $\siga^2$ and $\siggama^2$ terms are aligned with their corresponding projection directions, so the $\sigma_\alpha^2$ and $\sigma_\gamma^2$ terms remain. According to Lemma~\ref{lma1} and \ref{lma2}, the eigenvalue associated with these vectors is $\sige^2+m_U\sigma_\gamma^2+hm_U\sigma_\alpha^2$, and the multiplicity is $g - 1$. Likewise, vectors of the form $(\barbone_{g} \otimes \btaubeta{j})_{\text{row}} \circledast \barbone_{\bm}$ yield eigenvalue $\sige^2+m_U\sigma_\gamma^2+gm_U \sigma_\beta^2$. Since $\btaubeta{j}$  is orthogonal to  $\bone_h$, the $\sigma_\alpha^2$ term (involving $\barbJ{h}=\barbone_h\barbone_h^\top$) vanishes, while the $\sigb^2$ and $\siggama^2$ terms are preserved by the alignment of $\barbone_{g}$ and $\barbone_{\bm}$. The multiplicity is $h - 1$.
	    
	Finally, the fully aligned direction $(\barbone_{g} \otimes \barbone_{h})_{\text{row}} \circledast \barbone_{\bm}$ lies within the span of all four projection directions. Consequently, all terms in  (\ref{scalar-cov-random})  contribute to the eigenvalue, and by Lemma~\ref{lma1} and \ref{lma2}, the resulting eigenvalue is $\sige^2 + m_U\sigma_\gamma^2+hm_U \sigma_\alpha^2 + gm_U \sigma_\beta^2$, with multiplicity one.
	
\end{proof}
\subsection{Proof of Corollary~\ref{corollary1}}\label{proof2}
\begin{proof}
    
We have $m_U\barbI_{\bm}\check{\bV}\barbI_{\bm}=\bC^\top\bLambda_\lambda\bC$ and $\bC^\top\bC=\bI$ following Theorem~\ref{theom1}.
The inverse is
\begin{align*}
	(m_U\barbI_{\bm}\check{\bV}\barbI_{\bm})^{-1}=\bC^\top\bLambda_{1/\lambda}\bC,
\end{align*}
where 
\begin{align*}
	&\bLambda_{1/\lambda}\\&=
			\diag\bigl[
			1/\lambda_1,\, 1/\lambda_0\bI_{[m_{gh}-1:m_{gh}-1]},\, 1/\lambda_3\bI_{h-1},\, 1/\lambda_0\bI_{[\sum_{j=1}^{h-1}(m_{gj}-1):\sum_{j=1}^{h-1}(m_{gj}-1)]},\, 1/\lambda_5\bI_{g-1}, \\
			&\quad
			1/\lambda_0\bI_{[\sum_{i=1}^{g-1}(m_{ih}-1):\sum_{i=1}^{g-1}(m_{ih}-1)]},\, 1/\lambda_7\bI_{(g-1)(h-1)},\, 1/\lambda_0\bI_{[\sum_{i=1}^{g-1}\sum_{j=1}^{h-1}(m_{ij}-1):\sum_{i=1}^{g-1}\sum_{j=1}^{h-1}(m_{ij}-1)]}
			\bigr].
            \end{align*}
Thus
\begin{align*}
	(m_U\barbI_{\bm}\check{\bV}\barbI_{\bm})^{-1}=\bC^\top\bLambda_{1/\lambda}\bC
	=&\frac{1}{\lambda_1}\bC_1\bC_1^\top+\frac{1}{\lambda_3}\bC_3\bC_3^\top+\frac{1}{\lambda_5}\bC_5\bC_5^\top+\frac{1}{\lambda_7}\bC_7\bC_7^\top
    \\&+\frac{1}{\lambda_0}(\bC_2\bC_2^\top+\bC_4\bC_4^\top+\bC_6\bC_6^\top+\bC_8\bC_8^\top).
\end{align*} 
We have
\begin{align}\label{C1C1}
	\bC_1\bC_1^\top&=[(\barbone_{g}\otimes\barbone_{h})_{\text{row}}\circledast\barbone_{\bm}][(\barbone_{g}\otimes\barbone_{h})_{\text{row}}\circledast\barbone_{\bm}]^\top
	=(\barbJ{g}\otimes\barbJ{h})_{\text{cell}}\circledast\barbJ{\bm}.
\end{align}
Now, consider the product
\begin{align}\label{Cequation1}
	\bC^\top\bC= \bC_1\bC_1^\top+\bC_2\bC_2^\top+\bC_3\bC_3^\top+\bC_4\bC_4^\top+\bC_5\bC_5^\top+\bC_6\bC_6^\top+\bC_7\bC_7^\top+\bC_8\bC_8^\top=\bI_n.
\end{align}
Premultiplying (\ref{Cequation1}) by $(\bJ_g\otimes\bI_h)_{\text{cell}}\circledast\barbJ{\bm}$ and following Lemma~\ref{lma1}, we have
\begin{align*}
	& [(\bJ_g\otimes\bI_h)_{\text{cell}}\circledast\barbJ{\bm}]\bC_1\bC_1^\top=g\bC_1\bC_1^\top=g(\barbJ{g}\otimes\barbJ{h})_{\text{cell}}\circledast\barbJ{\bm},
	&& [(\bJ_g\otimes\bI_h)_{\text{cell}}\circledast\barbJ{\bm}]\bC_2\bC_2^\top=\bzero_{[n:n]},
	\\& [(\bJ_g\otimes\bI_h)_{\text{cell}}\circledast\barbJ{\bm}]\bC_3\bC_3^\top=g\bC_3\bC_3^\top,
	&& [(\bJ_g\otimes\bI_h)_{\text{cell}}\circledast\barbJ{\bm}]\bC_4\bC_4^\top=\bzero_{[n:n]},
	\\   & [(\bJ_g\otimes\bI_h)_{\text{cell}}\circledast\barbJ{\bm}]\bC_5\bC_5^\top=\bzero_{[n:n]},
	&& [(\bJ_g\otimes\bI_h)_{\text{cell}}\circledast\barbJ{\bm}]\bC_6\bC_6^\top=\bzero_{[n:n]},
	\\& [(\bJ_g\otimes\bI_h)_{\text{cell}}\circledast\barbJ{\bm}]\bC_7\bC_7^\top=\bzero_{[n:n]},
	&& [(\bJ_g\otimes\bI_h)_{\text{cell}}\circledast\barbJ{\bm}]\bC_8\bC_8^\top=\bzero_{[n:n]}.
\end{align*}
It then follows that
\begin{align}\label{C3C3}
	\bC_3\bC_3^\top=(\barbJ{g}\otimes\bI_{h})_{\text{cell}}\circledast\barbJ{\bm}-(\barbJ{g}\otimes\barbJ{h})_{\text{cell}}\circledast\barbJ{\bm}
	=\left[\barbJ{g}\otimes(\bI_h-\barbJ{h})\right]_{\text{cell}}\circledast\barbJ{\bm}.
\end{align}
Similarly, premultiplying (\ref{Cequation1}) by $(\bI_g\otimes\bJ_h)_{\text{cell}}\circledast\barbJ{\bm}$ and following Lemma~\ref{lma1}, we have
\begin{align*}
	& [(\bI_g\otimes\bJ_h)_{\text{cell}}\circledast\barbJ{\bm}]\bC_1\bC_1^\top=h\bC_1\bC_1^\top=h(\barbJ{g}\otimes\barbJ{h})_{\text{cell}}\circledast\barbJ{\bm},
	&& [(\bI_g\otimes\bJ_h)_{\text{cell}}\circledast\barbJ{\bm}]\bC_2\bC_2^\top=\bzero_{[n:n]},
	\\& [(\bI_g\otimes\bJ_h)_{\text{cell}}\circledast\barbJ{\bm}]\bC_3\bC_3^\top=\bzero_{[n:n]},
	&& [(\bI_g\otimes\bJ_h)_{\text{cell}}\circledast\barbJ{\bm}]\bC_4\bC_4^\top=\bzero_{[n:n]},
	\\   & [(\bI_g\otimes\bJ_h)_{\text{cell}}\circledast\barbJ{\bm}]\bC_5\bC_5^\top=h\bC_5\bC_5^\top,
	&& [(\bI_g\otimes\bJ_h)_{\text{cell}}\circledast\barbJ{\bm}]\bC_6\bC_6^\top=\bzero_{[n:n]},
	\\& [(\bI_g\otimes\bJ_h)_{\text{cell}}\circledast\barbJ{\bm}]\bC_7\bC_7^\top=\bzero_{[n:n]},
	&& [(\bI_g\otimes\bJ_h)_{\text{cell}}\circledast\barbJ{\bm}]\bC_8\bC_8^\top=\bzero_{[n:n]}.
\end{align*}
It then follows that
\begin{align}\label{C5C5}
	\bC_5\bC_5^\top=(\bI_g\otimes\barbJ{h})_{\text{cell}}\circledast\barbJ{\bm}-(\barbJ{g}\otimes\barbJ{h})_{\text{cell}}\circledast\barbJ{\bm}
	=\left[(\bI_g-\barbJ{g})\otimes\barbJ{h}\right]_{\text{cell}}\circledast\barbJ{\bm}.
\end{align}
In the same manner, premultiplying (\ref{Cequation1}) by $(\bI_g\otimes\bI_h)_{\text{cell}}\circledast\barbJ{\bm}$ and following Lemma~\ref{lma1}, we have
\begin{align*}
	& [(\bJ_g\otimes\bI_h)_{\text{cell}}\circledast\barbJ{\bm}]\bC_1\bC_1^\top=\bC_1\bC_1^\top,
	&& [(\bJ_g\otimes\bI_h)_{\text{cell}}\circledast\barbJ{\bm}]\bC_2\bC_2^\top=\bzero_{[n:n]},
	\\& [(\bJ_g\otimes\bI_h)_{\text{cell}}\circledast\barbJ{\bm}]\bC_3\bC_3^\top=\bC_3\bC_3^\top,
	&& [(\bJ_g\otimes\bI_h)_{\text{cell}}\circledast\barbJ{\bm}]\bC_4\bC_4^\top=\bzero_{[n:n]},
	\\   & [(\bJ_g\otimes\bI_h)_{\text{cell}}\circledast\barbJ{\bm}]\bC_5\bC_5^\top=\bC_5\bC_5^\top,
	&& [(\bJ_g\otimes\bI_h)_{\text{cell}}\circledast\barbJ{\bm}]\bC_6\bC_6^\top=\bzero_{[n:n]},
	\\& [(\bJ_g\otimes\bI_h)_{\text{cell}}\circledast\barbJ{\bm}]\bC_7\bC_7^\top=\bC_7\bC_7^\top,
	&& [(\bJ_g\otimes\bI_h)_{\text{cell}}\circledast\barbJ{\bm}]\bC_8\bC_8^\top=\bzero_{[n:n]}.
\end{align*}
Then combining (\ref{C1C1}), (\ref{C3C3}) and (\ref{C5C5}), we have 
\begin{align}
	\bC_7\bC_7^\top=&(\bI_{g}\otimes\bI_{h})_{\text{cell}}\circledast\barbJ{\bm}-\bC_1\bC_1^\top-\bC_3\bC_3^\top-\bC_5\bC_5^\top \nonumber
	\\
	=&(\bI_{g}\otimes\bI_{h})_{\text{cell}}\circledast\barbJ{\bm}-(\barbJ{g}\otimes\barbJ{h})_{\text{cell}}\circledast\barbJ{\bm}
	\nonumber\\& -\left[\barbJ{g}\otimes(\bI_h-\barbJ{h})\right]_{\text{cell}}\circledast\barbJ{\bm}-\left[(\bI_g-\barbJ{g})\otimes\barbJ{h}\right]_{\text{cell}}\circledast\barbJ{\bm}
	\nonumber\\
	=&
	\left[(\bI_g-\barbJ{g})\otimes(\bI_h-\barbJ{h})\right]_{\text{cell}}\circledast\barbJ{\bm}.\label{C7C7}
\end{align}
Hence
\begin{align}
	\bC_2\bC_2^\top+\bC_4\bC_4^\top+\bC_6\bC_6^\top+\bC_8\bC_8^\top 
	&=\bI_n-\bC_1\bC_1^\top-\bC_3\bC_3^\top-\bC_5\bC_5^\top-\bC_7\bC_7^\top\nonumber
    \\&=\bI_n-(\bI_{g}\otimes\bI_{h})_{\text{cell}}\circledast\barbJ{\bm}\nonumber
	\\&=
	(\bI_{g}\otimes\bI_{h})_{\text{cell}}\circledast(\bI_{\bm}-\barbJ{\bm}).
\end{align}
It follows that an explicit representation of $(m_U\barbI_{\bm}\check{\bV}\barbI_{\bm})^{-1}$ is then
\begin{align}
	(m_U\barbI_{\bm}\check{\bV}\barbI_{\bm})^{-1}
	=&\frac{1}{\lambda_1}(\barbJ{g}\otimes\barbJ{h})_{\text{cell}}\circledast\barbJ{\bm}
	+\frac{1}{\lambda_3}\left[\barbJ{g}\otimes(\bI_h-\barbJ{h})\right]_{\text{cell}}\circledast\barbJ{\bm} \nonumber
	\\&
	+\frac{1}{\lambda_5}\left[(\bI_g-\barbJ{g})\otimes\barbJ{h}\right]_{\text{cell}}\circledast\barbJ{\bm}
	+\frac{1}{\lambda_7}\left[(\bI_g-\barbJ{g})\otimes(\bI_h-\barbJ{h})\right]_{\text{cell}}\circledast\barbJ{\bm}\nonumber
    \\&
    +\frac{1}{\lambda_0}\left[\bI_n-(\bI_g\otimes\bI_h)_{\text{cell}}\circledast\barbJ{\bm}\right]
    .\label{Dm}
\end{align}
Let $\barbJ{a}^{0}=\bI_a$, $\barbJ{a}^1=\barbJ{a}$,  and
\begin{align*}
	& \delta_{00}=\frac{1}{\lambda_7}-\frac{1}{\lambda_0},\quad\delta_{01}=\frac{1}{\lambda_5}-\frac{1}{\lambda_7},\quad\delta_{10}=\frac{1}{\lambda_3}-\frac{1}{\lambda_7},
	\quad
	\delta_{11}=\frac{1}{\lambda_1}-\frac{1}{\lambda_3}-\frac{1}{\lambda_5}+\frac{1}{\lambda_7}.
\end{align*}
We can rewrite 
\begin{align*}
	(m_U\barbI_{\bm}\check{\bV}\barbI_{\bm})^{-1}=\frac{1}{\lambda_0}\bI_n+[\sum_{\bii=\bzero}^{\bone}\delta_{i_1i_2} (\barbJ{g}^{i_1}\otimes \barbJ{h}^{i_2})]_{\text{cell}}\circledast\barbJ{\bm},
\end{align*}
where the exponents, $i_1$ and $i_2$ are also used as subscripts to $\delta$ for notational convenience and readability. They are treated as components of the binary vector $\bii$, with the subscripts written in reverse natural order to allow interpretation of $\bii$ as binary numbers.
The summation $\sum_{\bii=\bzero}^{\bone}$ denotes the sum over the $4$ binary vectors from 00 to 11.

Following Lemma \ref{lma1}, we then have 
\begin{align*}
	\check{\bV}^{-1}&
	=\frac{m_U}{\lambda_0}\tildebI_{\bm}+m_U\barbI_{\bm}\{[\sum_{\bii=\bzero}^{\bone}\delta_{i_1i_2} (\barbJ{g}^{i_1}\otimes \barbJ{h}^{i_2})]_{\text{cell}}\circledast\barbJ{\bm}\}\barbI_{\bm}
    \\&
	=\frac{m_U}{\lambda_0}\tildebI_{\bm}+m_U[\sum_{\bii=\bzero}^{\bone}\delta_{i_1i_2} (\barbJ{g}^{i_1}\otimes \barbJ{h}^{i_2})]_{\text{cell}}\circledast\tildebJ_{\bm}.
\end{align*}
\end{proof}

\subsection{Proof of Theorem~\ref{theorem4}}\label{Proof Theorem4}
\begin{proof}
We write
\begin{align*}
\bV=\check{\bV}+\sige^2\bI_n-\frac{\sige^2}{m_U}\bI_{\bm}^M=\check{\bV}+\sige^2\bI_{\bm}^{\Delta},
\end{align*}
where $\bI_{\bm}^{\Delta}= \diag[(1 - m_{ij}/m_U)\bI_{m_{ij}}]$. Following the Woodbury matrix identity, we have
\begin{align*}
\bV^{-1}=(\check{\bV}+\sige^2\bI_{\bm}^{\Delta})^{-1}=\check{\bV}^{-1}-\check{\bV}^{-1}\Bigl(\frac{1}{\sige^2}\bI_{\bm}^{\Delta-1}+\check{\bV}^{-1}\Bigr)^{-1}\check{\bV}^{-1}.
\end{align*}
Recalling Corollary \ref{corollary1}, we rewrite \eqref{inverse tildeV} as
\begin{align*}
\check{\bV}^{-1}
=&\frac{m_U}{\lambda_0}\tildebI_{\bm}
+m_U\Bigl(\frac{1}{\lambda_7}-\frac{1}{\lambda_0}\Bigr)(\bI_g\otimes\bI_h)_{\text{cell}}\circledast\tildebJ_{\bm}
+m_U\Bigl(\frac{1}{\lambda_5}-\frac{1}{\lambda_7}\Bigr)(\barbJ{g}\otimes\bI_h)_{\text{cell}}\circledast\tildebJ_{\bm}
\\
&\quad
+m_U\Bigl(\frac{1}{\lambda_3}-\frac{1}{\lambda_7}\Bigr)(\bI_g\otimes\barbJ{h})_{\text{cell}}\circledast\tildebJ_{\bm}
+m_U\Bigl(\frac{1}{\lambda_1}-\frac{1}{\lambda_3}-\frac{1}{\lambda_5}+\frac{1}{\lambda_7}\Bigr)(\barbJ{g}\otimes\barbJ{h})_{\text{cell}}\circledast\tildebJ_{\bm}
\\
&=\bU_1+\mathcal{R},
\end{align*}
where
\begin{align}
    \bU_1
&=\frac{m_U}{\lambda_0}\tildebI_{\bm}
+m_U\Bigl(\frac{1}{\lambda_7}-\frac{1}{\lambda_0}\Bigr)(\bI_g\otimes\bI_h)_{\text{cell}}\circledast\tildebJ_{\bm}\nonumber
\\&
=\diag\left[\frac{m_U}{m_{ij}\sige^2}\bI_{m_{ij}}\right]
+\diag\left[\frac{m_U}{m_{ij}^2}\Bigl(\frac{1}{\lambda_7}-\frac{1}{\lambda_0}\Bigr)\bJ_{m_{ij}}\right],\label{U1}
\end{align}
and
\begin{align}
\mathcal{R}
=&\,m_U\Bigl(\frac{1}{\lambda_5}-\frac{1}{\lambda_7}\Bigr)(\barbJ{g}\otimes\bI_h)_{\text{cell}}\circledast\tildebJ_{\bm}
+m_U\Bigl(\frac{1}{\lambda_3}-\frac{1}{\lambda_7}\Bigr)(\bI_g\otimes\barbJ{h})_{\text{cell}}\circledast\tildebJ_{\bm}\nonumber
\\
&\quad
+m_U\Bigl(\frac{1}{\lambda_1}-\frac{1}{\lambda_3}-\frac{1}{\lambda_5}+\frac{1}{\lambda_7}\Bigr)(\barbJ{g}\otimes\barbJ{h})_{\text{cell}}\circledast\tildebJ_{\bm}.\label{R term}
\end{align}
To simplify the calculation, let
\begin{align*}
a_{ij}=\frac{m_U}{(m_U-m_{ij})\sige^2},\qquad
b_{ij}=\frac{m_U}{m_{ij}\sige^2},\qquad
c_{ij}=\frac{m_U}{m_{ij}^2}\Bigl(\frac{1}{\lambda_7}-\frac{1}{\lambda_0}\Bigr).
\end{align*}
We write
\begin{align*}
\frac{1}{\sige^2}\bI_{\bm}^{\Delta-1}+\check{\bV}^{-1}
&=\diag\left[\frac{m_U}{\sige^2(m_U-m_{ij})}\bI_{m_{ij}}\right]+\bU_1+\mathcal{R}
\\
&=\diag[a_{ij}\bI_{m_{ij}}]+\diag[b_{ij}\bI_{m_{ij}}+c_{ij}\bJ_{m_{ij}}]+\mathcal{R}
\\
&=\diag[(a_{ij}+b_{ij})\bI_{m_{ij}}+c_{ij}\bJ_{m_{ij}}]+\mathcal{R}
\\
&=\bU_2+\mathcal{R},
\end{align*}
where $\bU_2=\diag[(a_{ij}+b_{ij})\bI_{m_{ij}}+c_{ij}\bJ_{m_{ij}}]$. Hence
\begin{align*}
\Bigl(\frac{1}{\sige^2}\bI_{\bm}^{\Delta-1}+\check{\bV}^{-1}\Bigr)^{-1}
=(\bU_2+\mathcal{R})^{-1}=(\bI+\bU_2^{-1}\mathcal{R})^{-1}\bU_2^{-1}.
\end{align*}
We have
\begin{align}
    \bU_2^{-1}
&=\left(\diag[(a_{ij}+b_{ij})\bI_{m_{ij}}+c_{ij}\bJ_{m_{ij}}]\right)^{-1}\nonumber
\\
&=\diag\left[\{(a_{ij}+b_{ij})\bI_{m_{ij}}+c_{ij}\bJ_{m_{ij}}\}^{-1}\right]\nonumber
\\
&=\diag\left[\frac{1}{a_{ij}+b_{ij}}\left(\bI_{m_{ij}}-\frac{c_{ij}}{a_{ij}+b_{ij}+m_{ij}c_{ij}}\bJ_{m_{ij}}\right)\right].\label{U2inv}
\end{align}
Let $s(g,h)=\sqrt{\log\{\min(g,h)\}}$. As $g,h\to\infty$, we have $s(g,h)\to\infty$. Following the Neumann series identity, we write
\begin{align*}
(\bI+\bU_2^{-1}\mathcal{R})^{-1}=\bI+\lim_{g,h\to\infty}\sum_{l=1}^{s(g,h)}(-\bU_2^{-1}\mathcal{R})^l.
\end{align*}
Therefore,
\begin{align*}
\Bigl(\frac{1}{\sige^2}\bI_{\bm}^{\Delta-1}+\check{\bV}^{-1}\Bigr)^{-1}
=\bU_2^{-1}+\Bigl\{\lim_{g,h\to\infty}\sum_{l=1}^{s(g,h)}(-\bU_2^{-1}\mathcal{R})^l\Bigr\}\bU_2^{-1}.
\end{align*}
Combining the above equations, we rewrite $\bV^{-1}$ as
\begin{align*}
\bV^{-1}
&=\bU_1+\mathcal{R}-(\bU_1+\mathcal{R})\Bigl[\bU_2^{-1}+\Bigl\{\lim_{g,h\to\infty}\sum_{l=1}^{s(g,h)}(-\bU_2^{-1}\mathcal{R})^l\Bigr\}\bU_2^{-1}\Bigr](\bU_1+\mathcal{R})
\\
&=\bU_1-\bU_1\bU_2^{-1}\bU_1+\mathcal{R}-\bU_1\bU_2^{-1}\mathcal{R}
-\mathcal{R}\bU_2^{-1}\bU_1-\mathcal{R}\bU_2^{-1}\mathcal{R}
\\
&\quad
-\bU_1\Bigl\{\lim_{g,h\to\infty}\sum_{l=1}^{s(g,h)}(-\bU_2^{-1}\mathcal{R})^l\Bigr\}\bU_2^{-1}\bU_1
-\bU_1\Bigl\{\lim_{g,h\to\infty}\sum_{l=1}^{s(g,h)}(-\bU_2^{-1}\mathcal{R})^l\Bigr\}\bU_2^{-1}\mathcal{R}
\\
&\quad
-\mathcal{R}\Bigl\{\lim_{g,h\to\infty}\sum_{l=1}^{s(g,h)}(-\bU_2^{-1}\mathcal{R})^l\Bigr\}\bU_2^{-1}\bU_1
-\mathcal{R}\Bigl\{\lim_{g,h\to\infty}\sum_{l=1}^{s(g,h)}(-\bU_2^{-1}\mathcal{R})^l\Bigr\}\bU_2^{-1}\mathcal{R}
\\
&=\bU_1-\bU_1\bU_2^{-1}\bU_1+\mathcal{R}-\bU_1\bU_2^{-1}\mathcal{R}
-\mathcal{R}\bU_2^{-1}\bU_1-\mathcal{R}\bU_2^{-1}\mathcal{R}
\\
&\quad
-\lim_{g,h\to\infty}\sum_{l=1}^{s(g,h)}\bU_1(-\bU_2^{-1}\mathcal{R})^l\bU_2^{-1}\bU_1
+\lim_{g,h\to\infty}\sum_{l=1}^{s(g,h)}\bU_1(-\bU_2^{-1}\mathcal{R})^{l+1}
\\
&\quad
-\lim_{g,h\to\infty}\sum_{l=1}^{s(g,h)}\mathcal{R}(-\bU_2^{-1}\mathcal{R})^l\bU_2^{-1}\bU_1
+\lim_{g,h\to\infty}\sum_{l=1}^{s(g,h)}\mathcal{R}(-\bU_2^{-1}\mathcal{R})^{l+1}.
\end{align*}
We now show that all terms involving $\mathcal{R}$ converge to zero as $g,h \to \infty$.  
Since both $\bU_2^{-1}$ and $-\mathcal{R}$ are nonnegative matrices, and $\bU_2$ is block-diagonal with block positive coefficients depending on $m_{ij}$, we simplify the analysis by introducing a finite positive constant that bounds these coefficients uniformly.  
Following Lemma~\ref{lma matrix ineq}, this allows each block to be treated independently of $m_{ij}$, leading to
\begin{align*}
-\bU_2^{-1}\bR
\le \text{constant}\times
\Big[
(\barbJ{g}\otimes\bI_h)_{\text{cell}}\circledast\tildebJ_{\bm}
+(\bI_g\otimes\barbJ{h})_{\text{cell}}\circledast\tildebJ_{\bm}
+(\barbJ{g}\otimes\barbJ{h})_{\text{cell}}\circledast\tildebJ_{\bm}
\Big].
\end{align*}
Applying Lemma~\ref{lema: matrix kth product}, as $g,h \to \infty$ we obtain
\begin{align*}
\Big[
(\barbJ{g}\otimes\bI_h)_{\text{cell}}\circledast\tildebJ_{\bm}
+(\bI_g\otimes\barbJ{h})_{\text{cell}}\circledast\tildebJ_{\bm}
+(\barbJ{g}\otimes\barbJ{h})_{\text{cell}}\circledast\tildebJ_{\bm}
\Big]^l
=\bO_{[n:n]}\!\left(\frac{1}{\min(g,h)m_L^{l+1}}+\frac{3^{l}-2}{ghm_L^{l+1}}\right).
\end{align*}
Hence, by Lemmas~\ref{Sandwich bound} and~\ref{matrix remiander order}, all terms involving $\mathcal{R}$ vanish as $g,h \to \infty$.  
Therefore, we obtain the asymptotic approximation to $\bV^{-1}$.
\begin{align*}
\bV^{-1}=\bU_1-\bU_1\bU_2^{-1}\bU_1+\bO_{[n:n]}\left(\frac{1}{\min(g^{1-\varepsilon},h^{1-\varepsilon})m_L^2}\right),
\end{align*}
where $0\le \varepsilon<1$. After substituting $\bU_1$ and $\bU_2^{-1}$ by \eqref{U1} and \eqref{U2inv}, we have
\begin{align*}
&\bV^{-1}
\\
&=
\diag\left[b_{ij}\bI_{m_{ij}}+c_{ij}\bJ_{m_{ij}}\right]
-\diag\left[b_{ij}\bI_{m_{ij}}+c_{ij}\bJ_{m_{ij}}\right]
\\
&\qquad\times
\diag\left[\frac{1}{a_{ij}+b_{ij}}\left(\bI_{m_{ij}}-\frac{c_{ij}}{a_{ij}+b_{ij}+m_{ij}c_{ij}}\bJ_{m_{ij}}\right)\right]
\times
\diag\left[b_{ij}\bI_{m_{ij}}+c_{ij}\bJ_{m_{ij}}\right]
\\
&\qquad+\bO_{[n:n]}\left(\frac{1}{\min(g^{1-\varepsilon},h^{1-\varepsilon})m_L^2}\right)
\\
&=\diag\left[b_{ij}\bI_{m_{ij}}+c_{ij}\bJ_{m_{ij}}\right]
-\diag\left[\frac{b_{ij}}{a_{ij}+b_{ij}}\bI_{m_{ij}}+\frac{a_{ij}c_{ij}}{(a_{ij}+b_{ij})(a_{ij}+b_{ij}+m_{ij}c_{ij})}\bJ_{m_{ij}}\right]
\\
&\qquad\times
\diag\left[b_{ij}\bI_{m_{ij}}+c_{ij}\bJ_{m_{ij}}\right]
+\bO_{[n:n]}\left(\frac{1}{\min(g^{1-\varepsilon},h^{1-\varepsilon})m_L^2}\right)
\\
&=
\diag\left[b_{ij}\bI_{m_{ij}}+c_{ij}\bJ_{m_{ij}}\right]
-\diag\left[\frac{b^2_{ij}}{a_{ij}+b_{ij}}\bI_{m_{ij}}
+\frac{2a_{ij}b_{ij}c_{ij}+b_{ij}^2c_{ij}+m_{ij}a_{ij}c_{ij}^2+m_{ij}b_{ij}c_{ij}^2}{(a_{ij}+b_{ij})(a_{ij}+b_{ij}+m_{ij}c_{ij})}\bJ_{m_{ij}}\right]
\\
&\qquad
+\bO_{[n:n]}\left(\frac{1}{\min(g^{1-\varepsilon},h^{1-\varepsilon})m_L^2}\right)
\\
&=\diag\left[\frac{a_{ij}b_{ij}}{a_{ij}+b_{ij}}\bI_{m_{ij}}
+\frac{a_{ij}^2c_{ij}}{(a_{ij}+b_{ij})(a_{ij}+b_{ij}+m_{ij}c_{ij})}\bJ_{m_{ij}}\right]
+\bO_{[n:n]}\left(\frac{1}{\min(g^{1-\varepsilon},h^{1-\varepsilon})m_L^2}\right).
\end{align*}
Recall that
\begin{align*}
a_{ij}=\frac{m_U}{(m_U-m_{ij})\sige^2},\quad
b_{ij}=\frac{m_U}{m_{ij}\sige^2},\quad
c_{ij}=\frac{m_U}{m_{ij}^2}\Bigl(\frac{1}{\sige^2+m_U\siggama^2}-\frac{1}{\sige^2}\Bigr)
=\frac{-m_U^2\siggama^2}{m_{ij}^2\sige^2(\sige^2+m_U\siggama^2)}.
\end{align*}
We calculate that
\begin{align*}
&a_{ij}b_{ij}=\frac{m_U}{(m_U-m_{ij})\sige^2}\times\frac{m_U}{m_{ij}\sige^2}
=\frac{m_U^2}{m_{ij}(m_U-m_{ij})\sige^4},
\\
&a_{ij}+b_{ij}=\frac{m_U}{(m_U-m_{ij})\sige^2}+\frac{m_U}{m_{ij}\sige^2}
=\frac{m_U^2}{m_{ij}(m_U-m_{ij})\sige^2},
\end{align*}
and so
\begin{align*}
\frac{a_{ij}b_{ij}}{a_{ij}+b_{ij}}=\frac{1}{\sige^2}.
\end{align*}
We further calculate that
\begin{align*}
\frac{a_{ij}^2}{a_{ij}+b_{ij}}=\frac{m_{ij}}{(m_U-m_{ij})\sige^2},
\end{align*}
and so
\begin{align*}
a_{ij}+b_{ij}+m_{ij}c_{ij}
&=\frac{m_U^2(\sige^2+m_{ij}\siggama^2)}{m_{ij}(m_U-m_{ij})\sige^2(\sige^2+m_U\siggama^2)}.
\end{align*}
This gives
\begin{align*}
&\frac{a_{ij}^2c_{ij}}{(a_{ij}+b_{ij})(a_{ij}+b_{ij}+m_{ij}c_{ij})}
=\frac{a_{ij}^2}{a_{ij}+b_{ij}}\times\frac{1}{a_{ij}+b_{ij}+m_{ij}c_{ij}}\times c_{ij}
=\frac{-\siggama^2}{\sige^2(\sige^2+m_{ij}\siggama^2)}.
\end{align*}
Therefore,
\begin{align*}
\bV^{-1}
&=\diag\left[
\frac{a_{ij}b_{ij}}{a_{ij}+b_{ij}}\bI_{m_{ij}}
+\frac{a_{ij}^2c_{ij}}{(a_{ij}+b_{ij})(a_{ij}+b_{ij}+m_{ij}c_{ij})}\bJ_{m_{ij}}
\right]
+\bO_{[n:n]}\left(\frac{1}{\min(g^{1-\varepsilon},h^{1-\varepsilon})m_L^2}\right)
\\
&=\diag\left[
\frac{1}{\sige^2}\bI_{m_{ij}}
-\frac{\siggama^2}{\sige^2(\sige^2+m_{ij}\siggama^2)}\bJ_{m_{ij}}
\right]
+\bO_{[n:n]}\left(\frac{1}{\min(g^{1-\varepsilon},h^{1-\varepsilon})m_L^2}\right)
\\
&=\frac{1}{\sige^2}\bI_n-\frac{\siggama^2}{\sige^2}\diag\left[\frac{1}{\sige^2+m_{ij}\siggama^2}\bJ_{m_{ij}}\right]
+\bO_{[n:n]}\left(\frac{1}{\min(g^{1-\varepsilon},h^{1-\varepsilon})m_L^2}\right).
\end{align*}
\end{proof}

\subsection{Proof of Theorem~\ref{theorem2}}\label{proof3}
\begin{proof}

Following elementary expansion (see e.g.~\cite[Section 4.5]{jiang2010large}),
\begin{align*}
    \bV^{-1} &= \check{\bV}^{-1} + \bV^{-1} - \check{\bV}^{-1}
    = \check{\bV}^{-1} + \check{\bV}^{-1}( \check{\bV}-\bV )\bV^{-1},
\end{align*}
we obtain the matrix expansion \cite[Lemma 5.4]{das2004mean}
\begin{align*}
    \bV^{-1} = \left[ \sum_{l=0}^{r} \left\{ \check{\bV}^{-1}(\check{\bV}-\bV) \right\}^l \right] \check{\bV}^{-1}
    + \left\{ \check{\bV}^{-1}(\check{\bV}-\bV) \right\}^{r+1} \bV^{-1}, \quad \text{for } r=0,1,2,\ldots.
\end{align*}
By the spectral theorem \cite[Theorem 2.5.1]{horn2012matrix}, $\bD$ has an orthogonal decomposition $\bD = \bU \bLambda_D \bU^\top$ where $\bLambda_D = \diag(\lambda_{D1}, \ldots, \lambda_{Dn})$ with $\lambda_{Di} \ge 0$ and at most $\mathrm{Rank}(\bD)$ nonzero eigenvalues. It follows that
\begin{align*}
\bV =\sige^2\bI_n+\bD= \bU (\sige^2 \bI_n + \bLambda_D) \bU^\top
\end{align*}
has eigenvalues $\sige^2 + \lambda_{Di} > 0$ and is therefore positive definite \cite[Observation 7.1.4]{horn2012matrix}. Hence, $\bV^{-1}$ exists and satisfies
\begin{equation}\label{bound Vinv}
\| \bV^{-1} \|_2 = \max_i \left\{ \frac{1}{\sige^2 + \lambda_{Di}} \right\} \le \frac{1}{\sige^2},
\end{equation}
which implies that $\bV^{-1} = \bO_{[n:n]}(1/\sige^2)$, where $\bO_{[n:n]}(a)$ denotes an $n \times n$ matrix with entries of order $O(a)$.

We write 
\begin{align*}
\check{\bV}-\bV=\sige^2(\frac{1}{m_U}\bI_{\bm}^M-\bI_n)=-\sige^2\bI_{\bm}^{\Delta}, 
\end{align*}
where $\bI_{\bm}^{\Delta} = \diag((1 - m_{11}/m_U)\bI_{m_{11}}, \ldots, (1 - m_{gh}/m_U)\bI_{m_{gh}})$.
Then we have 
\begin{align*}
\check{\bV}^{-1}(\check{\bV}-\bV)=-\sige^2\check{\bV}^{-1}\bI_{\bm}^{\Delta}
=-m_U\tildebI_{\bm}\bI_{\bm}^{\Delta}-m_U\sige^2\left[\sum_{\bii=\bzero_{[2:1]}}^{\bone_{[2:1]}}\delta_{i_1i_2} (\barbJ{g}^{i_1}\otimes \barbJ{h}^{i_2})\right]_{\text{cell}}\circledast(\tildebJ_{\bm}\bI_{\bm}^{\Delta}).
\end{align*}
From the Khatri--Rao structure, we have 
\begin{equation}\label{JJORDER}
    \begin{split}
    &\delta_{00}(\bI_g \otimes \bI_h) = \bO_{[gh:gh]}(\sige^{-2}), 
\\&\delta_{10}(\barbJ{g} \otimes \bI_h) = \bO_{[gh:gh]}(g^{-1}(\sige^2+m_U\siggama^2)^{-1}), 
\\&
\delta_{10}(\bI_g \otimes \barbJ{h}) = \bO_{[gh:gh]}(h^{-1}(\sige^2+m_U\siggama^2)^{-1}),
\\&\delta_{10}(\barbJ{g} \otimes\barbJ{h})=\bO_{[gh:gh]}(g^{-1}h^{-1}(\sige^2+m_U\siggama^2)^{-1}),
    \end{split}
\end{equation}
and 
\begin{align}
\tildebJ_{\bm}\bI_{\bm}^{\Delta}=\bO_{[n:n]}(m_L^{-2}\Delta).\label{I-DELTA}
\end{align}
Suppose all variance components are finite and  $\Delta=(m_U-m_L)/m_U<0.5$ (i.e.~the data are not severely unbalanced), then we have
\begin{align*}
&\check{\bV}^{-1}(\check{\bV}-\bV) 
\\&
=-m_U\tildebI_{\bm}\bI_{\bm}^{\Delta}-
m_U\sige^2\left[\sum_{\bii=\bzero_{[2:1]}}^{\bone_{[2:1]}}\delta_{i_1i_2} (\barbJ{g}^{i_1}\otimes \barbJ{h}^{i_2})\right]_{\text{cell}}\circledast(\tildebJ_{\bm}\bI_{\bm}^{\Delta})
\\&=-m_U\tildebI_{\bm}\bI_{\bm}^{\Delta}-m_U(\bI_g\otimes\bI_h)_{\text{cell}}\circledast(\tildebJ_{\bm}\bI_{\bm}^{\Delta})+\bO_{[n:n]}\left(\frac{\Delta}{(1-\Delta)\min(g,h)m_L (1+m_U\siggama^2/\sige^2)}\right)
\\&=-m_U\tildebI_{\bm}\bI_{\bm}^{\Delta}+\bO_{[n:n]}\left(\frac{\Delta}{(1-\Delta)m_L}\right)+\bO_{[n:n]}\left(\frac{\Delta}{(1-\Delta)\min(g,h)m_L (1+m_U\siggama^2/\sige^2)}\right)
\\&=\bO_{[n:n]}\left(\frac{\Delta}{(1-\Delta)}\right).
\end{align*}
Consider the power expansion
\begin{align*}
\{\check{\bV}^{-1}(\check{\bV}-\bV)\}^l=\sum_{\bii=\bzero_{[l:1]}}^{\bone_{[l:1]}}\bF^{(i_1)}\cdots\bF^{(i_l)},
\end{align*}
where 
$\bF^{(0)}=m_U\tildebI_{\bm}\bI_{\bm}^{\Delta}$ and  $ \bF^{(1)}=m_U\sige^2\left[\sum_{\bii=\bzero}^{\bone}\delta_{i_1i_2} (\barbJ{g}^{i_1}\otimes \barbJ{h}^{i_2})\right]_{\text{cell}}\circledast(\tildebJ_{\bm}\bI_{\bm}^{\Delta})$.
We provide detailed computations for $l = 2$ and $l = 3$ using the properties of Khatri--Rao products in Lemma~\ref{lma1}.

For $l = 2$, we obtain
\begin{align*}
\left\{\check{\bV}^{-1}(\check{\bV}-\bV)\right\}^2 = \bF^{(0)} \bF^{(0)} + \bF^{(0)} \bF^{(1)} + \bF^{(1)} \bF^{(0)} + \bF^{(1)} \bF^{(1)},
\end{align*}
where
\begin{align*}
&\bF^{(0)} \bF^{(0)} = \left(m_U\tildebI_{\bm} \bI_{\bm}^{\Delta}\right)^2, \\
&\bF^{(0)} \bF^{(1)} = m_U \sige^2 \left[\sum_{\bii = \bzero_{[2:1]}}^{\bone_{[2:1]}} \delta_{i_1 i_2} \left( \barbJ{g}^{i_1} \otimes \barbJ{h}^{i_2} \right) \right]_{\text{cell}} \circledast \left( m_U\tildebI_{\bm} \bI_{\bm}^{\Delta} \tildebJ_{\bm} \bI_{\bm}^{\Delta} \right), \\
&\bF^{(1)} \bF^{(0)} = m_U \sige^2 \left[\sum_{\bii = \bzero_{[2:1]}}^{\bone} \delta_{i_1 i_2} \left( \barbJ{g}^{i_1} \otimes \barbJ{h}^{i_2} \right) \right]_{\text{cell}} \circledast \left(m_U \tildebJ_{\bm} \bI_{\bm}^{\Delta} \tildebI_{\bm} \bI_{\bm}^{\Delta} \right), \\
&\bF^{(1)} \bF^{(1)} = m_U^2 \sige^4 \left[ \sum_{\bii = \bzero_{[4:1]}}^{\bone_{[4:1]}} \delta_{i_1 i_2} \delta_{i_3 i_4} \left( \barbJ{g}^{i_1} \otimes \barbJ{h}^{i_2} \right) \bM^{\Delta} \left( \barbJ{g}^{i_3} \otimes \barbJ{h}^{i_4} \right) \right]_{\text{cell}} \circledast \left( \tildebJ_{\bm} \bI_{\bm}^{\Delta} \right),
\end{align*}
with
\begin{align*}
\bM^{\Delta} = \diag\left( \frac{m_U - m_{11}}{m_U m_{11}}, \ldots, \frac{m_U - m_{gh}}{m_U m_{gh}} \right) = \bO_{[gh:gh]} \left( \Delta m_L^{-1} \right).
\end{align*}
Assuming that the variance components are finite and $\Delta = (m_U - m_L)/m_U < 0.5$ (i.e.~the data are not severely unbalanced), and using (\ref{JJORDER}) and (\ref{I-DELTA}), and
$m_U\tildebI_{\bm}\bI_{\bm}^{\Delta}\tildebJ_{\bm}\bI_{\bm}^{\Delta}=\tildebJ_{\bm}\bI_{\bm}^{\Delta}m_U\tildebI_{\bm}\bI_{\bm}^{\Delta}=\bO_{[n:n]}(m_L^{-2}\Delta^2/(1-\Delta))$, we have 
\begin{align*}
    \bF^{(0)}\bF^{(0)}&=(m_U\tildebI_{\bm}\bI_{\bm}^{\Delta})^2=  \bO_{[n:n]}\left(\frac{\Delta^2}{(1-\Delta)^2}\right),
\\ \bF^{(0)}\bF^{(1)}&=m_U\sige^2\left[\sum_{\bii=\bzero_{[2:1]}}^{\bone_{[2:1]}}\delta_{i_1i_2} (\barbJ{g}^{i_1}\otimes \barbJ{h}^{i_2})\right]_{\text{cell}}\circledast(m_U\tildebI_{\bm}\bI_{\bm}^{\Delta}\tildebJ_{\bm}\bI_{\bm}^{\Delta})
 \\&=-m_U(\bI_g\otimes\bI_h)_{\text{cell}}\circledast(m_U\tildebI_{\bm}\bI_{\bm}^{\Delta}\tildebJ_{\bm}\bI_{\bm}^{\Delta})+   \bO_{[n:n]}\left(\frac{\Delta^2}{(1-\Delta)^2\min(g,h)m_L (1+m_U\siggama^2/\sige^2)}\right)
 \\&
 =\bO_{[n:n]}\left(\frac{\Delta^2}{(1-\Delta)^2m_L }\right) + \bO_{[n:n]}\left(\frac{\Delta^2}{(1-\Delta)^2\min(g,h)m_L (1+m_U\siggama^2/\sige^2)}\right)
 \\&
 =\bO_{[n:n]}\left(\frac{\Delta^2}{(1-\Delta)^2m_L }\right),
\\
\bF^{(1)}\bF^{(0)}&=m_U\sige^2\left[\sum_{\bii=\bzero_{[2:1]}}^{\bone_{[2:1]}}\delta_{i_1i_2} (\barbJ{g}^{i_1}\otimes \barbJ{h}^{i_2})\right]_{\text{cell}}\circledast(m_U\tildebJ_{\bm}\bI_{\bm}^{\Delta}\tildebI_{\bm}\bI_{\bm}^{\Delta})
 \\&=-m_U(\bI_g\otimes\bI_h)_{\text{cell}}\circledast(m_U\tildebI_{\bm}\bI_{\bm}^{\Delta}\tildebJ_{\bm}\bI_{\bm}^{\Delta})+   \bO_{[n:n]}\left(\frac{\Delta^2}{(1-\Delta)^2\min(g,h)m_L (1+m_U\siggama^2/\sige^2)}\right)
 \\&
 =\bO_{[n:n]}\left(\frac{\Delta^2}{(1-\Delta)^2m_L }\right) + \bO_{[n:n]}\left(\frac{\Delta^2}{(1-\Delta)^2\min(g,h)m_L (1+m_U\siggama^2/\sige^2)}\right)
 \\&
= \bO_{[n:n]}\left(\frac{\Delta^2}{(1-\Delta)^2m_L }\right),
\\
 \bF^{(1)}\bF^{(1)}&=m_U^2\sige^4
\left[
\sum_{\bii=\bzero_{[4:1]}}^{\bone_{[4:1]}}
\delta_{i_1 i_2} \delta_{i_3 i_4}
(\barbJ{g}^{i_1} \otimes \barbJ{h}^{i_2})\, \bM^{\Delta}\,
(\barbJ{g}^{i_3} \otimes \barbJ{h}^{i_4})
\right]_{\text{cell}} \circledast (\tildebJ_{\bm}\bI_{\bm}^{\Delta})
\\&
=-m_U^2(\bM^{\Delta})_{\text{cell}}\circledast (\tildebJ_{\bm}\bI_{\bm}^{\Delta})
+ \bO_{[n:n]}\left(\frac{\Delta^2}{(1-\Delta)^2\min(g^2,h^2) (1+m_U\siggama^2/\sige^2)^2}\right)
\\&
=\bO_{[n:n]}\left(\frac{\Delta^2}{(1+\Delta)^2m_L}\right)
+\bO_{[n:n]}\left(\frac{\Delta^2}{(1-\Delta)^2\min(g^2,h^2) (1+m_U\siggama^2/\sige^2)^2}\right)
\\&
=\bO_{[n:n]}\left(\frac{\Delta^2}{(1+\Delta)^2m_L}\right).
\end{align*}
Therefore,
we have 
\begin{align*}
\{\check{\bV}^{-1}(\check{\bV}-\bV)\}^2&= \bF^{(0)}\bF^{(0)}+\bO_{[n:n]}\left(\frac{\Delta^2}{(1+\Delta)^2m_L}\right)
=\bO_{[n:n]}\left(\frac{\Delta^2}{(1+\Delta)^2}\right).
\end{align*}

For $l=3$, similar arguments show that
\begin{align*}
\{\check{\bV}^{-1}(\check{\bV}-\bV)\}^3=& \bF^{(0)}\bF^{(0)}\bF^{(0)}+\bF^{(0)}\bF^{(0)}\bF^{(1)}
+\bF^{(0)}\bF^{(1)}\bF^{(0)}
+\bF^{(0)}\bF^{(1)}\bF^{(1)}\\&
+\bF^{(1)}\bF^{(0)}\bF^{(0)}
+\bF^{(1)}\bF^{(0)}\bF^{(1)}
+\bF^{(1)}\bF^{(1)}\bF^{(0)}
+\bF^{(1)}\bF^{(1)}\bF^{(1)},
\end{align*}
where 
\begin{align*}
&\bF^{(0)}\bF^{(0)}\bF^{(0)}=-(m_U\tildebI_{\bm}\bI_{\bm}^{\Delta})^3,
\\&
\bF^{(0)}\bF^{(0)}\bF^{(1)}
=-m_U\sige^2\left[\sum_{\bii=\bzero_{[2:1]}}^{\bone_{[2:1]}}\delta_{i_1i_2} (\barbJ{g}^{i_1}\otimes \barbJ{h}^{i_2})\right]_{\text{cell}}\circledast\{(m_U\tildebI_{\bm}\bI_{\bm}^{\Delta})^2\tildebJ_{\bm}\bI_{\bm}^{\Delta}\},
\\&
\bF^{(0)}\bF^{(1)}\bF^{(0)}
=-m_U\sige^2\left[\sum_{\bii=\bzero_{[2:1]}}^{\bone_{[2:1]}}\delta_{i_1i_2} (\barbJ{g}^{i_1}\otimes \barbJ{h}^{i_2})\right]_{\text{cell}}\circledast(m_U^2\tildebI_{\bm}\bI_{\bm}^{\Delta}\tildebJ_{\bm}\bI_{\bm}^{\Delta}\tildebI_{\bm}\bI_{\bm}^{\Delta}),
\\&
\bF^{(0)}\bF^{(1)}\bF^{(1)}
=-m_U^2\sige^4
\left[
\sum_{\bii=\bzero_{[4:1]}}^{\bone_{[4:1]}}
\delta_{i_1 i_2} \delta_{i_3 i_4}
(\barbJ{g}^{i_1} \otimes \barbJ{h}^{i_2})\, \bM^{\Delta}\,
(\barbJ{g}^{i_3} \otimes \barbJ{h}^{i_4})
\right]_{\text{cell}} \circledast (m_U\tildebI_{\bm}\bI_{\bm}^{\Delta}\tildebJ_{\bm}\bI_{\bm}^{\Delta}),
\\&
\bF^{(1)}\bF^{(0)}\bF^{(0)}
=-m_U\sige^2\left[\sum_{\bii=\bzero_{[2:1]}}^{\bone_{[2:1]}}\delta_{i_1i_2} (\barbJ{g}^{i_1}\otimes \barbJ{h}^{i_2})\right]_{\text{cell}}\circledast\{\tildebJ_{\bm}\bI_{\bm}^{\Delta}(m_U\tildebI_{\bm}\bI_{\bm}^{\Delta})^2\},
\\&
\bF^{(1)}\bF^{(0)}\bF^{(1)}
=-m_U^2\sige^4
\left[
\sum_{\bii=\bzero_{[4:1]}}^{\bone_{[4:1]}}
\delta_{i_1 i_2} \delta_{i_3 i_4}
(\barbJ{g}^{i_1} \otimes \barbJ{h}^{i_2})\, \bM^{\Delta^2}\,
(\barbJ{g}^{i_3} \otimes \barbJ{h}^{i_4})
\right]_{\text{cell}} \circledast (\tildebJ_{\bm}\bI_{\bm}^{\Delta}),
\\&
\bF^{(1)}\bF^{(1)}\bF^{(0)}
=-m_U^2\sige^4
\left[
\sum_{\bii=\bzero_{[4:1]}}^{\bone_{[4:1]}}
\delta_{i_1 i_2} \delta_{i_3 i_4}
(\barbJ{g}^{i_1} \otimes \barbJ{h}^{i_2})\, \bM^{\Delta}\,
(\barbJ{g}^{i_3} \otimes \barbJ{h}^{i_4})
\right]_{\text{cell}} \circledast (m_U\tildebJ_{\bm}\bI_{\bm}^{\Delta}\tildebI_{\bm}\bI_{\bm}^{\Delta}),
\\&
\bF^{(1)}\bF^{(1)}\bF^{(1)}
=-m_U^3\sige^6
\left[
\sum_{\bii=\bzero_{[6:1]}}^{\bone_{[6:1]}}
\delta_{i_1 i_2} \delta_{i_3 i_4}
(\barbJ{g}^{i_1} \otimes \barbJ{h}^{i_2})\, \bM^{\Delta}\,
(\barbJ{g}^{i_3} \otimes \barbJ{h}^{i_4})\bM^{\Delta}\,
(\barbJ{g}^{i_5} \otimes \barbJ{h}^{i_6})
\right]_{\text{cell}} \circledast (\tildebJ_{\bm}\bI_{\bm}^{\Delta}),
\end{align*}
with 
\begin{align*}
\bM^{\Delta^2}&=\diag((m_U-m_{11})^2m_U/(m_U^2m_Lm_{11}),\ldots,(m_U-m_{gh})^2m_U/(m_U^2m_Lm_{gh}))\\&=\bO_{[gh:gh]}(\Delta^2m_U m_L^{-2}).
\end{align*}
Suppose the variance components are finite and  $\Delta=(m_U-m_L)/m_U<0.5$ (i.e.~the data are not severely unbalanced), 
following (\ref{JJORDER}) and (\ref{I-DELTA}), 
$(m_U\tildebI_{\bm}\bI_{\bm}^{\Delta})^2\tildebJ_{\bm}\bI_{\bm}^{\Delta}=\tildebJ_{\bm}\bI_{\bm}^{\Delta}(m_U\tildebI_{\bm}\bI_{\bm}^{\Delta})^2=m_U\tildebI_{\bm}\bI_{\bm}^{\Delta}\tildebJ_{\bm}\bI_{\bm}^{\Delta}m_U\tildebI_{\bm}\bI_{\bm}^{\Delta}=\bO_{[n:n]}(m_L^{-2}\Delta^3/(1-\Delta)^2)$, we have 
\begin{align*}
\bF^{(i_1)}\bF^{(i_2)}\bF^{(i_3)}=\begin{cases}
     \bO_{[n:n]}\left(\frac{\Delta^3}{(1-\Delta)^3}\right), \quad \text{if} \quad i_1=i_2=i_3=0,\\
    \bO_{[n:n]}\left(\frac{\Delta^3}{(1-\Delta)^3m_L}\right),\quad \text{otherwise.}
\end{cases}
\end{align*}
The above detailed computations for $l = 2$ and $l = 3$, show that each term involving $\bF^{(1)}$ contributes an additional $\bO(m_L^{-1})$ factor. In particular, for $l = 2$, each summand is shown to be of order $\bO_{[n:n]}(\Delta^2 / (1 - \Delta)^2)$ or smaller. For $l = 3$, the dominating terms scale as $\bO_{[n:n]}(\Delta^3 / (1 - \Delta)^3)$.

Thus, combining the above and (\ref{bound Vinv}), we conclude that for any $r \geq 1$,
\begin{align*}
  \left\{ \check{\bV}^{-1}(\check{\bV}-\bV) \right\}^r \bV^{-1} = \bO_{[n:n]}\left( \frac{\Delta^{r+1}}{(1 - \Delta)^{r+1}} \right).
\end{align*}

We write truncated terms at $r=1$ and $r=2$,
\begin{equation}\label{aproxl1}
\begin{split}
&\check{\bV}^{-1}(\check{\bV}-\bV)\check{\bV}^{-1}
=-\frac{1}{\sige^2}(\bF^{(0)} \bF^{(0)} + \bF^{(0)} \bF^{(1)} + \bF^{(1)} \bF^{(0)} + \bF^{(1)} \bF^{(1)})\bI_{\bm}^{\Delta^{-1}},
\end{split}
\end{equation}
and
\begin{equation}\label{aproxl2}
\begin{split}
&\left\{\check{\bV}^{-1}(\check{\bV}-\bV)\right\}^2\check{\bV}^{-1}
\\&= -\frac{1}{\sige^2}\big(\bF^{(0)}\bF^{(0)}\bF^{(0)}+\bF^{(0)}\bF^{(0)}\bF^{(1)}
+\bF^{(0)}\bF^{(1)}\bF^{(0)}
+\bF^{(0)}\bF^{(1)}\bF^{(1)}\\&\quad
+\bF^{(1)}\bF^{(0)}\bF^{(0)}
+\bF^{(1)}\bF^{(0)}\bF^{(1)}
+\bF^{(1)}\bF^{(1)}\bF^{(0)}
+\bF^{(1)}\bF^{(1)}\bF^{(1)}\big)\bI_{\bm}^{\Delta^{-1}}.
\end{split}
\end{equation}

\end{proof}
\subsection{Proof of Theorem~\ref{theorem3}}  \label{proof4}
\begin{proof}
We have $\bV = \check{\bV} + \sigma_e^2 \bI_{\bm}^{\Delta}$, 
where $\bI_{\bm}^{\Delta} = \diag((1 - m_{11}/m_U)\bI_{m_{11}}, \ldots, (1 - m_{gh}/m_U)\bI_{m_{gh}})$.  We can decompose $\sigma_e^2 \bI_{\bm}^{\Delta}$ as a sum of rank-one matrices
\begin{align*}
\sigma_e^2 \bI_{\bm}^{\Delta} = \sum_{i=1}^{g} \sum_{j=1}^{h} \sum_{k=1}^{m_{ij}} \bE_{ijk} 
= \sum_{i=1}^{g} \sum_{j=1}^{h} \sum_{k=1}^{m_{ij}} \sigma_e^2 \left(1 - \frac{m_{ij}}{m_U} \right) \bI_{n(ijk)}.
\end{align*}
We write $\bI_{n(ijk)} = \bu_{(ijk)} \bu_{(ijk)}^\top$, where $\bu_{(ijk)}$ is an $n \times 1$ vector with all entries zero except for a one at the $(ijk)$th position. Then we have the reproducing property
\begin{align*}
\bu_{(ijk)}^\top \check{\bV}^{-1} \bu_{(ijk)} 
= \tr\left( \bu_{(ijk)}^\top \check{\bV}^{-1} \bu_{(ijk)} \right) 
= \tr\left( \check{\bV}^{-1} \bu_{(ijk)} \bu_{(ijk)}^\top \right) 
= \tr\left( \check{\bV}^{-1} \bI_{(ijk)} \right).
\end{align*}
To prove the result, we first write $\bW_{112} = \check{\bV} + \bE_{111}$. Since $\bW_{112}$ is nonsingular, applying the Sherman--Morrison formula \citep{henderson1981deriving} yields
\begin{equation}\label{W112}
\begin{split}
\bW_{112}^{-1}
&= (\check{\bV} + \bE_{111})^{-1} \\
&= \left[\check{\bV} + \sigma_e^2 \left(1 - \frac{m_{ij}}{m_U} \right) \bI_{n(ijk)}\right]^{-1} \\
&= \left[\check{\bV} + \sigma_e^2 \left(1 - \frac{m_{ij}}{m_U} \right) \bu_{(ijk)} \bu_{(ijk)}^\top \right]^{-1} \\
&= \check{\bV}^{-1} - \frac{1}{1 + \sige^2(1-m_{ij}/m_U)\bu_{(ijk)}^\top \check{\bV}^{-1} \bu_{(ijk)}} \check{\bV}^{-1} \bE_{111} \check{\bV}^{-1} \\
&= \check{\bV}^{-1} - \frac{1}{1 + \tr(\check{\bV}^{-1} \bE_{111} )} \check{\bV}^{-1} \bE_{111} \check{\bV}^{-1} \\
&= \bW_{111}^{-1} - \kappa_{111} \bW_{111}^{-1} \bE_{111} \bW_{111}^{-1},
\end{split}
\end{equation}
where $\kappa_{111} = 1 / \left[1 + \tr(\check{\bV}^{-1} \bI_{(111)})\right]$ and $\bW_{111} = \check{\bV}$.
We have now expressed $\bW_{112}^{-1}$ in terms of $\check{\bV}^{-1}$. Next, since $\bW_{113} = \check{\bV} + \bE_{111} + \bE_{112}$ and both $\bW_{112}$ and $\bW_{113}$ are nonsingular, we apply the same formula again
\begin{align*}
\bW_{113}^{-1} = (\bW_{112} + \bE_{112})^{-1} = \bW_{112}^{-1} - \kappa_{112} \bW_{112}^{-1} \bE_{112} \bW_{112}^{-1},
\end{align*}
where $\kappa_{112} = 1 / \left[1 + \tr(\bW_{112}^{-1} \bE_{112})\right]$, and $\bW_{112}^{-1}$ comes from equation~\eqref{W112}.
Continuing this process $n$ times, where $n$ is the rank of $\bI_{\bm}^{\Delta}$, we obtain
\begin{align*}
\bW_{gh\,m_{gh}+1}^{-1} = \bW_{gh\,m_{gh}}^{-1} - \kappa_{gh\,m_{gh}} \bW_{gh\,m_{gh}}^{-1} \bE_{gh\,m_{gh}} \bW_{gh\,m_{gh}}^{-1}.
\end{align*}
Since $\bW_{gh\,m_{gh}+1} = \bV$, the theorem is proved.
\end{proof}

\section{Lemmas}\label{appendix:lema}
The remaining properties of the Khatri–Rao product are given in Section~\ref{appendix:KR lemma}, and the supporting lemmas used in the proof of Theorem~\ref{theorem4} are collected in Section~\ref{appendix:lemma-theorem}.
Proofs are provided in the Supplementary Material.

\subsection{Lemmas for Khatri-Rao Products}\label{appendix:KR lemma}

\begin{lemma}\label{lma2}
   Let $\ba, \bd \in \mathbb{R}^g$, $\bb, \be \in \mathbb{R}^h$, and $\bc, \bbf \in \mathbb{R}^n$ be vectors partitioned into $gh$ subvectors. Let $\bP, \bP_1 \in \mathbb{R}^{g \times g}$, $\bQ, \bQ_1 \in \mathbb{R}^{h \times h}$, and $\bR, \bR_1 \in \mathbb{R}^{n \times n}$, where $\bR$ and $\bR_1$ are block matrices of the form $\bR = (\bR_{[m_{ij}:m_{i'j'}]})$ and $\bR_1 = (\bR_{1[m_{ij}:m_{i'j'}]})$, with each block $\bR_{[m_{ij}:m_{i'j'}]}$ and $\bR_{1[m_{ij}:m_{i'j'}]}$ being an $m_{ij} \times m_{i'j'}$ matrix located at the $(ih - h + j,\; i'h - h + j')$ block position. Then the following equalities hold.
    \begin{enumerate}
		\item {Bilinearity:}
		\begin{align*}
			&(\bP\otimes (\bQ \pm \bQ_1))_{\mathrm{cell}} \circledast \bR   
			=   (\bP\otimes \bQ)_{\mathrm{cell}} \circledast \bR   
			\pm   (\bP\otimes \bQ_1)_{\mathrm{cell}} \circledast \bR,   
			\\&
			((\bP \pm \bP_1)\otimes \bQ)_{\mathrm{cell}} \circledast \bR   
			=   (\bP\otimes \bQ)_{\mathrm{cell}} \circledast \bR   
			\pm  (\bP_1\otimes \bQ)_{\mathrm{cell}} \circledast \bR,  
			\\&
			(\bP\otimes \bQ)_{\mathrm{cell}} \circledast (\bR \pm \bR_1)   
			=
			(\bP\otimes \bQ)_{\mathrm{cell}} \circledast \bR   
			\pm  (\bP\otimes \bQ)_{\mathrm{cell}} \circledast \bR_1.
		\end{align*}
		
		\item {Inner Products and Matrix Products:}
		\begin{align*}
			&\left[ (\ba \otimes \bb)_{\mathrm{row}} \circledast \bc \right]^\top 
			\left[ (\bd \otimes \be)_{\mathrm{row}} \circledast \bbf \right]
			= \sum_{i=1}^g \sum_{j=1}^h 
			\left( a_i b_j d_i e_j \sum_{k=1}^{m_{ij}} c_{ijk} f_{ijk} \right),
			\\&
			\left[ (\ba \otimes \bb)_{\mathrm{row}} \circledast \bc \right] 
			\left[ (\bd \otimes \be)_{\mathrm{row}} \circledast \bbf \right]^\top
			= (\ba \bd^\top \otimes \bb \be^\top)_{\mathrm{cell}} \circledast (\bc \bbf^\top),
			\\&
			\left[ (\ba \otimes \bb)_{\mathrm{row}} \circledast \bc \right]^\top 
			\left[(\bP\otimes \bQ)_{\mathrm{cell}} \circledast \bR\right]
			= \col\left[\sum_{s=1}^g\sum_{t=1}^h a_i b_j p_{si} q_{tj} \bc_{st}^\top \bR_{[m_{st} : m_{ij}]}\right]_{i=1,j=1}^{g,h},
			\\&
			\left[ (\ba \otimes \bb)_{\mathrm{row}} \circledast \bc \right]^\top
			\left[(\bP\otimes \bQ)_{\mathrm{cell}} \circledast \bR\right]
			\left[ (\bd \otimes \be)_{\mathrm{row}} \circledast \bbf \right]
			\\&=\sum_{s=1}^g\sum_{t=1}^h\sum_{i=1}^g\sum_{j=1}^h a_i b_j d_s e_t p_{is} q_{jt} \bc_{ij}^\top \bR_{[m_{ij} : m_{st}]} \bbf_{st}.
		\end{align*}
\end{enumerate}
\end{lemma}
\subsection{Lemmas for Theorem~\ref{theorem4}}\label{appendix:lemma-theorem}
\begin{lemma}\label{lma matrix ineq}
Let $\bA,\bB, \bC \in \mathbb{R}^{n\times n}$ be $n\times n$ matrices with nonnegative entries. Suppose
$\bC \ge \bA \quad \text{entrywise}.$
Then
$\bA \bB \le \bC \bB \quad \text{entrywise}.$
Moreover, $\bA \bB < \bC \bB$ entrywise holds if and only if for each row $i$ of $\bC-\bA$ there exists some index $k$ with $\bC_{ik}>\bA_{ik}$ and, for every column $j$, at least one such $k$ satisfies $\bB_{kj}>0$. 
\end{lemma}

\begin{lemma}\label{lema: matrix kth product}
	As $g, h\to \infty$, for $l=1,2,\ldots$,  we have 
	\begin{align*}
&		\big[(\barbJ{g}\otimes\bI_h)_{\text{cell}}\circledast\tildebJ_{\bm}
		+(\bI_g\otimes\barbJ{h})_{\text{cell}}\circledast\tildebJ_{\bm}
		+(\barbJ{g}\otimes\barbJ{h})_{\text{cell}}\circledast\tildebJ_{\bm}\big]^l
\\&
\le 
\frac{1}{m_L^{l-1}}\left[(\barbJ{g}\otimes\bI_h)_{\text{cell}}\circledast\tildebJ_{\bm}\right]
	+\frac{1}{m_L^{l-1}}\left[(\bI_{g}\otimes\barbJ{h})_{\text{cell}}\circledast\tildebJ_{\bm}\right]
	+ \frac{3^{l}-2}{m_L^{l-1}}\left[(\barbJ{g}\otimes\barbJ{h})_{\text{cell}}\circledast\tildebJ_{\bm}\right]
        \\&=
		\bO_{[n:n]}\left(\frac{1}{\min(g,h) m_L^{l+1}}+\frac{3^{l}-2}{ghm_L^{l+1}}\right),
	\end{align*}
where $\bO_{[n:n]}(a)$ denotes an $n \times n$ matrix with all entries of order $O(a)$.
\end{lemma}

\begin{lemma}\label{Sandwich bound}
Let $1<d_2<\infty$ and $g> 1$. For every $\varepsilon\in(0,1)$ there exists $G_\varepsilon<\infty$ such that, 
\begin{align*}
g^{-1-\varepsilon}\ \le   \frac{ d_2^{\sqrt{\log (g)}}}{g} \ \le\ g^{-1+\varepsilon}, \quad
\text{for all $g\ge G_\varepsilon$.}
\end{align*}
\end{lemma}

 \begin{lemma}\label{matrix remiander order}
Let 
$\bU_1=\diag[b_{ij}\bI_{m_{ij}}+c_{ij}\bJ_{m_{ij}}]$ and 
$\bU_2=\diag[(a_{ij}+b_{ij})\bI_{m_{ij}}+c_{ij}\bJ_{m_{ij}}]$, 
where 
$a_{ij}=m_U/[(m_U-m_{ij})\sige^2]$, 
$b_{ij}=m_U/(m_{ij}\sige^2)$, and 
$c_{ij}=m_U/m_{ij}(1/(\sige^2+m_U\siggama^2)-1/\sige^2)$, 
for $i=1,\ldots,g$ and $j=1,\ldots,h$. 
Let $s(g,h)=\sqrt{\log\{\min(g,h)\}}$. 
Define
\begin{align*}
\mathcal{R}
=& m_U(\frac{1}{\lambda_5}-\frac{1}{\lambda_7})(\barbJ{g}\otimes\bI_h)_{\text{cell}}\circledast\tildebJ_{\bm}
        +m_U(\frac{1}{\lambda_3}-\frac{1}{\lambda_7})(\bI_g\otimes\barbJ{h})_{\text{cell}}\circledast\tildebJ_{\bm}
\\&        +m_U(\frac{1}{\lambda_1}-\frac{1}{\lambda_3}-\frac{1}{\lambda_5}+\frac{1}{\lambda_7})(\barbJ{g}\otimes\barbJ{h})_{\text{cell}}\circledast\tildebJ_{\bm}.    
\end{align*}
Then, as $g,h\to\infty$, we have
\begin{align*}
&\mathcal{R}=\bO_{[n:n]}\left(\frac{1}{\min(g,h)m_L^2}\right),
&&
\sum_{l=1}^{s(g,h)} \bU_1(-\bU_2^{-1}\mathcal{R})^{l+1}=
\bO_{[n:n]}\left(
\frac{1}{\min(g^{1-\varepsilon},h^{1-\varepsilon})m_L^2}\right),
\\&
\bU_1\bU_2^{-1}\mathcal{R}=\bO_{[n:n]}\left(\frac{1}{\min(g,h)m_L^2}\right),
&&
\sum_{l=1}^{s(g,h)}\bU_1  (-\bU_2^{-1}\mathcal{R})^l\bU_2^{-1}\bU_1=
\bO_{[n:n]}\left(
\frac{1}{\min(g^{1-\varepsilon},h^{1-\varepsilon})m_L^2}\right),
\\&
\mathcal{R}\bU_2^{-1}\bU_1=\bO_{[n:n]}\left(\frac{1}{\min(g,h)m_L^2}\right),
&&
\sum_{l=1}^{s(g,h)}-\mathcal{R} (-\bU_2^{-1}\mathcal{R})^{l+1}=\bO_{[n:n]}\left(
\frac{1}{\min(g^{1-\varepsilon},h^{1-\varepsilon})m_L^3}
\right),
\\&
\mathcal{R}\bU_2^{-1}\mathcal{R}=\bO_{[n:n]}\left(\frac{1}{\min(g,h)m_L^3}\right),
&&
\sum_{l=1}^{s(g,h)}-\mathcal{R} (-\bU_2^{-1}\mathcal{R})^l\bU_2^{-1}\bU_1=\bO_{[n:n]}\left(
\frac{1}{\min(g^{1-\varepsilon},h^{1-\varepsilon})m_L^3}
\right),
\end{align*}
where $\varepsilon\in [0,1)$.
 \end{lemma}

\section{Examples}\label{Example}   
    Assume that $g=2$ and $h=3$ and $\bm=[m_{11},m_{12}, m_{13}, m_{21}, m_{22}, m_{23}]^\top$. Then we can write the following
	\begin{align*}
		(\bI_2\otimes\bone_3)_{\text{row}}\circledast\bone_{\bm}=
		\begin{bmatrix}
			1&0\\\hline
			1&0\\\hline
			1&0\\\hline
			0&1\\ \hline
			0&1\\\hline 
			0&1\\
		\end{bmatrix}\circledast
		\begin{bmatrix}
			\bone_{m_{11}}\\
			\bone_{m_{12}}\\
			\bone_{m_{13}}\\
			\bone_{m_{21}}\\
			\bone_{m_{22}}\\
			\bone_{m_{23}}
		\end{bmatrix}
		=
		\begin{bmatrix}
			[1&0]\otimes\bone_{m_{11}}\\
			[1&0]\otimes\bone_{m_{12}}\\
			[1&0]\otimes\bone_{m_{13}}\\
			[0&1]\otimes\bone_{m_{21}}\\
			[0&1]\otimes\bone_{m_{22}}\\
			[0&1]\otimes\bone_{m_{23}}\\
		\end{bmatrix}
		=\begin{bmatrix}
			\bone_{m_{11}}&\bzero_{m_{11}}\\
			\bone_{m_{12}}&\bzero_{m_{12}}\\
			\bone_{m_{13}}&\bzero_{m_{13}}\\
			\bzero_{m_{21}}&\bone_{m_{21}}\\
			\bzero_{m_{22}}&\bone_{m_{22}}\\
			\bzero_{m_{23}}&\bone_{m_{23}}\\
		\end{bmatrix}.
	\end{align*}
	and
	\begin{align*}
		&\bZ_1\bZ_1^\top\\&=(\bI_2\otimes\bJ_3)_{\text{cell}}\circledast\bJ_{\bm}\\
		&=\begin{bmatrix}
			1&1&1&0&0&0\\
			1&1&1&0&0&0\\
			1&1&1&0&0&0\\
			0&0&0&1&1&1\\
			0&0&0&1&1&1\\
			0&0&0&1&1&1\\
		\end{bmatrix}
		\circledast\begin{bmatrix}
			\bJ_{[m_{11}:m_{11}]}
			&\bJ_{[m_{11}:m_{12}]}
			&\bJ_{[m_{11}:m_{13}]}
			&\bJ_{[m_{11}:m_{21}]}
			&\bJ_{[m_{11}:m_{22}]}
			&\bJ_{[m_{11}:m_{23}]}
			\\
			\bJ_{[m_{12}:m_{11}]}
			&\bJ_{[m_{12}:m_{12}]}
			&\bJ_{[m_{12}:m_{13}]}
			&\bJ_{[m_{12}:m_{21}]}
			&\bJ_{[m_{12}:m_{22}]}
			&\bJ_{[m_{12}:m_{23}]}
			\\
			\bJ_{[m_{13}:m_{11}]}
			&\bJ_{[m_{13}:m_{12}]}
			&\bJ_{[m_{13}:m_{13}]}
			&\bJ_{[m_{13}:m_{21}]}
			&\bJ_{[m_{13}:m_{22}]}
			&\bJ_{[m_{13}:m_{23}]}
			\\
			\bJ_{[m_{21}:m_{11}]}
			&\bJ_{[m_{21}:m_{12}]}
			&\bJ_{[m_{21}:m_{13}]}
			&\bJ_{[m_{21}:m_{21}]}
			&\bJ_{[m_{21}:m_{22}]}
			&\bJ_{[m_{21}:m_{23}]}
			\\
			\bJ_{[m_{22}:m_{11}]}
			&\bJ_{[m_{22}:m_{12}]}
			&\bJ_{[m_{22}:m_{13}]}
			&\bJ_{[m_{22}:m_{21}]}
			&\bJ_{[m_{22}:m_{22}]}
			&\bJ_{[m_{22}:m_{23}]}
			\\
			\bJ_{[m_{23}:m_{11}]}
			&\bJ_{[m_{23}:m_{12}]}
			&\bJ_{[m_{23}:m_{13}]}
			&\bJ_{[m_{23}:m_{21}]}
			&\bJ_{[m_{23}:m_{22}]}
			&\bJ_{[m_{23}:m_{23}]}
		\end{bmatrix}\\&
		=
		\begin{bmatrix}
			1\otimes\bJ_{[m_{11}:m_{11}]}
			&1\otimes\bJ_{[m_{11}:m_{12}]}
			&1\otimes\bJ_{[m_{11}:m_{13}]}
			&0\otimes\bJ_{[m_{11}:m_{21}]}
			&0\otimes\bJ_{[m_{11}:m_{22}]}
			&0\otimes\bJ_{[m_{11}:m_{23}]}
			\\
			1\otimes\bJ_{[m_{12}:m_{11}]}
			&1\otimes\bJ_{[m_{12}:m_{12}]}
			&1\otimes\bJ_{[m_{12}:m_{13}]}
			&0\otimes\bJ_{[m_{12}:m_{21}]}
			&0\otimes\bJ_{[m_{12}:m_{22}]}
			&0\otimes\bJ_{[m_{12}:m_{23}]}
			\\
			1\otimes\bJ_{[m_{13}:m_{11}]}
			&1\otimes\bJ_{[m_{13}:m_{12}]}
			&1\otimes\bJ_{[m_{13}:m_{13}]}
			&0\otimes\bJ_{[m_{13}:m_{21}]}
			&0\otimes\bJ_{[m_{13}:m_{22}]}
			&0\otimes\bJ_{[m_{13}:m_{23}]}
			\\
			0\otimes\bJ_{[m_{21}:m_{11}]}
			&0\otimes\bJ_{[m_{21}:m_{12}]}
			&0\otimes\bJ_{[m_{21}:m_{13}]}
			&1\otimes\bJ_{[m_{21}:m_{21}]}
			&1\otimes\bJ_{[m_{21}:m_{22}]}
			&1\otimes\bJ_{[m_{21}:m_{23}]}
			\\
			0\otimes\bJ_{[m_{22}:m_{11}]}
			&0\otimes\bJ_{[m_{22}:m_{12}]}
			&0\otimes\bJ_{[m_{22}:m_{13}]}
			&1\otimes\bJ_{[m_{22}:m_{21}]}
			&1\otimes\bJ_{[m_{22}:m_{22}]}
			&1\otimes\bJ_{[m_{22}:m_{23}]}
			\\
			0\otimes\bJ_{[m_{23}:m_{11}]}
			&0\otimes\bJ_{[m_{23}:m_{12}]}
			&0\otimes\bJ_{[m_{23}:m_{13}]}
			&1\otimes\bJ_{[m_{23}:m_{21}]}
			&1\otimes\bJ_{[m_{23}:m_{22}]}
			&1\otimes\bJ_{[m_{23}:m_{23}]}
		\end{bmatrix}\\&
		=
		\begin{bmatrix}
			\bJ_{[m_{11}:m_{11}]}
			&\bJ_{[m_{11}:m_{12}]}
			&\bJ_{[m_{11}:m_{13}]}
			&\bzero_{[m_{11}:m_{21}]}
			&\bzero_{[m_{11}:m_{22}]}
			&\bzero_{[m_{11}:m_{23}]}
			\\
			\bJ_{[m_{12}:m_{11}]}
			&\bJ_{[m_{12}:m_{12}]}
			&\bJ_{[m_{12}:m_{13}]}
			&\bzero_{[m_{12}:m_{21}]}
			&\bzero_{[m_{12}:m_{22}]}
			&\bzero_{[m_{12}:m_{23}]}
			\\
			\bJ_{[m_{13}:m_{11}]}
			&\bJ_{[m_{13}:m_{12}]}
			&\bJ_{[m_{13}:m_{13}]}
			&\bzero_{[m_{13}:m_{21}]}
			&\bzero_{[m_{13}:m_{22}]}
			&\bzero_{[m_{13}:m_{23}]}
			\\
			\bzero_{[m_{21}:m_{11}]}
			&\bzero_{[m_{21}:m_{12}]}
			&\bzero_{[m_{21}:m_{13}]}
			&\bJ_{[m_{21}:m_{21}]}
			&\bJ_{[m_{21}:m_{22}]}
			&\bJ_{[m_{21}:m_{23}]}
			\\
			\bzero_{[m_{22}:m_{11}]}
			&\bzero_{[m_{22}:m_{12}]}
			&\bzero_{[m_{22}:m_{13}]}
			&\bJ_{[m_{22}:m_{21}]}
			&\bJ_{[m_{22}:m_{22}]}
			&\bJ_{[m_{22}:m_{23}]}
			\\
			\bzero_{[m_{23}:m_{11}]}
			&\bzero_{[m_{23}:m_{12}]}
			&\bzero_{[m_{23}:m_{13}]}
			&\bJ_{[m_{23}:m_{21}]}
			&\bJ_{[m_{23}:m_{22}]}
			&\bJ_{[m_{23}:m_{23}]}
		\end{bmatrix}.
	\end{align*}

\end{appendix}

\begin{funding}
ZL, SAS and AHW are respectively supported by the Australian Research Council Discovery Projects DE260101297, DP220103269 and DP230101908.
\end{funding}

\begin{supplement}
Comprehensive calculations and detailed proofs support the findings presented in the main manuscript.
\end{supplement}

	\bibliographystyle{imsart-number}
\bibliography{mythesisbib}


\newpage
 \section*{Supplementary Material}

\subsection*{Proof of Lemma 1}
\begin{proof}
Since the properties of the Khatri–Rao product follow directly from their definition, we illustrate the expressions with a simple example. Suppose $g=2$ and $h=3$, and let $\bm=(m_{11},m_{12},m_{13},m_{21},m_{22},m_{23})^\top$. Then we write
\begin{align*}
\ba=\begin{bmatrix}a_1\\a_2\end{bmatrix},\quad
\bb=\begin{bmatrix}b_1\\ b_2\\ b_3\end{bmatrix},\quad
\bP=\begin{bmatrix}p_{11}&p_{12}\\ p_{21}&p_{22}\end{bmatrix},\quad
\bR=\begin{bmatrix}r_{11}&r_{12}\\ r_{21}&r_{22}\end{bmatrix},\quad
\bQ=\begin{bmatrix}
q_{11}&q_{12}&q_{13}\\
q_{21}&q_{22}&q_{23}\\
q_{31}&q_{32}&q_{33}
\end{bmatrix},\quad
\bS=\begin{bmatrix}
s_{11}&s_{12}&s_{13}\\
s_{21}&s_{22}&s_{23}\\
s_{31}&s_{32}&s_{33}
\end{bmatrix}.    
\end{align*}
The following 15 matrices are computed using the Khatri–Rao product.
\begin{align*}
 1).\quad &   (\ba\otimes\bb)_{\text{row}}\circledast\barbone_{\bm}
=\left(\begin{bmatrix}
    a_1\\a_2
\end{bmatrix}\otimes
\begin{bmatrix}
    b_1\\b_2\\ b_3
\end{bmatrix}\right)_{\text{row}}\circledast
\begin{bmatrix}
\frac{1}{\sqrt{m_{11}}}\bone_{m_{11}}
\\ \frac{1}{\sqrt{m_{12}}}\bone_{m_{12}}
\\ \frac{1}{\sqrt{m_{13}}}\bone_{m_{13}}
\\ \frac{1}{\sqrt{m_{21}}}\bone_{m_{21}}
\\ \frac{1}{\sqrt{m_{22}}}\bone_{m_{22}}
\\ \frac{1}{\sqrt{m_{23}}}\bone_{m_{23}}
\end{bmatrix}
=\begin{bmatrix}
    a_1b_1\\\hline
    a_1b_2\\\hline
    a_1b3\\\hline
    a_2b_1\\\hline
    a_2b_2\\\hline
    a_2b3
\end{bmatrix}_{\text{row}}\circledast
\begin{bmatrix}
\frac{1}{\sqrt{m_{11}}}\bone_{m_{11}}
\\ \frac{1}{\sqrt{m_{12}}}\bone_{m_{12}}
\\ \frac{1}{\sqrt{m_{13}}}\bone_{m_{13}}
\\ \frac{1}{\sqrt{m_{21}}}\bone_{m_{21}}
\\ \frac{1}{\sqrt{m_{22}}}\bone_{m_{22}}
\\ \frac{1}{\sqrt{m_{23}}}\bone_{m_{23}}
\end{bmatrix}
=\begin{bmatrix}
\frac{a_1b_1}{\sqrt{m_{11}}}\bone_{m_{11}}
\\ \frac{a_1b_2}{\sqrt{m_{12}}}\bone_{m_{12}}
\\ \frac{a_1b_3}{\sqrt{m_{13}}}\bone_{m_{13}}
\\ \frac{a_2b_1}{\sqrt{m_{21}}}\bone_{m_{21}}
\\ \frac{a_2b_2}{\sqrt{m_{22}}}\bone_{m_{22}}
\\ \frac{a_2b_3}{\sqrt{m_{23}}}\bone_{m_{23}}
\end{bmatrix}.
\end{align*}

\begin{align*}
 2).\quad  &(\bP\otimes\bQ)_{\text{cell}}\circledast\barbJ{\bm}
\\&=
\left(\begin{bmatrix}
p_{11}&p_{12}\\
p_{21}&p_{22}
\end{bmatrix}
\otimes
\begin{bmatrix}
q_{11}&q_{12}&q_{13}\\
q_{21}&q_{22}&q_{23}\\
q_{31}&q_{32}&q_{33}\\
\end{bmatrix}\right)_{\text{cell}}
\\&\quad
\circledast
\resizebox{1.1\textwidth}{!}{$
\begin{bmatrix}
	\frac{1}{m_{11}}\bJ_{[m_{11}:m_{11}]}
&\frac{1}{\sqrt{m_{11}m_{12}}}\bJ_{[m_{11}:m_{12}]}
&\frac{1}{\sqrt{m_{11}m_{13}}}\bJ_{[m_{11}:m_{13}]}
&\frac{1}{\sqrt{m_{11}m_{21}}}\bJ_{[m_{11}:m_{21}]}
&\frac{1}{\sqrt{m_{11}m_{22}}}\bJ_{[m_{11}:m_{22}]}
&\frac{1}{\sqrt{m_{11}m_{23}}}\bJ_{[m_{11}:m_{23}]}
			\\
\frac{1}{\sqrt{m_{11}m_{12}}}\bJ_{[m_{12}:m_{11}]}
&\frac{1}{m_{12}}\bJ_{[m_{12}:m_{12}]}
&\frac{1}{\sqrt{m_{12}m_{13}}}\bJ_{[m_{12}:m_{13}]}
&\frac{1}{\sqrt{m_{12}m_{21}}}\bJ_{[m_{12}:m_{21}]}
&\frac{1}{\sqrt{m_{12}m_{22}}}\bJ_{[m_{12}:m_{22}]}
&\frac{1}{\sqrt{m_{12}m_{23}}}\bJ_{[m_{12}:m_{23}]}
			\\
\frac{1}{\sqrt{m_{11}m_{13}}}\bJ_{[m_{13}:m_{11}]}
&\frac{1}{\sqrt{m_{12}m_{13}}}\bJ_{[m_{13}:m_{12}]}
&\frac{1}{m_{13}}\bJ_{[m_{13}:m_{13}]}
&\frac{1}{\sqrt{m_{13}m_{21}}}\bJ_{[m_{13}:m_{21}]}
&\frac{1}{\sqrt{m_{13}m_{22}}}\bJ_{[m_{13}:m_{22}]}
&\frac{1}{\sqrt{m_{13}m_{23}}}\bJ_{[m_{13}:m_{23}]}
			\\
\frac{1}{\sqrt{m_{11}m_{21}}}\bJ_{[m_{21}:m_{11}]}
&\frac{1}{\sqrt{m_{12}m_{21}}}\bJ_{[m_{21}:m_{12}]}
&\frac{1}{\sqrt{m_{13}m_{21}}}\bJ_{[m_{21}:m_{13}]}
&\frac{1}{m_{21}}\bJ_{[m_{21}:m_{21}]}
&\frac{1}{\sqrt{m_{21}m_{22}}}\bJ_{[m_{21}:m_{22}]}
&\frac{1}{\sqrt{m_{21}m_{23}}}\bJ_{[m_{21}:m_{23}]}
			\\
\frac{1}{\sqrt{m_{11}m_{22}}}\bJ_{[m_{22}:m_{11}]}
&\frac{1}{\sqrt{m_{12}m_{22}}}\bJ_{[m_{22}:m_{12}]}
&\frac{1}{\sqrt{m_{13}m_{22}}}\bJ_{[m_{22}:m_{13}]}
&\frac{1}{\sqrt{m_{21}m_{22}}}\bJ_{[m_{22}:m_{21}]}
&\frac{1}{m_{22}}\bJ_{[m_{22}:m_{22}]}
&\frac{1}{\sqrt{m_{22}m_{23}}}\bJ_{[m_{22}:m_{23}]}
			\\
\frac{1}{\sqrt{m_{11}m_{23}}}\bJ_{[m_{23}:m_{11}]}
&\frac{1}{\sqrt{m_{12}m_{23}}}\bJ_{[m_{23}:m_{12}]}
&\frac{1}{\sqrt{m_{13}m_{23}}}\bJ_{[m_{23}:m_{13}]}
&\frac{1}{\sqrt{m_{21}m_{23}}}\bJ_{[m_{23}:m_{21}]}
&\frac{1}{\sqrt{m_{22}m_{23}}}\bJ_{[m_{23}:m_{22}]}
&\frac{1}{m_{23}}\bJ_{[m_{23}:m_{23}]}
		\end{bmatrix}$
        }
\\&
=\begin{bmatrix}
p_{11}q_{11}&p_{11}q_{12}&p_{11}q_{13}&p_{12}q_{11}&p_{12}q_{12}&p_{12}q_{13}\\
p_{11}q_{21}&p_{11}q_{22}&p_{11}q_{23}&p_{12}q_{21}&p_{12}q_{22}&p_{12}q_{23}\\
p_{11}q_{31}&p_{11}q_{32}&p_{11}q_{33}&p_{12}q_{31}&p_{12}q_{32}&p_{12}q_{33}\\
p_{21}q_{11}&p_{21}q_{12}&p_{21}q_{13}&p_{22}q_{11}&p_{22}q_{12}&p_{22}q_{13}\\
p_{21}q_{21}&p_{21}q_{22}&p_{21}q_{23}&p_{22}q_{21}&p_{22}q_{22}&p_{22}q_{23}\\
p_{21}q_{31}&p_{21}q_{32}&p_{21}q_{33}&p_{22}q_{31}&p_{22}q_{32}&p_{22}q_{33}\\
\end{bmatrix}_{\text{cell}}
\\&\quad
\circledast
\resizebox{1.1\textwidth}{!}{$
\begin{bmatrix}
	\frac{1}{m_{11}}\bJ_{[m_{11}:m_{11}]}
&\frac{1}{\sqrt{m_{11}m_{12}}}\bJ_{[m_{11}:m_{12}]}
&\frac{1}{\sqrt{m_{11}m_{13}}}\bJ_{[m_{11}:m_{13}]}
&\frac{1}{\sqrt{m_{11}m_{21}}}\bJ_{[m_{11}:m_{21}]}
&\frac{1}{\sqrt{m_{11}m_{22}}}\bJ_{[m_{11}:m_{22}]}
&\frac{1}{\sqrt{m_{11}m_{23}}}\bJ_{[m_{11}:m_{23}]}
			\\
\frac{1}{\sqrt{m_{11}m_{12}}}\bJ_{[m_{12}:m_{11}]}
&\frac{1}{m_{12}}\bJ_{[m_{12}:m_{12}]}
&\frac{1}{\sqrt{m_{12}m_{13}}}\bJ_{[m_{12}:m_{13}]}
&\frac{1}{\sqrt{m_{12}m_{21}}}\bJ_{[m_{12}:m_{21}]}
&\frac{1}{\sqrt{m_{12}m_{22}}}\bJ_{[m_{12}:m_{22}]}
&\frac{1}{\sqrt{m_{12}m_{23}}}\bJ_{[m_{12}:m_{23}]}
			\\
\frac{1}{\sqrt{m_{11}m_{13}}}\bJ_{[m_{13}:m_{11}]}
&\frac{1}{\sqrt{m_{12}m_{13}}}\bJ_{[m_{13}:m_{12}]}
&\frac{1}{m_{13}}\bJ_{[m_{13}:m_{13}]}
&\frac{1}{\sqrt{m_{13}m_{21}}}\bJ_{[m_{13}:m_{21}]}
&\frac{1}{\sqrt{m_{13}m_{22}}}\bJ_{[m_{13}:m_{22}]}
&\frac{1}{\sqrt{m_{13}m_{23}}}\bJ_{[m_{13}:m_{23}]}
			\\
\frac{1}{\sqrt{m_{11}m_{21}}}\bJ_{[m_{21}:m_{11}]}
&\frac{1}{\sqrt{m_{12}m_{21}}}\bJ_{[m_{21}:m_{12}]}
&\frac{1}{\sqrt{m_{13}m_{21}}}\bJ_{[m_{21}:m_{13}]}
&\frac{1}{m_{21}}\bJ_{[m_{21}:m_{21}]}
&\frac{1}{\sqrt{m_{21}m_{22}}}\bJ_{[m_{21}:m_{22}]}
&\frac{1}{\sqrt{m_{21}m_{23}}}\bJ_{[m_{21}:m_{23}]}
			\\
\frac{1}{\sqrt{m_{11}m_{22}}}\bJ_{[m_{22}:m_{11}]}
&\frac{1}{\sqrt{m_{12}m_{22}}}\bJ_{[m_{22}:m_{12}]}
&\frac{1}{\sqrt{m_{13}m_{22}}}\bJ_{[m_{22}:m_{13}]}
&\frac{1}{\sqrt{m_{21}m_{22}}}\bJ_{[m_{22}:m_{21}]}
&\frac{1}{m_{22}}\bJ_{[m_{22}:m_{22}]}
&\frac{1}{\sqrt{m_{22}m_{23}}}\bJ_{[m_{22}:m_{23}]}
			\\
\frac{1}{\sqrt{m_{11}m_{23}}}\bJ_{[m_{23}:m_{11}]}
&\frac{1}{\sqrt{m_{12}m_{23}}}\bJ_{[m_{23}:m_{12}]}
&\frac{1}{\sqrt{m_{13}m_{23}}}\bJ_{[m_{23}:m_{13}]}
&\frac{1}{\sqrt{m_{21}m_{23}}}\bJ_{[m_{23}:m_{21}]}
&\frac{1}{\sqrt{m_{22}m_{23}}}\bJ_{[m_{23}:m_{22}]}
&\frac{1}{m_{23}}\bJ_{[m_{23}:m_{23}]}
		\end{bmatrix}$
        }
\\&
=
\resizebox{1.1\textwidth}{!}{$
\begin{bmatrix}
\frac{p_{11}q_{11}}{m_{11}}\bJ_{[m_{11}:m_{11}]}
&\frac{p_{11}q_{12}}{\sqrt{m_{11}m_{12}}}\bJ_{[m_{11}:m_{12}]}
&\frac{p_{11}q_{13}}{\sqrt{m_{11}m_{13}}}\bJ_{[m_{11}:m_{13}]}
&\frac{p_{12}q_{11}}{\sqrt{m_{11}m_{21}}}\bJ_{[m_{11}:m_{21}]}
&\frac{p_{12}q_{12}}{\sqrt{m_{11}m_{22}}}\bJ_{[m_{11}:m_{22}]}
&\frac{p_{12}q_{13}}{\sqrt{m_{11}m_{23}}}\bJ_{[m_{11}:m_{23}]}
\\[4pt]
\frac{p_{11}q_{21}}{\sqrt{m_{12}m_{11}}}\bJ_{[m_{12}:m_{11}]}
&\frac{p_{11}q_{22}}{m_{12}}\bJ_{[m_{12}:m_{12}]}
&\frac{p_{11}q_{23}}{\sqrt{m_{12}m_{13}}}\bJ_{[m_{12}:m_{13}]}
&\frac{p_{12}q_{21}}{\sqrt{m_{12}m_{21}}}\bJ_{[m_{12}:m_{21}]}
&\frac{p_{12}q_{22}}{\sqrt{m_{12}m_{22}}}\bJ_{[m_{12}:m_{22}]}
&\frac{p_{12}q_{23}}{\sqrt{m_{12}m_{23}}}\bJ_{[m_{12}:m_{23}]}
\\[4pt]
\frac{p_{11}q_{31}}{\sqrt{m_{13}m_{11}}}\bJ_{[m_{13}:m_{11}]}
&\frac{p_{11}q_{32}}{\sqrt{m_{13}m_{12}}}\bJ_{[m_{13}:m_{12}]}
&\frac{p_{11}q_{33}}{m_{13}}\bJ_{[m_{13}:m_{13}]}
&\frac{p_{12}q_{31}}{\sqrt{m_{13}m_{21}}}\bJ_{[m_{13}:m_{21}]}
&\frac{p_{12}q_{32}}{\sqrt{m_{13}m_{22}}}\bJ_{[m_{13}:m_{22}]}
&\frac{p_{12}q_{33}}{\sqrt{m_{13}m_{23}}}\bJ_{[m_{13}:m_{23}]}
\\[6pt]
\frac{p_{21}q_{11}}{\sqrt{m_{21}m_{11}}}\bJ_{[m_{21}:m_{11}]}
&\frac{p_{21}q_{12}}{\sqrt{m_{21}m_{12}}}\bJ_{[m_{21}:m_{12}]}
&\frac{p_{21}q_{13}}{\sqrt{m_{21}m_{13}}}\bJ_{[m_{21}:m_{13}]}
&\frac{p_{22}q_{11}}{m_{21}}\bJ_{[m_{21}:m_{21}]}
&\frac{p_{22}q_{12}}{\sqrt{m_{21}m_{22}}}\bJ_{[m_{21}:m_{22}]}
&\frac{p_{22}q_{13}}{\sqrt{m_{21}m_{23}}}\bJ_{[m_{21}:m_{23}]}
\\[4pt]
\frac{p_{21}q_{21}}{\sqrt{m_{22}m_{11}}}\bJ_{[m_{22}:m_{11}]}
&\frac{p_{21}q_{22}}{\sqrt{m_{22}m_{12}}}\bJ_{[m_{22}:m_{12}]}
&\frac{p_{21}q_{23}}{\sqrt{m_{22}m_{13}}}\bJ_{[m_{22}:m_{13}]}
&\frac{p_{22}q_{21}}{\sqrt{m_{22}m_{21}}}\bJ_{[m_{22}:m_{21}]}
&\frac{p_{22}q_{22}}{m_{22}}\bJ_{[m_{22}:m_{22}]}
&\frac{p_{22}q_{23}}{\sqrt{m_{22}m_{23}}}\bJ_{[m_{22}:m_{23}]}
\\[4pt]
\frac{p_{21}q_{31}}{\sqrt{m_{23}m_{11}}}\bJ_{[m_{23}:m_{11}]}
&\frac{p_{21}q_{32}}{\sqrt{m_{23}m_{12}}}\bJ_{[m_{23}:m_{12}]}
&\frac{p_{21}q_{33}}{\sqrt{m_{23}m_{13}}}\bJ_{[m_{23}:m_{13}]}
&\frac{p_{22}q_{31}}{\sqrt{m_{23}m_{21}}}\bJ_{[m_{23}:m_{21}]}
&\frac{p_{22}q_{32}}{\sqrt{m_{23}m_{22}}}\bJ_{[m_{23}:m_{22}]}
&\frac{p_{22}q_{33}}{m_{23}}\bJ_{[m_{23}:m_{23}]}
\end{bmatrix}$
}.
\end{align*}

\begin{align*}
3). \quad &(\bR\otimes\bS)_{\text{cell}}\circledast\barbJ{\bm}
\\&
=
\resizebox{1.1\textwidth}{!}{$
\begin{bmatrix}
\frac{r_{11}s_{11}}{m_{11}}\bJ_{[m_{11}:m_{11}]}
&\frac{r_{11}s_{12}}{\sqrt{m_{11}m_{12}}}\bJ_{[m_{11}:m_{12}]}
&\frac{r_{11}s_{13}}{\sqrt{m_{11}m_{13}}}\bJ_{[m_{11}:m_{13}]}
&\frac{r_{12}s_{11}}{\sqrt{m_{11}m_{21}}}\bJ_{[m_{11}:m_{21}]}
&\frac{r_{12}s_{12}}{\sqrt{m_{11}m_{22}}}\bJ_{[m_{11}:m_{22}]}
&\frac{r_{12}s_{13}}{\sqrt{m_{11}m_{23}}}\bJ_{[m_{11}:m_{23}]}
\\[4pt]
\frac{r_{11}s_{21}}{\sqrt{m_{12}m_{11}}}\bJ_{[m_{12}:m_{11}]}
&\frac{r_{11}s_{22}}{m_{12}}\bJ_{[m_{12}:m_{12}]}
&\frac{r_{11}s_{23}}{\sqrt{m_{12}m_{13}}}\bJ_{[m_{12}:m_{13}]}
&\frac{r_{12}s_{21}}{\sqrt{m_{12}m_{21}}}\bJ_{[m_{12}:m_{21}]}
&\frac{r_{12}s_{22}}{\sqrt{m_{12}m_{22}}}\bJ_{[m_{12}:m_{22}]}
&\frac{r_{12}s_{23}}{\sqrt{m_{12}m_{23}}}\bJ_{[m_{12}:m_{23}]}
\\[4pt]
\frac{r_{11}s_{31}}{\sqrt{m_{13}m_{11}}}\bJ_{[m_{13}:m_{11}]}
&\frac{r_{11}s_{32}}{\sqrt{m_{13}m_{12}}}\bJ_{[m_{13}:m_{12}]}
&\frac{r_{11}s_{33}}{m_{13}}\bJ_{[m_{13}:m_{13}]}
&\frac{r_{12}s_{31}}{\sqrt{m_{13}m_{21}}}\bJ_{[m_{13}:m_{21}]}
&\frac{r_{12}s_{32}}{\sqrt{m_{13}m_{22}}}\bJ_{[m_{13}:m_{22}]}
&\frac{r_{12}s_{33}}{\sqrt{m_{13}m_{23}}}\bJ_{[m_{13}:m_{23}]}
\\[6pt]
\frac{r_{21}s_{11}}{\sqrt{m_{21}m_{11}}}\bJ_{[m_{21}:m_{11}]}
&\frac{r_{21}s_{12}}{\sqrt{m_{21}m_{12}}}\bJ_{[m_{21}:m_{12}]}
&\frac{r_{21}s_{13}}{\sqrt{m_{21}m_{13}}}\bJ_{[m_{21}:m_{13}]}
&\frac{r_{22}s_{11}}{m_{21}}\bJ_{[m_{21}:m_{21}]}
&\frac{r_{22}s_{12}}{\sqrt{m_{21}m_{22}}}\bJ_{[m_{21}:m_{22}]}
&\frac{r_{22}s_{13}}{\sqrt{m_{21}m_{23}}}\bJ_{[m_{21}:m_{23}]}
\\[4pt]
\frac{r_{21}s_{21}}{\sqrt{m_{22}m_{11}}}\bJ_{[m_{22}:m_{11}]}
&\frac{r_{21}s_{22}}{\sqrt{m_{22}m_{12}}}\bJ_{[m_{22}:m_{12}]}
&\frac{r_{21}s_{23}}{\sqrt{m_{22}m_{13}}}\bJ_{[m_{22}:m_{13}]}
&\frac{r_{22}s_{21}}{\sqrt{m_{22}m_{21}}}\bJ_{[m_{22}:m_{21}]}
&\frac{r_{22}s_{22}}{m_{22}}\bJ_{[m_{22}:m_{22}]}
&\frac{r_{22}s_{23}}{\sqrt{m_{22}m_{23}}}\bJ_{[m_{22}:m_{23}]}
\\[4pt]
\frac{r_{21}s_{31}}{\sqrt{m_{23}m_{11}}}\bJ_{[m_{23}:m_{11}]}
&\frac{r_{21}s_{32}}{\sqrt{m_{23}m_{12}}}\bJ_{[m_{23}:m_{12}]}
&\frac{r_{21}s_{33}}{\sqrt{m_{23}m_{13}}}\bJ_{[m_{23}:m_{13}]}
&\frac{r_{22}s_{31}}{\sqrt{m_{23}m_{21}}}\bJ_{[m_{23}:m_{21}]}
&\frac{r_{22}s_{32}}{\sqrt{m_{23}m_{22}}}\bJ_{[m_{23}:m_{22}]}
&\frac{r_{22}s_{33}}{m_{23}}\bJ_{[m_{23}:m_{23}]}
\end{bmatrix}$
}.
\end{align*}

\begin{align*}
 4).\quad  &(\bP\otimes\bQ)_{\text{cell}}\circledast\tildebJ_{\bm}
\\& =
 \resizebox{1.1\textwidth}{!}{$
\begin{bmatrix}
\frac{p_{11}q_{11}}{m_{11}^2}\bJ_{[m_{11}:m_{11}]}
&\frac{p_{11}q_{12}}{m_{11}m_{12}}\bJ_{[m_{11}:m_{12}]}
&\frac{p_{11}q_{13}}{m_{11}m_{13}}\bJ_{[m_{11}:m_{13}]}
&\frac{p_{12}q_{11}}{m_{11}m_{21}}\bJ_{[m_{11}:m_{21}]}
&\frac{p_{12}q_{12}}{m_{11}m_{22}}\bJ_{[m_{11}:m_{22}]}
&\frac{p_{12}q_{13}}{m_{11}m_{23}}\bJ_{[m_{11}:m_{23}]}
\\[4pt]
\frac{p_{11}q_{21}}{m_{12}m_{11}}\bJ_{[m_{12}:m_{11}]}
&\frac{p_{11}q_{22}}{m_{12}^2}\bJ_{[m_{12}:m_{12}]}
&\frac{p_{11}q_{23}}{m_{12}m_{13}}\bJ_{[m_{12}:m_{13}]}
&\frac{p_{12}q_{21}}{m_{12}m_{21}}\bJ_{[m_{12}:m_{21}]}
&\frac{p_{12}q_{22}}{m_{12}m_{22}}\bJ_{[m_{12}:m_{22}]}
&\frac{p_{12}q_{23}}{m_{12}m_{23}}\bJ_{[m_{12}:m_{23}]}
\\[4pt]
\frac{p_{11}q_{31}}{m_{13}m_{11}}\bJ_{[m_{13}:m_{11}]}
&\frac{p_{11}q_{32}}{m_{13}m_{12}}\bJ_{[m_{13}:m_{12}]}
&\frac{p_{11}q_{33}}{m_{13}^2}\bJ_{[m_{13}:m_{13}]}
&\frac{p_{12}q_{31}}{m_{13}m_{21}}\bJ_{[m_{13}:m_{21}]}
&\frac{p_{12}q_{32}}{m_{13}m_{22}}\bJ_{[m_{13}:m_{22}]}
&\frac{p_{12}q_{33}}{m_{13}m_{23}}\bJ_{[m_{13}:m_{23}]}
\\[6pt]
\frac{p_{21}q_{11}}{m_{21}m_{11}}\bJ_{[m_{21}:m_{11}]}
&\frac{p_{21}q_{12}}{m_{21}m_{12}}\bJ_{[m_{21}:m_{12}]}
&\frac{p_{21}q_{13}}{m_{21}m_{13}}\bJ_{[m_{21}:m_{13}]}
&\frac{p_{22}q_{11}}{m_{21}^2}\bJ_{[m_{21}:m_{21}]}
&\frac{p_{22}q_{12}}{m_{21}m_{22}}\bJ_{[m_{21}:m_{22}]}
&\frac{p_{22}q_{13}}{m_{21}m_{23}}\bJ_{[m_{21}:m_{23}]}
\\[4pt]
\frac{p_{21}q_{21}}{m_{22}m_{11}}\bJ_{[m_{22}:m_{11}]}
&\frac{p_{21}q_{22}}{m_{22}m_{12}}\bJ_{[m_{22}:m_{12}]}
&\frac{p_{21}q_{23}}{m_{22}m_{13}}\bJ_{[m_{22}:m_{13}]}
&\frac{p_{22}q_{21}}{m_{22}m_{21}}\bJ_{[m_{22}:m_{21}]}
&\frac{p_{22}q_{22}}{m_{22}^2}\bJ_{[m_{22}:m_{22}]}
&\frac{p_{22}q_{23}}{m_{22}m_{23}}\bJ_{[m_{22}:m_{23}]}
\\[4pt]
\frac{p_{21}q_{31}}{m_{23}m_{11}}\bJ_{[m_{23}:m_{11}]}
&\frac{p_{21}q_{32}}{m_{23}m_{12}}\bJ_{[m_{23}:m_{12}]}
&\frac{p_{21}q_{33}}{m_{23}m_{13}}\bJ_{[m_{23}:m_{13}]}
&\frac{p_{22}q_{31}}{m_{23}m_{21}}\bJ_{[m_{23}:m_{21}]}
&\frac{p_{22}q_{32}}{m_{23}m_{22}}\bJ_{[m_{23}:m_{22}]}
&\frac{p_{22}q_{33}}{m_{23}^2}\bJ_{[m_{23}:m_{23}]}
\end{bmatrix}$
}.
\end{align*}

\begin{align*}
 5).\quad  &(\bR\otimes\bS)_{\text{cell}}\circledast\tildebJ_{\bm}
\\& =
 \resizebox{1.1\textwidth}{!}{$
\begin{bmatrix}
\frac{r_{11}s_{11}}{m_{11}^2}\bJ_{[m_{11}:m_{11}]}
&\frac{r_{11}s_{12}}{m_{11}m_{12}}\bJ_{[m_{11}:m_{12}]}
&\frac{r_{11}s_{13}}{m_{11}m_{13}}\bJ_{[m_{11}:m_{13}]}
&\frac{r_{12}s_{11}}{m_{11}m_{21}}\bJ_{[m_{11}:m_{21}]}
&\frac{r_{12}s_{12}}{m_{11}m_{22}}\bJ_{[m_{11}:m_{22}]}
&\frac{r_{12}s_{13}}{m_{11}m_{23}}\bJ_{[m_{11}:m_{23}]}
\\[4pt]
\frac{r_{11}s_{21}}{m_{12}m_{11}}\bJ_{[m_{12}:m_{11}]}
&\frac{r_{11}s_{22}}{m_{12}^2}\bJ_{[m_{12}:m_{12}]}
&\frac{r_{11}s_{23}}{m_{12}m_{13}}\bJ_{[m_{12}:m_{13}]}
&\frac{r_{12}s_{21}}{m_{12}m_{21}}\bJ_{[m_{12}:m_{21}]}
&\frac{r_{12}s_{22}}{m_{12}m_{22}}\bJ_{[m_{12}:m_{22}]}
&\frac{r_{12}s_{23}}{m_{12}m_{23}}\bJ_{[m_{12}:m_{23}]}
\\[4pt]
\frac{r_{11}s_{31}}{m_{13}m_{11}}\bJ_{[m_{13}:m_{11}]}
&\frac{r_{11}s_{32}}{m_{13}m_{12}}\bJ_{[m_{13}:m_{12}]}
&\frac{r_{11}s_{33}}{m_{13}^2}\bJ_{[m_{13}:m_{13}]}
&\frac{r_{12}s_{31}}{m_{13}m_{21}}\bJ_{[m_{13}:m_{21}]}
&\frac{r_{12}s_{32}}{m_{13}m_{22}}\bJ_{[m_{13}:m_{22}]}
&\frac{r_{12}s_{33}}{m_{13}m_{23}}\bJ_{[m_{13}:m_{23}]}
\\[6pt]
\frac{r_{21}s_{11}}{m_{21}m_{11}}\bJ_{[m_{21}:m_{11}]}
&\frac{r_{21}s_{12}}{m_{21}m_{12}}\bJ_{[m_{21}:m_{12}]}
&\frac{r_{21}s_{13}}{m_{21}m_{13}}\bJ_{[m_{21}:m_{13}]}
&\frac{r_{22}s_{11}}{m_{21}^2}\bJ_{[m_{21}:m_{21}]}
&\frac{r_{22}s_{12}}{m_{21}m_{22}}\bJ_{[m_{21}:m_{22}]}
&\frac{r_{22}s_{13}}{m_{21}m_{23}}\bJ_{[m_{21}:m_{23}]}
\\[4pt]
\frac{r_{21}s_{21}}{m_{22}m_{11}}\bJ_{[m_{22}:m_{11}]}
&\frac{r_{21}s_{22}}{m_{22}m_{12}}\bJ_{[m_{22}:m_{12}]}
&\frac{r_{21}s_{23}}{m_{22}m_{13}}\bJ_{[m_{22}:m_{13}]}
&\frac{r_{22}s_{21}}{m_{22}m_{21}}\bJ_{[m_{22}:m_{21}]}
&\frac{r_{22}s_{22}}{m_{22}^2}\bJ_{[m_{22}:m_{22}]}
&\frac{r_{22}s_{23}}{m_{22}m_{23}}\bJ_{[m_{22}:m_{23}]}
\\[4pt]
\frac{r_{21}s_{31}}{m_{23}m_{11}}\bJ_{[m_{23}:m_{11}]}
&\frac{r_{21}s_{32}}{m_{23}m_{12}}\bJ_{[m_{23}:m_{12}]}
&\frac{r_{21}s_{33}}{m_{23}m_{13}}\bJ_{[m_{23}:m_{13}]}
&\frac{r_{22}s_{31}}{m_{23}m_{21}}\bJ_{[m_{23}:m_{21}]}
&\frac{r_{22}s_{32}}{m_{23}m_{22}}\bJ_{[m_{23}:m_{22}]}
&\frac{r_{22}s_{33}}{m_{23}^2}\bJ_{[m_{23}:m_{23}]}
\end{bmatrix}$
}.
\end{align*}

\begin{align*}
6).\quad  &\tildebJ_{\bm}\bI_{\bm}^f
\\&
=\resizebox{1.1\textwidth}{!}{$
\begin{bmatrix}
\frac{1}{m_{11}^2}\bJ_{[m_{11}:m_{11}]}
&\frac{1}{m_{11}m_{12}}\bJ_{[m_{11}:m_{12}]}
&\frac{1}{m_{11}m_{13}}\bJ_{[m_{11}:m_{13}]}
&\frac{1}{m_{11}m_{21}}\bJ_{[m_{11}:m_{21}]}
&\frac{1}{m_{11}m_{22}}\bJ_{[m_{11}:m_{22}]}
&\frac{1}{m_{11}m_{23}}\bJ_{[m_{11}:m_{23}]}
\\[4pt]
\frac{1}{m_{12}m_{11}}\bJ_{[m_{12}:m_{11}]}
&\frac{1}{m_{12}^2}\bJ_{[m_{12}:m_{12}]}
&\frac{1}{m_{12}m_{13}}\bJ_{[m_{12}:m_{13}]}
&\frac{1}{m_{12}m_{21}}\bJ_{[m_{12}:m_{21}]}
&\frac{1}{m_{12}m_{22}}\bJ_{[m_{12}:m_{22}]}
&\frac{1}{m_{12}m_{23}}\bJ_{[m_{12}:m_{23}]}
\\[4pt]
\frac{1}{m_{13}m_{11}}\bJ_{[m_{13}:m_{11}]}
&\frac{1}{m_{13}m_{12}}\bJ_{[m_{13}:m_{12}]}
&\frac{1}{m_{13}^2}\bJ_{[m_{13}:m_{13}]}
&\frac{1}{m_{13}m_{21}}\bJ_{[m_{13}:m_{21}]}
&\frac{1}{m_{13}m_{22}}\bJ_{[m_{13}:m_{22}]}
&\frac{1}{m_{13}m_{23}}\bJ_{[m_{13}:m_{23}]}
\\[6pt]
\frac{1}{m_{21}m_{11}}\bJ_{[m_{21}:m_{11}]}
&\frac{1}{m_{21}m_{12}}\bJ_{[m_{21}:m_{12}]}
&\frac{1}{m_{21}m_{13}}\bJ_{[m_{21}:m_{13}]}
&\frac{1}{m_{21}^2}\bJ_{[m_{21}:m_{21}]}
&\frac{1}{m_{21}m_{22}}\bJ_{[m_{21}:m_{22}]}
&\frac{1}{m_{21}m_{23}}\bJ_{[m_{21}:m_{23}]}
\\[4pt]
\frac{1}{m_{22}m_{11}}\bJ_{[m_{22}:m_{11}]}
&\frac{1}{m_{22}m_{12}}\bJ_{[m_{22}:m_{12}]}
&\frac{1}{m_{22}m_{13}}\bJ_{[m_{22}:m_{13}]}
&\frac{1}{m_{22}m_{21}}\bJ_{[m_{22}:m_{21}]}
&\frac{1}{m_{22}^2}\bJ_{[m_{22}:m_{22}]}
&\frac{1}{m_{22}m_{23}}\bJ_{[m_{22}:m_{23}]}
\\[4pt]
\frac{1}{m_{23}m_{11}}\bJ_{[m_{23}:m_{11}]}
&\frac{1}{m_{23}m_{12}}\bJ_{[m_{23}:m_{12}]}
&\frac{1}{m_{23}m_{13}}\bJ_{[m_{23}:m_{13}]}
&\frac{1}{m_{23}m_{21}}\bJ_{[m_{23}:m_{21}]}
&\frac{1}{m_{23}m_{22}}\bJ_{[m_{23}:m_{22}]}
&\frac{1}{m_{23}^2}\bJ_{[m_{23}:m_{23}]}
\end{bmatrix}$
}
\\&\quad
\times
\begin{bmatrix}
f(m_{11})\bI_{m_{11}}& &&&&\\
&f(m_{12})\bI_{m_{12}}&&&&\\
&&f(m_{13})\bI_{m_{13}}&&&\\
&&&f(m_{21})\bI_{m_{21}}&&\\
&&&&f(m_{22})\bI_{m_{22}}&\\
&&&&&f(m_{23})\bI_{m_{23}}\\
\end{bmatrix}
\\&
=
\resizebox{1.1\textwidth}{!}{$
\begin{bmatrix}
\frac{f(m_{11})}{m_{11}^2}\bJ_{[m_{11}:m_{11}]}
&\frac{f(m_{12})}{m_{11}m_{12}}\bJ_{[m_{11}:m_{12}]}
&\frac{f(m_{13})}{m_{11}m_{13}}\bJ_{[m_{11}:m_{13}]}
&\frac{f(m_{21})}{m_{11}m_{21}}\bJ_{[m_{11}:m_{21}]}
&\frac{f(m_{22})}{m_{11}m_{22}}\bJ_{[m_{11}:m_{22}]}
&\frac{f(m_{23})}{m_{11}m_{23}}\bJ_{[m_{11}:m_{23}]}
\\[4pt]
\frac{f(m_{11})}{m_{12}m_{11}}\bJ_{[m_{12}:m_{11}]}
&\frac{f(m_{12})}{m_{12}^2}\bJ_{[m_{12}:m_{12}]}
&\frac{f(m_{13})}{m_{12}m_{13}}\bJ_{[m_{12}:m_{13}]}
&\frac{f(m_{21})}{m_{12}m_{21}}\bJ_{[m_{12}:m_{21}]}
&\frac{f(m_{22})}{m_{12}m_{22}}\bJ_{[m_{12}:m_{22}]}
&\frac{f(m_{23})}{m_{12}m_{23}}\bJ_{[m_{12}:m_{23}]}
\\[4pt]
\frac{f(m_{11})}{m_{13}m_{11}}\bJ_{[m_{13}:m_{11}]}
&\frac{f(m_{12})}{m_{13}m_{12}}\bJ_{[m_{13}:m_{12}]}
&\frac{f(m_{13})}{m_{13}^2}\bJ_{[m_{13}:m_{13}]}
&\frac{f(m_{21})}{m_{13}m_{21}}\bJ_{[m_{13}:m_{21}]}
&\frac{f(m_{22})}{m_{13}m_{22}}\bJ_{[m_{13}:m_{22}]}
&\frac{f(m_{23})}{m_{13}m_{23}}\bJ_{[m_{13}:m_{23}]}
\\[6pt]
\frac{f(m_{11})}{m_{21}m_{11}}\bJ_{[m_{21}:m_{11}]}
&\frac{f(m_{12})}{m_{21}m_{12}}\bJ_{[m_{21}:m_{12}]}
&\frac{f(m_{13})}{m_{21}m_{13}}\bJ_{[m_{21}:m_{13}]}
&\frac{f(m_{21})}{m_{21}^2}\bJ_{[m_{21}:m_{21}]}
&\frac{f(m_{22})}{m_{21}m_{22}}\bJ_{[m_{21}:m_{22}]}
&\frac{f(m_{23})}{m_{21}m_{23}}\bJ_{[m_{21}:m_{23}]}
\\[4pt]
\frac{f(m_{11})}{m_{22}m_{11}}\bJ_{[m_{22}:m_{11}]}
&\frac{f(m_{12})}{m_{22}m_{12}}\bJ_{[m_{22}:m_{12}]}
&\frac{f(m_{13})}{m_{22}m_{13}}\bJ_{[m_{22}:m_{13}]}
&\frac{f(m_{21})}{m_{22}m_{21}}\bJ_{[m_{22}:m_{21}]}
&\frac{f(m_{22})}{m_{22}^2}\bJ_{[m_{22}:m_{22}]}
&\frac{f(m_{23})}{m_{22}m_{23}}\bJ_{[m_{22}:m_{23}]}
\\[4pt]
\frac{f(m_{11})}{m_{23}m_{11}}\bJ_{[m_{23}:m_{11}]}
&\frac{f(m_{12})}{m_{23}m_{12}}\bJ_{[m_{23}:m_{12}]}
&\frac{f(m_{13})}{m_{23}m_{13}}\bJ_{[m_{23}:m_{13}]}
&\frac{f(m_{21})}{m_{23}m_{21}}\bJ_{[m_{23}:m_{21}]}
&\frac{f(m_{22})}{m_{23}m_{22}}\bJ_{[m_{23}:m_{22}]}
&\frac{f(m_{23})}{m_{23}^2}\bJ_{[m_{23}:m_{23}]}
\end{bmatrix}$
}.
\end{align*}

\begin{align*}
7).\quad  &\bI_{\bm}^f\tildebJ_{\bm}
\\&
=
\begin{bmatrix}
f(m_{11})\bI_{m_{11}}& &&&&\\
&f(m_{12})\bI_{m_{12}}&&&&\\
&&f(m_{13})\bI_{m_{13}}&&&\\
&&&f(m_{21})\bI_{m_{21}}&&\\
&&&&f(m_{22})\bI_{m_{22}}&\\
&&&&&f(m_{23})\bI_{m_{23}}\\
\end{bmatrix}
\\&\quad
\times
\resizebox{1.1\textwidth}{!}{$
\begin{bmatrix}
\frac{1}{m_{11}^2}\bJ_{[m_{11}:m_{11}]}
&\frac{1}{m_{11}m_{12}}\bJ_{[m_{11}:m_{12}]}
&\frac{1}{m_{11}m_{13}}\bJ_{[m_{11}:m_{13}]}
&\frac{1}{m_{11}m_{21}}\bJ_{[m_{11}:m_{21}]}
&\frac{1}{m_{11}m_{22}}\bJ_{[m_{11}:m_{22}]}
&\frac{1}{m_{11}m_{23}}\bJ_{[m_{11}:m_{23}]}
\\[4pt]
\frac{1}{m_{12}m_{11}}\bJ_{[m_{12}:m_{11}]}
&\frac{1}{m_{12}^2}\bJ_{[m_{12}:m_{12}]}
&\frac{1}{m_{12}m_{13}}\bJ_{[m_{12}:m_{13}]}
&\frac{1}{m_{12}m_{21}}\bJ_{[m_{12}:m_{21}]}
&\frac{1}{m_{12}m_{22}}\bJ_{[m_{12}:m_{22}]}
&\frac{1}{m_{12}m_{23}}\bJ_{[m_{12}:m_{23}]}
\\[4pt]
\frac{1}{m_{13}m_{11}}\bJ_{[m_{13}:m_{11}]}
&\frac{1}{m_{13}m_{12}}\bJ_{[m_{13}:m_{12}]}
&\frac{1}{m_{13}^2}\bJ_{[m_{13}:m_{13}]}
&\frac{1}{m_{13}m_{21}}\bJ_{[m_{13}:m_{21}]}
&\frac{1}{m_{13}m_{22}}\bJ_{[m_{13}:m_{22}]}
&\frac{1}{m_{13}m_{23}}\bJ_{[m_{13}:m_{23}]}
\\[6pt]
\frac{1}{m_{21}m_{11}}\bJ_{[m_{21}:m_{11}]}
&\frac{1}{m_{21}m_{12}}\bJ_{[m_{21}:m_{12}]}
&\frac{1}{m_{21}m_{13}}\bJ_{[m_{21}:m_{13}]}
&\frac{1}{m_{21}^2}\bJ_{[m_{21}:m_{21}]}
&\frac{1}{m_{21}m_{22}}\bJ_{[m_{21}:m_{22}]}
&\frac{1}{m_{21}m_{23}}\bJ_{[m_{21}:m_{23}]}
\\[4pt]
\frac{1}{m_{22}m_{11}}\bJ_{[m_{22}:m_{11}]}
&\frac{1}{m_{22}m_{12}}\bJ_{[m_{22}:m_{12}]}
&\frac{1}{m_{22}m_{13}}\bJ_{[m_{22}:m_{13}]}
&\frac{1}{m_{22}m_{21}}\bJ_{[m_{22}:m_{21}]}
&\frac{1}{m_{22}^2}\bJ_{[m_{22}:m_{22}]}
&\frac{1}{m_{22}m_{23}}\bJ_{[m_{22}:m_{23}]}
\\[4pt]
\frac{1}{m_{23}m_{11}}\bJ_{[m_{23}:m_{11}]}
&\frac{1}{m_{23}m_{12}}\bJ_{[m_{23}:m_{12}]}
&\frac{1}{m_{23}m_{13}}\bJ_{[m_{23}:m_{13}]}
&\frac{1}{m_{23}m_{21}}\bJ_{[m_{23}:m_{21}]}
&\frac{1}{m_{23}m_{22}}\bJ_{[m_{23}:m_{22}]}
&\frac{1}{m_{23}^2}\bJ_{[m_{23}:m_{23}]}
\end{bmatrix}$
}
\\&
=
\resizebox{1.1\textwidth}{!}{$
\begin{bmatrix}
\frac{f(m_{11})}{m_{11}^2}\bJ_{[m_{11}:m_{11}]}
&\frac{f(m_{11})}{m_{11}m_{12}}\bJ_{[m_{11}:m_{12}]}
&\frac{f(m_{11})}{m_{11}m_{13}}\bJ_{[m_{11}:m_{13}]}
&\frac{f(m_{11})}{m_{11}m_{21}}\bJ_{[m_{11}:m_{21}]}
&\frac{f(m_{11})}{m_{11}m_{22}}\bJ_{[m_{11}:m_{22}]}
&\frac{f(m_{11})}{m_{11}m_{23}}\bJ_{[m_{11}:m_{23}]}
\\[4pt]
\frac{f(m_{12})}{m_{12}m_{11}}\bJ_{[m_{12}:m_{11}]}
&\frac{f(m_{12})}{m_{12}^2}\bJ_{[m_{12}:m_{12}]}
&\frac{f(m_{12})}{m_{12}m_{13}}\bJ_{[m_{12}:m_{13}]}
&\frac{f(m_{12})}{m_{12}m_{21}}\bJ_{[m_{12}:m_{21}]}
&\frac{f(m_{12})}{m_{12}m_{22}}\bJ_{[m_{12}:m_{22}]}
&\frac{f(m_{12})}{m_{12}m_{23}}\bJ_{[m_{12}:m_{23}]}
\\[4pt]
\frac{f(m_{13})}{m_{13}m_{11}}\bJ_{[m_{13}:m_{11}]}
&\frac{f(m_{13})}{m_{13}m_{12}}\bJ_{[m_{13}:m_{12}]}
&\frac{f(m_{13})}{m_{13}^2}\bJ_{[m_{13}:m_{13}]}
&\frac{f(m_{13})}{m_{13}m_{21}}\bJ_{[m_{13}:m_{21}]}
&\frac{f(m_{13})}{m_{13}m_{22}}\bJ_{[m_{13}:m_{22}]}
&\frac{f(m_{13})}{m_{13}m_{23}}\bJ_{[m_{13}:m_{23}]}
\\[6pt]
\frac{f(m_{21})}{m_{21}m_{11}}\bJ_{[m_{21}:m_{11}]}
&\frac{f(m_{21})}{m_{21}m_{12}}\bJ_{[m_{21}:m_{12}]}
&\frac{f(m_{21})}{m_{21}m_{13}}\bJ_{[m_{21}:m_{13}]}
&\frac{f(m_{21})}{m_{21}^2}\bJ_{[m_{21}:m_{21}]}
&\frac{f(m_{21})}{m_{21}m_{22}}\bJ_{[m_{21}:m_{22}]}
&\frac{f(m_{21})}{m_{21}m_{23}}\bJ_{[m_{21}:m_{23}]}
\\[4pt]
\frac{f(m_{22})}{m_{22}m_{11}}\bJ_{[m_{22}:m_{11}]}
&\frac{f(m_{22})}{m_{22}m_{12}}\bJ_{[m_{22}:m_{12}]}
&\frac{f(m_{22})}{m_{22}m_{13}}\bJ_{[m_{22}:m_{13}]}
&\frac{f(m_{22})}{m_{22}m_{21}}\bJ_{[m_{22}:m_{21}]}
&\frac{f(m_{22})}{m_{22}^2}\bJ_{[m_{22}:m_{22}]}
&\frac{f(m_{22})}{m_{22}m_{23}}\bJ_{[m_{22}:m_{23}]}
\\[4pt]
\frac{f(m_{23})}{m_{23}m_{11}}\bJ_{[m_{23}:m_{11}]}
&\frac{f(m_{23})}{m_{23}m_{12}}\bJ_{[m_{23}:m_{12}]}
&\frac{f(m_{23})}{m_{23}m_{13}}\bJ_{[m_{23}:m_{13}]}
&\frac{f(m_{23})}{m_{23}m_{21}}\bJ_{[m_{23}:m_{21}]}
&\frac{f(m_{23})}{m_{23}m_{22}}\bJ_{[m_{23}:m_{22}]}
&\frac{f(m_{23})}{m_{23}^2}\bJ_{[m_{23}:m_{23}]}
\end{bmatrix}$
}.
\end{align*}

\begin{align*}
 8).\quad  &(\bP\otimes\bQ)_{\text{cell}}\circledast\tildebJ_{\bm}\bI_{\bm}^f
\\& =
\resizebox{1.1\textwidth}{!}{$
\begin{bmatrix}
\frac{p_{11}q_{11}\,f(m_{11})}{m_{11}^2}\bJ_{[m_{11}:m_{11}]}
&\frac{p_{11}q_{12}\,f(m_{12})}{m_{11}m_{12}}\bJ_{[m_{11}:m_{12}]}
&\frac{p_{11}q_{13}\,f(m_{13})}{m_{11}m_{13}}\bJ_{[m_{11}:m_{13}]}
&\frac{p_{12}q_{11}\,f(m_{21})}{m_{11}m_{21}}\bJ_{[m_{11}:m_{21}]}
&\frac{p_{12}q_{12}\,f(m_{22})}{m_{11}m_{22}}\bJ_{[m_{11}:m_{22}]}
&\frac{p_{12}q_{13}\,f(m_{23})}{m_{11}m_{23}}\bJ_{[m_{11}:m_{23}]}
\\[4pt]
\frac{p_{11}q_{21}\,f(m_{11})}{m_{12}m_{11}}\bJ_{[m_{12}:m_{11}]}
&\frac{p_{11}q_{22}\,f(m_{12})}{m_{12}^2}\bJ_{[m_{12}:m_{12}]}
&\frac{p_{11}q_{23}\,f(m_{13})}{m_{12}m_{13}}\bJ_{[m_{12}:m_{13}]}
&\frac{p_{12}q_{21}\,f(m_{21})}{m_{12}m_{21}}\bJ_{[m_{12}:m_{21}]}
&\frac{p_{12}q_{22}\,f(m_{22})}{m_{12}m_{22}}\bJ_{[m_{12}:m_{22}]}
&\frac{p_{12}q_{23}\,f(m_{23})}{m_{12}m_{23}}\bJ_{[m_{12}:m_{23}]}
\\[4pt]
\frac{p_{11}q_{31}\,f(m_{11})}{m_{13}m_{11}}\bJ_{[m_{13}:m_{11}]}
&\frac{p_{11}q_{32}\,f(m_{12})}{m_{13}m_{12}}\bJ_{[m_{13}:m_{12}]}
&\frac{p_{11}q_{33}\,f(m_{13})}{m_{13}^2}\bJ_{[m_{13}:m_{13}]}
&\frac{p_{12}q_{31}\,f(m_{21})}{m_{13}m_{21}}\bJ_{[m_{13}:m_{21}]}
&\frac{p_{12}q_{32}\,f(m_{22})}{m_{13}m_{22}}\bJ_{[m_{13}:m_{22}]}
&\frac{p_{12}q_{33}\,f(m_{23})}{m_{13}m_{23}}\bJ_{[m_{13}:m_{23}]}
\\[6pt]
\frac{p_{21}q_{11}\,f(m_{11})}{m_{21}m_{11}}\bJ_{[m_{21}:m_{11}]}
&\frac{p_{21}q_{12}\,f(m_{12})}{m_{21}m_{12}}\bJ_{[m_{21}:m_{12}]}
&\frac{p_{21}q_{13}\,f(m_{13})}{m_{21}m_{13}}\bJ_{[m_{21}:m_{13}]}
&\frac{p_{22}q_{11}\,f(m_{21})}{m_{21}^2}\bJ_{[m_{21}:m_{21}]}
&\frac{p_{22}q_{12}\,f(m_{22})}{m_{21}m_{22}}\bJ_{[m_{21}:m_{22}]}
&\frac{p_{22}q_{13}\,f(m_{23})}{m_{21}m_{23}}\bJ_{[m_{21}:m_{23}]}
\\[4pt]
\frac{p_{21}q_{21}\,f(m_{11})}{m_{22}m_{11}}\bJ_{[m_{22}:m_{11}]}
&\frac{p_{21}q_{22}\,f(m_{12})}{m_{22}m_{12}}\bJ_{[m_{22}:m_{12}]}
&\frac{p_{21}q_{23}\,f(m_{13})}{m_{22}m_{13}}\bJ_{[m_{22}:m_{13}]}
&\frac{p_{22}q_{21}\,f(m_{21})}{m_{22}m_{21}}\bJ_{[m_{22}:m_{21}]}
&\frac{p_{22}q_{22}\,f(m_{22})}{m_{22}^2}\bJ_{[m_{22}:m_{22}]}
&\frac{p_{22}q_{23}\,f(m_{23})}{m_{22}m_{23}}\bJ_{[m_{22}:m_{23}]}
\\[4pt]
\frac{p_{21}q_{31}\,f(m_{11})}{m_{23}m_{11}}\bJ_{[m_{23}:m_{11}]}
&\frac{p_{21}q_{32}\,f(m_{12})}{m_{23}m_{12}}\bJ_{[m_{23}:m_{12}]}
&\frac{p_{21}q_{33}\,f(m_{13})}{m_{23}m_{13}}\bJ_{[m_{23}:m_{13}]}
&\frac{p_{22}q_{31}\,f(m_{21})}{m_{23}m_{21}}\bJ_{[m_{23}:m_{21}]}
&\frac{p_{22}q_{32}\,f(m_{22})}{m_{23}m_{22}}\bJ_{[m_{23}:m_{22}]}
&\frac{p_{22}q_{33}\,f(m_{23})}{m_{23}^2}\bJ_{[m_{23}:m_{23}]}
\end{bmatrix}$
}.
\end{align*}

\begin{align*}
 9).\quad  &(\bP\otimes\bQ)_{\text{cell}}\circledast \bI_{\bm}^f\tildebJ_{\bm}
\\& =
\resizebox{1.1\textwidth}{!}{$
\begin{bmatrix}
\frac{f(m_{11})\,p_{11}q_{11}}{m_{11}^2}\bJ_{[m_{11}:m_{11}]}
&\frac{f(m_{11})\,p_{11}q_{12}}{m_{11}m_{12}}\bJ_{[m_{11}:m_{12}]}
&\frac{f(m_{11})\,p_{11}q_{13}}{m_{11}m_{13}}\bJ_{[m_{11}:m_{13}]}
&\frac{f(m_{11})\,p_{12}q_{11}}{m_{11}m_{21}}\bJ_{[m_{11}:m_{21}]}
&\frac{f(m_{11})\,p_{12}q_{12}}{m_{11}m_{22}}\bJ_{[m_{11}:m_{22}]}
&\frac{f(m_{11})\,p_{12}q_{13}}{m_{11}m_{23}}\bJ_{[m_{11}:m_{23}]}
\\[4pt]
\frac{f(m_{12})\,p_{11}q_{21}}{m_{12}m_{11}}\bJ_{[m_{12}:m_{11}]}
&\frac{f(m_{12})\,p_{11}q_{22}}{m_{12}^2}\bJ_{[m_{12}:m_{12}]}
&\frac{f(m_{12})\,p_{11}q_{23}}{m_{12}m_{13}}\bJ_{[m_{12}:m_{13}]}
&\frac{f(m_{12})\,p_{12}q_{21}}{m_{12}m_{21}}\bJ_{[m_{12}:m_{21}]}
&\frac{f(m_{12})\,p_{12}q_{22}}{m_{12}m_{22}}\bJ_{[m_{12}:m_{22}]}
&\frac{f(m_{12})\,p_{12}q_{23}}{m_{12}m_{23}}\bJ_{[m_{12}:m_{23}]}
\\[4pt]
\frac{f(m_{13})\,p_{11}q_{31}}{m_{13}m_{11}}\bJ_{[m_{13}:m_{11}]}
&\frac{f(m_{13})\,p_{11}q_{32}}{m_{13}m_{12}}\bJ_{[m_{13}:m_{12}]}
&\frac{f(m_{13})\,p_{11}q_{33}}{m_{13}^2}\bJ_{[m_{13}:m_{13}]}
&\frac{f(m_{13})\,p_{12}q_{31}}{m_{13}m_{21}}\bJ_{[m_{13}:m_{21}]}
&\frac{f(m_{13})\,p_{12}q_{32}}{m_{13}m_{22}}\bJ_{[m_{13}:m_{22}]}
&\frac{f(m_{13})\,p_{12}q_{33}}{m_{13}m_{23}}\bJ_{[m_{13}:m_{23}]}
\\[6pt]
\frac{f(m_{21})\,p_{21}q_{11}}{m_{21}m_{11}}\bJ_{[m_{21}:m_{11}]}
&\frac{f(m_{21})\,p_{21}q_{12}}{m_{21}m_{12}}\bJ_{[m_{21}:m_{12}]}
&\frac{f(m_{21})\,p_{21}q_{13}}{m_{21}m_{13}}\bJ_{[m_{21}:m_{13}]}
&\frac{f(m_{21})\,p_{22}q_{11}}{m_{21}^2}\bJ_{[m_{21}:m_{21}]}
&\frac{f(m_{21})\,p_{22}q_{12}}{m_{21}m_{22}}\bJ_{[m_{21}:m_{22}]}
&\frac{f(m_{21})\,p_{22}q_{13}}{m_{21}m_{23}}\bJ_{[m_{21}:m_{23}]}
\\[4pt]
\frac{f(m_{22})\,p_{21}q_{21}}{m_{22}m_{11}}\bJ_{[m_{22}:m_{11}]}
&\frac{f(m_{22})\,p_{21}q_{22}}{m_{22}m_{12}}\bJ_{[m_{22}:m_{12}]}
&\frac{f(m_{22})\,p_{21}q_{23}}{m_{22}m_{13}}\bJ_{[m_{22}:m_{13}]}
&\frac{f(m_{22})\,p_{22}q_{21}}{m_{22}m_{21}}\bJ_{[m_{22}:m_{21}]}
&\frac{f(m_{22})\,p_{22}q_{22}}{m_{22}^2}\bJ_{[m_{22}:m_{22}]}
&\frac{f(m_{22})\,p_{22}q_{23}}{m_{22}m_{23}}\bJ_{[m_{22}:m_{23}]}
\\[4pt]
\frac{f(m_{23})\,p_{21}q_{31}}{m_{23}m_{11}}\bJ_{[m_{23}:m_{11}]}
&\frac{f(m_{23})\,p_{21}q_{32}}{m_{23}m_{12}}\bJ_{[m_{23}:m_{12}]}
&\frac{f(m_{23})\,p_{21}q_{33}}{m_{23}m_{13}}\bJ_{[m_{23}:m_{13}]}
&\frac{f(m_{23})\,p_{22}q_{31}}{m_{23}m_{21}}\bJ_{[m_{23}:m_{21}]}
&\frac{f(m_{23})\,p_{22}q_{32}}{m_{23}m_{22}}\bJ_{[m_{23}:m_{22}]}
&\frac{f(m_{23})\,p_{22}q_{33}}{m_{23}^2}\bJ_{[m_{23}:m_{23}]}
\end{bmatrix}$
}.
\end{align*}

\begin{align*}
10).\quad&[(\ba\otimes\bb)_{\text{row}}\circledast\barbone_{\bm}]^\top
[(\bP\otimes\bQ)_{\text{cell}}\circledast\barbJ{\bm}]
\\&=
\begin{bmatrix}
\frac{a_1b_1}{\sqrt{m_{11}}}\bone_{m_{11}}^{\!\top} &
\frac{a_1b_2}{\sqrt{m_{12}}}\bone_{m_{12}}^{\!\top} &
\frac{a_1b_3}{\sqrt{m_{13}}}\bone_{m_{13}}^{\!\top} &
\frac{a_2b_1}{\sqrt{m_{21}}}\bone_{m_{21}}^{\!\top} &
\frac{a_2b_2}{\sqrt{m_{22}}}\bone_{m_{22}}^{\!\top} &
\frac{a_2b_3}{\sqrt{m_{23}}}\bone_{m_{23}}^{\!\top}
\end{bmatrix}
\\& \times
\resizebox{1.1\textwidth}{!}{$
\begin{bmatrix}
\frac{p_{11}q_{11}}{m_{11}}\bJ_{[m_{11}:m_{11}]}
&\frac{p_{11}q_{12}}{\sqrt{m_{11}m_{12}}}\bJ_{[m_{11}:m_{12}]}
&\frac{p_{11}q_{13}}{\sqrt{m_{11}m_{13}}}\bJ_{[m_{11}:m_{13}]}
&\frac{p_{12}q_{11}}{\sqrt{m_{11}m_{21}}}\bJ_{[m_{11}:m_{21}]}
&\frac{p_{12}q_{12}}{\sqrt{m_{11}m_{22}}}\bJ_{[m_{11}:m_{22}]}
&\frac{p_{12}q_{13}}{\sqrt{m_{11}m_{23}}}\bJ_{[m_{11}:m_{23}]}
\\[4pt]
\frac{p_{11}q_{21}}{\sqrt{m_{12}m_{11}}}\bJ_{[m_{12}:m_{11}]}
&\frac{p_{11}q_{22}}{m_{12}}\bJ_{[m_{12}:m_{12}]}
&\frac{p_{11}q_{23}}{\sqrt{m_{12}m_{13}}}\bJ_{[m_{12}:m_{13}]}
&\frac{p_{12}q_{21}}{\sqrt{m_{12}m_{21}}}\bJ_{[m_{12}:m_{21}]}
&\frac{p_{12}q_{22}}{\sqrt{m_{12}m_{22}}}\bJ_{[m_{12}:m_{22}]}
&\frac{p_{12}q_{23}}{\sqrt{m_{12}m_{23}}}\bJ_{[m_{12}:m_{23}]}
\\[4pt]
\frac{p_{11}q_{31}}{\sqrt{m_{13}m_{11}}}\bJ_{[m_{13}:m_{11}]}
&\frac{p_{11}q_{32}}{\sqrt{m_{13}m_{12}}}\bJ_{[m_{13}:m_{12}]}
&\frac{p_{11}q_{33}}{m_{13}}\bJ_{[m_{13}:m_{13}]}
&\frac{p_{12}q_{31}}{\sqrt{m_{13}m_{21}}}\bJ_{[m_{13}:m_{21}]}
&\frac{p_{12}q_{32}}{\sqrt{m_{13}m_{22}}}\bJ_{[m_{13}:m_{22}]}
&\frac{p_{12}q_{33}}{\sqrt{m_{13}m_{23}}}\bJ_{[m_{13}:m_{23}]}
\\[6pt]
\frac{p_{21}q_{11}}{\sqrt{m_{21}m_{11}}}\bJ_{[m_{21}:m_{11}]}
&\frac{p_{21}q_{12}}{\sqrt{m_{21}m_{12}}}\bJ_{[m_{21}:m_{12}]}
&\frac{p_{21}q_{13}}{\sqrt{m_{21}m_{13}}}\bJ_{[m_{21}:m_{13}]}
&\frac{p_{22}q_{11}}{m_{21}}\bJ_{[m_{21}:m_{21}]}
&\frac{p_{22}q_{12}}{\sqrt{m_{21}m_{22}}}\bJ_{[m_{21}:m_{22}]}
&\frac{p_{22}q_{13}}{\sqrt{m_{21}m_{23}}}\bJ_{[m_{21}:m_{23}]}
\\[4pt]
\frac{p_{21}q_{21}}{\sqrt{m_{22}m_{11}}}\bJ_{[m_{22}:m_{11}]}
&\frac{p_{21}q_{22}}{\sqrt{m_{22}m_{12}}}\bJ_{[m_{22}:m_{12}]}
&\frac{p_{21}q_{23}}{\sqrt{m_{22}m_{13}}}\bJ_{[m_{22}:m_{13}]}
&\frac{p_{22}q_{21}}{\sqrt{m_{22}m_{21}}}\bJ_{[m_{22}:m_{21}]}
&\frac{p_{22}q_{22}}{m_{22}}\bJ_{[m_{22}:m_{22}]}
&\frac{p_{22}q_{23}}{\sqrt{m_{22}m_{23}}}\bJ_{[m_{22}:m_{23}]}
\\[4pt]
\frac{p_{21}q_{31}}{\sqrt{m_{23}m_{11}}}\bJ_{[m_{23}:m_{11}]}
&\frac{p_{21}q_{32}}{\sqrt{m_{23}m_{12}}}\bJ_{[m_{23}:m_{12}]}
&\frac{p_{21}q_{33}}{\sqrt{m_{23}m_{13}}}\bJ_{[m_{23}:m_{13}]}
&\frac{p_{22}q_{31}}{\sqrt{m_{23}m_{21}}}\bJ_{[m_{23}:m_{21}]}
&\frac{p_{22}q_{32}}{\sqrt{m_{23}m_{22}}}\bJ_{[m_{23}:m_{22}]}
&\frac{p_{22}q_{33}}{m_{23}}\bJ_{[m_{23}:m_{23}]}
\end{bmatrix}$
}
\\&
=\sum_{i=1}^2\sum_{j=1}^3
\begin{bmatrix}
\frac{a_ib_jp_{i1}q_{j1}}{\sqrt{m_{11}}}\bone_{m_{11}}^{\!\top} &
\frac{a_ib_jp_{i1}q_{j2}}{\sqrt{m_{12}}}\bone_{m_{12}}^{\!\top} &
\frac{a_ib_jp_{i1}q_{j3}}{\sqrt{m_{13}}}\bone_{m_{13}}^{\!\top} &
\frac{a_ib_jp_{i2}q_{j1}}{\sqrt{m_{21}}}\bone_{m_{21}}^{\!\top} &
\frac{a_ib_jp_{i2}q_{j2}}{\sqrt{m_{22}}}\bone_{m_{22}}^{\!\top} &
\frac{a_ib_jp_{i2}q_{j3}}{\sqrt{m_{23}}}\bone_{m_{23}}^{\!\top}
\end{bmatrix}.
\end{align*}

\begin{align*}
11). \quad&(\ba^\top\bP\otimes\bb^\top\bQ)_{\text{col}}\circledast\barbone_{\bm}^\top
\\&
=\left\lbrace\left(\begin{bmatrix}
a_1&a_2
\end{bmatrix}\times
\begin{bmatrix}
p_{11}&p_{12}\\
p_{21}&p_{22}
\end{bmatrix}\right)
\otimes
\left(
\begin{bmatrix}
b_1&b_2&b_3
\end{bmatrix}\times
\begin{bmatrix}
q_{11}&q_{12}&q_{13}\\
q_{21}&q_{22}&q_{23}\\
q_{31}&q_{32}&q_{33}\\
\end{bmatrix}
\right)\right\rbrace_{\text{col}}
\\&\qquad
\circledast
\begin{bmatrix}
\frac{1}{\sqrt{m_{11}}}\bone_{m_{11}}^{\!\top} &
\frac{1}{\sqrt{m_{12}}}\bone_{m_{12}}^{\!\top} &
\frac{1}{\sqrt{m_{13}}}\bone_{m_{13}}^{\!\top} &
\frac{1}{\sqrt{m_{21}}}\bone_{m_{21}}^{\!\top} &
\frac{1}{\sqrt{m_{22}}}\bone_{m_{22}}^{\!\top} &
\frac{1}{\sqrt{m_{23}}}\bone_{m_{23}}^{\!\top}
\end{bmatrix}
\\&
=\left\lbrace\left( 
\sum_{i=1}^2\begin{bmatrix}
    a_ip_{i1}&a_ip_{i2}
\end{bmatrix}
\right)\otimes
\left(
\sum_{j=1}^3\begin{bmatrix}
    b_jq_{j1}&b_jq_{j2}&b_jq_{j3}
\end{bmatrix}
\right)\right\rbrace_{\text{col}}
\\&\qquad
\circledast
\begin{bmatrix}
\frac{1}{\sqrt{m_{11}}}\bone_{m_{11}}^{\!\top} &
\frac{1}{\sqrt{m_{12}}}\bone_{m_{12}}^{\!\top} &
\frac{1}{\sqrt{m_{13}}}\bone_{m_{13}}^{\!\top} &
\frac{1}{\sqrt{m_{21}}}\bone_{m_{21}}^{\!\top} &
\frac{1}{\sqrt{m_{22}}}\bone_{m_{22}}^{\!\top} &
\frac{1}{\sqrt{m_{23}}}\bone_{m_{23}}^{\!\top}
\end{bmatrix}
\\&
=\sum_{i=1}^2\sum_{j=1}^3
\begin{bmatrix}
  a_ib_jp_{i1}q_{j1}&a_ib_jp_{i1}q_{j2}&a_ib_jp_{i1}q_{j3}  
 & a_ib_jp_{i2}q_{j1}&a_ib_jp_{i2}q_{j2}&a_ib_jp_{i2}q_{j3}  
\end{bmatrix}_{\text{col}}
\\&\qquad
\circledast
\begin{bmatrix}
\frac{1}{\sqrt{m_{11}}}\bone_{m_{11}}^{\!\top} &
\frac{1}{\sqrt{m_{12}}}\bone_{m_{12}}^{\!\top} &
\frac{1}{\sqrt{m_{13}}}\bone_{m_{13}}^{\!\top} &
\frac{1}{\sqrt{m_{21}}}\bone_{m_{21}}^{\!\top} &
\frac{1}{\sqrt{m_{22}}}\bone_{m_{22}}^{\!\top} &
\frac{1}{\sqrt{m_{23}}}\bone_{m_{23}}^{\!\top}
\end{bmatrix}
\\&
=\sum_{i=1}^2\sum_{j=1}^3
\begin{bmatrix}
\frac{a_ib_jp_{i1}q_{j1}}{\sqrt{m_{11}}}\bone_{m_{11}}^{\!\top} &
\frac{a_ib_jp_{i1}q_{j2}}{\sqrt{m_{12}}}\bone_{m_{12}}^{\!\top} &
\frac{a_ib_jp_{i1}q_{j3}}{\sqrt{m_{13}}}\bone_{m_{13}}^{\!\top} &
\frac{a_ib_jp_{i2}q_{j1}}{\sqrt{m_{21}}}\bone_{m_{21}}^{\!\top} &
\frac{a_ib_jp_{i2}q_{j2}}{\sqrt{m_{22}}}\bone_{m_{22}}^{\!\top} &
\frac{a_ib_jp_{i2}q_{j3}}{\sqrt{m_{23}}}\bone_{m_{23}}^{\!\top}
\end{bmatrix}.
\end{align*}
Here we find 10) and 11) are the same, namely
$[(\ba\otimes\bb)_{\text{row}}\circledast\barbone_{\bm}]^\top
[(\bP\otimes\bQ)_{\text{cell}}\circledast\barbJ{\bm}]=
(\ba^\top\bP\otimes\bb^\top\bQ)_{\text{col}}\circledast\barbone_{\bm}^\top$.

\begin{align*}
12).\quad &
[(\bP\otimes\bQ)_{\text{cell}}\circledast\barbJ{\bm}]
[(\bR\otimes\bS)_{\text{cell}}\circledast\barbJ{\bm}]
\\&=
\resizebox{1.1\textwidth}{!}{$
\begin{bmatrix}
\frac{p_{11}q_{11}}{m_{11}}\bJ_{[m_{11}:m_{11}]}
&\frac{p_{11}q_{12}}{\sqrt{m_{11}m_{12}}}\bJ_{[m_{11}:m_{12}]}
&\frac{p_{11}q_{13}}{\sqrt{m_{11}m_{13}}}\bJ_{[m_{11}:m_{13}]}
&\frac{p_{12}q_{11}}{\sqrt{m_{11}m_{21}}}\bJ_{[m_{11}:m_{21}]}
&\frac{p_{12}q_{12}}{\sqrt{m_{11}m_{22}}}\bJ_{[m_{11}:m_{22}]}
&\frac{p_{12}q_{13}}{\sqrt{m_{11}m_{23}}}\bJ_{[m_{11}:m_{23}]}
\\[4pt]
\frac{p_{11}q_{21}}{\sqrt{m_{12}m_{11}}}\bJ_{[m_{12}:m_{11}]}
&\frac{p_{11}q_{22}}{m_{12}}\bJ_{[m_{12}:m_{12}]}
&\frac{p_{11}q_{23}}{\sqrt{m_{12}m_{13}}}\bJ_{[m_{12}:m_{13}]}
&\frac{p_{12}q_{21}}{\sqrt{m_{12}m_{21}}}\bJ_{[m_{12}:m_{21}]}
&\frac{p_{12}q_{22}}{\sqrt{m_{12}m_{22}}}\bJ_{[m_{12}:m_{22}]}
&\frac{p_{12}q_{23}}{\sqrt{m_{12}m_{23}}}\bJ_{[m_{12}:m_{23}]}
\\[4pt]
\frac{p_{11}q_{31}}{\sqrt{m_{13}m_{11}}}\bJ_{[m_{13}:m_{11}]}
&\frac{p_{11}q_{32}}{\sqrt{m_{13}m_{12}}}\bJ_{[m_{13}:m_{12}]}
&\frac{p_{11}q_{33}}{m_{13}}\bJ_{[m_{13}:m_{13}]}
&\frac{p_{12}q_{31}}{\sqrt{m_{13}m_{21}}}\bJ_{[m_{13}:m_{21}]}
&\frac{p_{12}q_{32}}{\sqrt{m_{13}m_{22}}}\bJ_{[m_{13}:m_{22}]}
&\frac{p_{12}q_{33}}{\sqrt{m_{13}m_{23}}}\bJ_{[m_{13}:m_{23}]}
\\[6pt]
\frac{p_{21}q_{11}}{\sqrt{m_{21}m_{11}}}\bJ_{[m_{21}:m_{11}]}
&\frac{p_{21}q_{12}}{\sqrt{m_{21}m_{12}}}\bJ_{[m_{21}:m_{12}]}
&\frac{p_{21}q_{13}}{\sqrt{m_{21}m_{13}}}\bJ_{[m_{21}:m_{13}]}
&\frac{p_{22}q_{11}}{m_{21}}\bJ_{[m_{21}:m_{21}]}
&\frac{p_{22}q_{12}}{\sqrt{m_{21}m_{22}}}\bJ_{[m_{21}:m_{22}]}
&\frac{p_{22}q_{13}}{\sqrt{m_{21}m_{23}}}\bJ_{[m_{21}:m_{23}]}
\\[4pt]
\frac{p_{21}q_{21}}{\sqrt{m_{22}m_{11}}}\bJ_{[m_{22}:m_{11}]}
&\frac{p_{21}q_{22}}{\sqrt{m_{22}m_{12}}}\bJ_{[m_{22}:m_{12}]}
&\frac{p_{21}q_{23}}{\sqrt{m_{22}m_{13}}}\bJ_{[m_{22}:m_{13}]}
&\frac{p_{22}q_{21}}{\sqrt{m_{22}m_{21}}}\bJ_{[m_{22}:m_{21}]}
&\frac{p_{22}q_{22}}{m_{22}}\bJ_{[m_{22}:m_{22}]}
&\frac{p_{22}q_{23}}{\sqrt{m_{22}m_{23}}}\bJ_{[m_{22}:m_{23}]}
\\[4pt]
\frac{p_{21}q_{31}}{\sqrt{m_{23}m_{11}}}\bJ_{[m_{23}:m_{11}]}
&\frac{p_{21}q_{32}}{\sqrt{m_{23}m_{12}}}\bJ_{[m_{23}:m_{12}]}
&\frac{p_{21}q_{33}}{\sqrt{m_{23}m_{13}}}\bJ_{[m_{23}:m_{13}]}
&\frac{p_{22}q_{31}}{\sqrt{m_{23}m_{21}}}\bJ_{[m_{23}:m_{21}]}
&\frac{p_{22}q_{32}}{\sqrt{m_{23}m_{22}}}\bJ_{[m_{23}:m_{22}]}
&\frac{p_{22}q_{33}}{m_{23}}\bJ_{[m_{23}:m_{23}]}
\end{bmatrix}$
}
\\&
\quad
\times
\resizebox{1.1\textwidth}{!}{$
\begin{bmatrix}
\frac{r_{11}s_{11}}{m_{11}}\bJ_{[m_{11}:m_{11}]}
&\frac{r_{11}s_{12}}{\sqrt{m_{11}m_{12}}}\bJ_{[m_{11}:m_{12}]}
&\frac{r_{11}s_{13}}{\sqrt{m_{11}m_{13}}}\bJ_{[m_{11}:m_{13}]}
&\frac{r_{12}s_{11}}{\sqrt{m_{11}m_{21}}}\bJ_{[m_{11}:m_{21}]}
&\frac{r_{12}s_{12}}{\sqrt{m_{11}m_{22}}}\bJ_{[m_{11}:m_{22}]}
&\frac{r_{12}s_{13}}{\sqrt{m_{11}m_{23}}}\bJ_{[m_{11}:m_{23}]}
\\[4pt]
\frac{r_{11}s_{21}}{\sqrt{m_{12}m_{11}}}\bJ_{[m_{12}:m_{11}]}
&\frac{r_{11}s_{22}}{m_{12}}\bJ_{[m_{12}:m_{12}]}
&\frac{r_{11}s_{23}}{\sqrt{m_{12}m_{13}}}\bJ_{[m_{12}:m_{13}]}
&\frac{r_{12}s_{21}}{\sqrt{m_{12}m_{21}}}\bJ_{[m_{12}:m_{21}]}
&\frac{r_{12}s_{22}}{\sqrt{m_{12}m_{22}}}\bJ_{[m_{12}:m_{22}]}
&\frac{r_{12}s_{23}}{\sqrt{m_{12}m_{23}}}\bJ_{[m_{12}:m_{23}]}
\\[4pt]
\frac{r_{11}s_{31}}{\sqrt{m_{13}m_{11}}}\bJ_{[m_{13}:m_{11}]}
&\frac{r_{11}s_{32}}{\sqrt{m_{13}m_{12}}}\bJ_{[m_{13}:m_{12}]}
&\frac{r_{11}s_{33}}{m_{13}}\bJ_{[m_{13}:m_{13}]}
&\frac{r_{12}s_{31}}{\sqrt{m_{13}m_{21}}}\bJ_{[m_{13}:m_{21}]}
&\frac{r_{12}s_{32}}{\sqrt{m_{13}m_{22}}}\bJ_{[m_{13}:m_{22}]}
&\frac{r_{12}s_{33}}{\sqrt{m_{13}m_{23}}}\bJ_{[m_{13}:m_{23}]}
\\[6pt]
\frac{r_{21}s_{11}}{\sqrt{m_{21}m_{11}}}\bJ_{[m_{21}:m_{11}]}
&\frac{r_{21}s_{12}}{\sqrt{m_{21}m_{12}}}\bJ_{[m_{21}:m_{12}]}
&\frac{r_{21}s_{13}}{\sqrt{m_{21}m_{13}}}\bJ_{[m_{21}:m_{13}]}
&\frac{r_{22}s_{11}}{m_{21}}\bJ_{[m_{21}:m_{21}]}
&\frac{r_{22}s_{12}}{\sqrt{m_{21}m_{22}}}\bJ_{[m_{21}:m_{22}]}
&\frac{r_{22}s_{13}}{\sqrt{m_{21}m_{23}}}\bJ_{[m_{21}:m_{23}]}
\\[4pt]
\frac{r_{21}s_{21}}{\sqrt{m_{22}m_{11}}}\bJ_{[m_{22}:m_{11}]}
&\frac{r_{21}s_{22}}{\sqrt{m_{22}m_{12}}}\bJ_{[m_{22}:m_{12}]}
&\frac{r_{21}s_{23}}{\sqrt{m_{22}m_{13}}}\bJ_{[m_{22}:m_{13}]}
&\frac{r_{22}s_{21}}{\sqrt{m_{22}m_{21}}}\bJ_{[m_{22}:m_{21}]}
&\frac{r_{22}s_{22}}{m_{22}}\bJ_{[m_{22}:m_{22}]}
&\frac{r_{22}s_{23}}{\sqrt{m_{22}m_{23}}}\bJ_{[m_{22}:m_{23}]}
\\[4pt]
\frac{r_{21}s_{31}}{\sqrt{m_{23}m_{11}}}\bJ_{[m_{23}:m_{11}]}
&\frac{r_{21}s_{32}}{\sqrt{m_{23}m_{12}}}\bJ_{[m_{23}:m_{12}]}
&\frac{r_{21}s_{33}}{\sqrt{m_{23}m_{13}}}\bJ_{[m_{23}:m_{13}]}
&\frac{r_{22}s_{31}}{\sqrt{m_{23}m_{21}}}\bJ_{[m_{23}:m_{21}]}
&\frac{r_{22}s_{32}}{\sqrt{m_{23}m_{22}}}\bJ_{[m_{23}:m_{22}]}
&\frac{r_{22}s_{33}}{m_{23}}\bJ_{[m_{23}:m_{23}]}
\end{bmatrix}$
}
\\&
=
\resizebox{1.1\textwidth}{!}{$
\begin{bmatrix}
\frac{(\bP\bR)_{11}(\bQ\bS)_{11}}{m_{11}}\bJ_{[m_{11}:m_{11}]}
&\frac{(\bP\bR)_{11}(\bQ\bS)_{12}}{\sqrt{m_{11}m_{12}}}\bJ_{[m_{11}:m_{12}]}
&\frac{(\bP\bR)_{11}(\bQ\bS)_{13}}{\sqrt{m_{11}m_{13}}}\bJ_{[m_{11}:m_{13}]}
&\frac{(\bP\bR)_{12}(\bQ\bS)_{11}}{\sqrt{m_{11}m_{21}}}\bJ_{[m_{11}:m_{21}]}
&\frac{(\bP\bR)_{12}(\bQ\bS)_{12}}{\sqrt{m_{11}m_{22}}}\bJ_{[m_{11}:m_{22}]}
&\frac{(\bP\bR)_{12}(\bQ\bS)_{13}}{\sqrt{m_{11}m_{23}}}\bJ_{[m_{11}:m_{23}]}
\\[4pt]
\frac{(\bP\bR)_{11}(\bQ\bS)_{21}}{\sqrt{m_{12}m_{11}}}\bJ_{[m_{12}:m_{11}]}
&\frac{(\bP\bR)_{11}(\bQ\bS)_{22}}{m_{12}}\bJ_{[m_{12}:m_{12}]}
&\frac{(\bP\bR)_{11}(\bQ\bS)_{23}}{\sqrt{m_{12}m_{13}}}\bJ_{[m_{12}:m_{13}]}
&\frac{(\bP\bR)_{12}(\bQ\bS)_{21}}{\sqrt{m_{12}m_{21}}}\bJ_{[m_{12}:m_{21}]}
&\frac{(\bP\bR)_{12}(\bQ\bS)_{22}}{\sqrt{m_{12}m_{22}}}\bJ_{[m_{12}:m_{22}]}
&\frac{(\bP\bR)_{12}(\bQ\bS)_{23}}{\sqrt{m_{12}m_{23}}}\bJ_{[m_{12}:m_{23}]}
\\[4pt]
\frac{(\bP\bR)_{11}(\bQ\bS)_{31}}{\sqrt{m_{13}m_{11}}}\bJ_{[m_{13}:m_{11}]}
&\frac{(\bP\bR)_{11}(\bQ\bS)_{32}}{\sqrt{m_{13}m_{12}}}\bJ_{[m_{13}:m_{12}]}
&\frac{(\bP\bR)_{11}(\bQ\bS)_{33}}{m_{13}}\bJ_{[m_{13}:m_{13}]}
&\frac{(\bP\bR)_{12}(\bQ\bS)_{31}}{\sqrt{m_{13}m_{21}}}\bJ_{[m_{13}:m_{21}]}
&\frac{(\bP\bR)_{12}(\bQ\bS)_{32}}{\sqrt{m_{13}m_{22}}}\bJ_{[m_{13}:m_{22}]}
&\frac{(\bP\bR)_{12}(\bQ\bS)_{33}}{\sqrt{m_{13}m_{23}}}\bJ_{[m_{13}:m_{23}]}
\\[6pt]
\frac{(\bP\bR)_{21}(\bQ\bS)_{11}}{\sqrt{m_{21}m_{11}}}\bJ_{[m_{21}:m_{11}]}
&\frac{(\bP\bR)_{21}(\bQ\bS)_{12}}{\sqrt{m_{21}m_{12}}}\bJ_{[m_{21}:m_{12}]}
&\frac{(\bP\bR)_{21}(\bQ\bS)_{13}}{\sqrt{m_{21}m_{13}}}\bJ_{[m_{21}:m_{13}]}
&\frac{(\bP\bR)_{22}(\bQ\bS)_{11}}{m_{21}}\bJ_{[m_{21}:m_{21}]}
&\frac{(\bP\bR)_{22}(\bQ\bS)_{12}}{\sqrt{m_{21}m_{22}}}\bJ_{[m_{21}:m_{22}]}
&\frac{(\bP\bR)_{22}(\bQ\bS)_{13}}{\sqrt{m_{21}m_{23}}}\bJ_{[m_{21}:m_{23}]}
\\[4pt]
\frac{(\bP\bR)_{21}(\bQ\bS)_{21}}{\sqrt{m_{22}m_{11}}}\bJ_{[m_{22}:m_{11}]}
&\frac{(\bP\bR)_{21}(\bQ\bS)_{22}}{\sqrt{m_{22}m_{12}}}\bJ_{[m_{22}:m_{12}]}
&\frac{(\bP\bR)_{21}(\bQ\bS)_{23}}{\sqrt{m_{22}m_{13}}}\bJ_{[m_{22}:m_{13}]}
&\frac{(\bP\bR)_{22}(\bQ\bS)_{21}}{\sqrt{m_{22}m_{21}}}\bJ_{[m_{22}:m_{21}]}
&\frac{(\bP\bR)_{22}(\bQ\bS)_{22}}{m_{22}}\bJ_{[m_{22}:m_{22}]}
&\frac{(\bP\bR)_{22}(\bQ\bS)_{23}}{\sqrt{m_{22}m_{23}}}\bJ_{[m_{22}:m_{23}]}
\\[4pt]
\frac{(\bP\bR)_{21}(\bQ\bS)_{31}}{\sqrt{m_{23}m_{11}}}\bJ_{[m_{23}:m_{11}]}
&\frac{(\bP\bR)_{21}(\bQ\bS)_{32}}{\sqrt{m_{23}m_{12}}}\bJ_{[m_{23}:m_{12}]}
&\frac{(\bP\bR)_{21}(\bQ\bS)_{33}}{\sqrt{m_{23}m_{13}}}\bJ_{[m_{23}:m_{13}]}
&\frac{(\bP\bR)_{22}(\bQ\bS)_{31}}{\sqrt{m_{23}m_{21}}}\bJ_{[m_{23}:m_{21}]}
&\frac{(\bP\bR)_{22}(\bQ\bS)_{32}}{\sqrt{m_{23}m_{22}}}\bJ_{[m_{23}:m_{22}]}
&\frac{(\bP\bR)_{22}(\bQ\bS)_{33}}{m_{23}}\bJ_{[m_{23}:m_{23}]}
\end{bmatrix}$
},
\end{align*}
with
\begin{align*}
(\bP\bR)_{ab} = \sum_{k=1}^{2} p_{a k} r_{k b},
\qquad
(\bQ\bS)_{tu}  = \sum_{k=1}^{3} q_{t k} s_{k u}.
\end{align*}

Therefore, we can write 
\begin{align*}
    [(\bP\otimes\bQ)_{\text{cell}}\circledast\barbJ{\bm}]
[(\bR\otimes\bS)_{\text{cell}}\circledast\barbJ{\bm}]&=[(\bP\bR)\otimes(\bQ\bS)]_{\text{cell}}\circledast\barbJ{\bm}.
\end{align*}



\begin{align*}
13).\quad &
[(\bP\otimes\bQ)_{\text{cell}}\circledast\tildebJ_{\bm}]
[(\bR\otimes\bS)_{\text{cell}}\circledast\tildebJ_{\bm}]
\\&=
\resizebox{1.1\textwidth}{!}{$
\begin{bmatrix}
\frac{p_{11}q_{11}}{m_{11}^2}\bJ_{[m_{11}:m_{11}]}
&\frac{p_{11}q_{12}}{m_{11}m_{12}}\bJ_{[m_{11}:m_{12}]}
&\frac{p_{11}q_{13}}{m_{11}m_{13}}\bJ_{[m_{11}:m_{13}]}
&\frac{p_{12}q_{11}}{m_{11}m_{21}}\bJ_{[m_{11}:m_{21}]}
&\frac{p_{12}q_{12}}{m_{11}m_{22}}\bJ_{[m_{11}:m_{22}]}
&\frac{p_{12}q_{13}}{m_{11}m_{23}}\bJ_{[m_{11}:m_{23}]}
\\[4pt]
\frac{p_{11}q_{21}}{m_{12}m_{11}}\bJ_{[m_{12}:m_{11}]}
&\frac{p_{11}q_{22}}{m_{12}^2}\bJ_{[m_{12}:m_{12}]}
&\frac{p_{11}q_{23}}{m_{12}m_{13}}\bJ_{[m_{12}:m_{13}]}
&\frac{p_{12}q_{21}}{m_{12}m_{21}}\bJ_{[m_{12}:m_{21}]}
&\frac{p_{12}q_{22}}{m_{12}m_{22}}\bJ_{[m_{12}:m_{22}]}
&\frac{p_{12}q_{23}}{m_{12}m_{23}}\bJ_{[m_{12}:m_{23}]}
\\[4pt]
\frac{p_{11}q_{31}}{m_{13}m_{11}}\bJ_{[m_{13}:m_{11}]}
&\frac{p_{11}q_{32}}{m_{13}m_{12}}\bJ_{[m_{13}:m_{12}]}
&\frac{p_{11}q_{33}}{m_{13}^2}\bJ_{[m_{13}:m_{13}]}
&\frac{p_{12}q_{31}}{m_{13}m_{21}}\bJ_{[m_{13}:m_{21}]}
&\frac{p_{12}q_{32}}{m_{13}m_{22}}\bJ_{[m_{13}:m_{22}]}
&\frac{p_{12}q_{33}}{m_{13}m_{23}}\bJ_{[m_{13}:m_{23}]}
\\[6pt]
\frac{p_{21}q_{11}}{m_{21}m_{11}}\bJ_{[m_{21}:m_{11}]}
&\frac{p_{21}q_{12}}{m_{21}m_{12}}\bJ_{[m_{21}:m_{12}]}
&\frac{p_{21}q_{13}}{m_{21}m_{13}}\bJ_{[m_{21}:m_{13}]}
&\frac{p_{22}q_{11}}{m_{21}^2}\bJ_{[m_{21}:m_{21}]}
&\frac{p_{22}q_{12}}{m_{21}m_{22}}\bJ_{[m_{21}:m_{22}]}
&\frac{p_{22}q_{13}}{m_{21}m_{23}}\bJ_{[m_{21}:m_{23}]}
\\[4pt]
\frac{p_{21}q_{21}}{m_{22}m_{11}}\bJ_{[m_{22}:m_{11}]}
&\frac{p_{21}q_{22}}{m_{22}m_{12}}\bJ_{[m_{22}:m_{12}]}
&\frac{p_{21}q_{23}}{m_{22}m_{13}}\bJ_{[m_{22}:m_{13}]}
&\frac{p_{22}q_{21}}{m_{22}m_{21}}\bJ_{[m_{22}:m_{21}]}
&\frac{p_{22}q_{22}}{m_{22}^2}\bJ_{[m_{22}:m_{22}]}
&\frac{p_{22}q_{23}}{m_{22}m_{23}}\bJ_{[m_{22}:m_{23}]}
\\[4pt]
\frac{p_{21}q_{31}}{m_{23}m_{11}}\bJ_{[m_{23}:m_{11}]}
&\frac{p_{21}q_{32}}{m_{23}m_{12}}\bJ_{[m_{23}:m_{12}]}
&\frac{p_{21}q_{33}}{m_{23}m_{13}}\bJ_{[m_{23}:m_{13}]}
&\frac{p_{22}q_{31}}{m_{23}m_{21}}\bJ_{[m_{23}:m_{21}]}
&\frac{p_{22}q_{32}}{m_{23}m_{22}}\bJ_{[m_{23}:m_{22}]}
&\frac{p_{22}q_{33}}{m_{23}^2}\bJ_{[m_{23}:m_{23}]}
\end{bmatrix}$
}
\\&\quad\times
 \resizebox{1.1\textwidth}{!}{$
\begin{bmatrix}
\frac{r_{11}s_{11}}{m_{11}^2}\bJ_{[m_{11}:m_{11}]}
&\frac{r_{11}s_{12}}{m_{11}m_{12}}\bJ_{[m_{11}:m_{12}]}
&\frac{r_{11}s_{13}}{m_{11}m_{13}}\bJ_{[m_{11}:m_{13}]}
&\frac{r_{12}s_{11}}{m_{11}m_{21}}\bJ_{[m_{11}:m_{21}]}
&\frac{r_{12}s_{12}}{m_{11}m_{22}}\bJ_{[m_{11}:m_{22}]}
&\frac{r_{12}s_{13}}{m_{11}m_{23}}\bJ_{[m_{11}:m_{23}]}
\\[4pt]
\frac{r_{11}s_{21}}{m_{12}m_{11}}\bJ_{[m_{12}:m_{11}]}
&\frac{r_{11}s_{22}}{m_{12}^2}\bJ_{[m_{12}:m_{12}]}
&\frac{r_{11}s_{23}}{m_{12}m_{13}}\bJ_{[m_{12}:m_{13}]}
&\frac{r_{12}s_{21}}{m_{12}m_{21}}\bJ_{[m_{12}:m_{21}]}
&\frac{r_{12}s_{22}}{m_{12}m_{22}}\bJ_{[m_{12}:m_{22}]}
&\frac{r_{12}s_{23}}{m_{12}m_{23}}\bJ_{[m_{12}:m_{23}]}
\\[4pt]
\frac{r_{11}s_{31}}{m_{13}m_{11}}\bJ_{[m_{13}:m_{11}]}
&\frac{r_{11}s_{32}}{m_{13}m_{12}}\bJ_{[m_{13}:m_{12}]}
&\frac{r_{11}s_{33}}{m_{13}^2}\bJ_{[m_{13}:m_{13}]}
&\frac{r_{12}s_{31}}{m_{13}m_{21}}\bJ_{[m_{13}:m_{21}]}
&\frac{r_{12}s_{32}}{m_{13}m_{22}}\bJ_{[m_{13}:m_{22}]}
&\frac{r_{12}s_{33}}{m_{13}m_{23}}\bJ_{[m_{13}:m_{23}]}
\\[6pt]
\frac{r_{21}s_{11}}{m_{21}m_{11}}\bJ_{[m_{21}:m_{11}]}
&\frac{r_{21}s_{12}}{m_{21}m_{12}}\bJ_{[m_{21}:m_{12}]}
&\frac{r_{21}s_{13}}{m_{21}m_{13}}\bJ_{[m_{21}:m_{13}]}
&\frac{r_{22}s_{11}}{m_{21}^2}\bJ_{[m_{21}:m_{21}]}
&\frac{r_{22}s_{12}}{m_{21}m_{22}}\bJ_{[m_{21}:m_{22}]}
&\frac{r_{22}s_{13}}{m_{21}m_{23}}\bJ_{[m_{21}:m_{23}]}
\\[4pt]
\frac{r_{21}s_{21}}{m_{22}m_{11}}\bJ_{[m_{22}:m_{11}]}
&\frac{r_{21}s_{22}}{m_{22}m_{12}}\bJ_{[m_{22}:m_{12}]}
&\frac{r_{21}s_{23}}{m_{22}m_{13}}\bJ_{[m_{22}:m_{13}]}
&\frac{r_{22}s_{21}}{m_{22}m_{21}}\bJ_{[m_{22}:m_{21}]}
&\frac{r_{22}s_{22}}{m_{22}^2}\bJ_{[m_{22}:m_{22}]}
&\frac{r_{22}s_{23}}{m_{22}m_{23}}\bJ_{[m_{22}:m_{23}]}
\\[4pt]
\frac{r_{21}s_{31}}{m_{23}m_{11}}\bJ_{[m_{23}:m_{11}]}
&\frac{r_{21}s_{32}}{m_{23}m_{12}}\bJ_{[m_{23}:m_{12}]}
&\frac{r_{21}s_{33}}{m_{23}m_{13}}\bJ_{[m_{23}:m_{13}]}
&\frac{r_{22}s_{31}}{m_{23}m_{21}}\bJ_{[m_{23}:m_{21}]}
&\frac{r_{22}s_{32}}{m_{23}m_{22}}\bJ_{[m_{23}:m_{22}]}
&\frac{r_{22}s_{33}}{m_{23}^2}\bJ_{[m_{23}:m_{23}]}
\end{bmatrix}$
}
\\&
=\resizebox{1.1\textwidth}{!}{$
\begin{bmatrix}
\frac{\Gamma_{11}^{\,11}}{m_{11}^2}\bJ_{[m_{11}:m_{11}]}
&\frac{\Gamma_{12}^{\,12}}{m_{11}m_{12}}\bJ_{[m_{11}:m_{12}]}
&\frac{\Gamma_{12}^{\,13}}{m_{11}m_{13}}\bJ_{[m_{11}:m_{13}]}
&\frac{\Gamma_{12}^{\,11}}{m_{11}m_{21}}\bJ_{[m_{11}:m_{21}]}
&\frac{\Gamma_{12}^{\,12}}{m_{11}m_{22}}\bJ_{[m_{11}:m_{22}]}
&\frac{\Gamma_{12}^{\,13}}{m_{11}m_{23}}\bJ_{[m_{11}:m_{23}]}
\\[4pt]
\frac{\Gamma_{11}^{\,21}}{m_{12}m_{11}}\bJ_{[m_{12}:m_{11}]}
&\frac{\Gamma_{11}^{\,22}}{m_{12}^2}\bJ_{[m_{12}:m_{12}]}
&\frac{\Gamma_{11}^{\,23}}{m_{12}m_{13}}\bJ_{[m_{12}:m_{13}]}
&\frac{\Gamma_{12}^{\,21}}{m_{12}m_{21}}\bJ_{[m_{12}:m_{21}]}
&\frac{\Gamma_{12}^{\,22}}{m_{12}m_{22}}\bJ_{[m_{12}:m_{22}]}
&\frac{\Gamma_{12}^{\,23}}{m_{12}m_{23}}\bJ_{[m_{12}:m_{23}]}
\\[4pt]
\frac{\Gamma_{11}^{\,31}}{m_{13}m_{11}}\bJ_{[m_{13}:m_{11}]}
&\frac{\Gamma_{11}^{\,32}}{m_{13}m_{12}}\bJ_{[m_{13}:m_{12}]}
&\frac{\Gamma_{11}^{\,33}}{m_{13}^2}\bJ_{[m_{13}:m_{13}]}
&\frac{\Gamma_{12}^{\,31}}{m_{13}m_{21}}\bJ_{[m_{13}:m_{21}]}
&\frac{\Gamma_{12}^{\,32}}{m_{13}m_{22}}\bJ_{[m_{13}:m_{22}]}
&\frac{\Gamma_{12}^{\,33}}{m_{13}m_{23}}\bJ_{[m_{13}:m_{23}]}
\\[6pt]
\frac{\Gamma_{21}^{\,11}}{m_{21}m_{11}}\bJ_{[m_{21}:m_{11}]}
&\frac{\Gamma_{22}^{\,12}}{m_{21}m_{12}}\bJ_{[m_{21}:m_{12}]}
&\frac{\Gamma_{22}^{\,13}}{m_{21}m_{13}}\bJ_{[m_{21}:m_{13}]}
&\frac{\Gamma_{22}^{\,11}}{m_{21}^2}\bJ_{[m_{21}:m_{21}]}
&\frac{\Gamma_{22}^{\,12}}{m_{21}m_{22}}\bJ_{[m_{21}:m_{22}]}
&\frac{\Gamma_{22}^{\,13}}{m_{21}m_{23}}\bJ_{[m_{21}:m_{23}]}
\\[4pt]
\frac{\Gamma_{21}^{\,21}}{m_{22}m_{11}}\bJ_{[m_{22}:m_{11}]}
&\frac{\Gamma_{22}^{\,22}}{m_{22}m_{12}}\bJ_{[m_{22}:m_{12}]}
&\frac{\Gamma_{22}^{\,23}}{m_{22}m_{13}}\bJ_{[m_{22}:m_{13}]}
&\frac{\Gamma_{22}^{\,21}}{m_{22}m_{21}}\bJ_{[m_{22}:m_{21}]}
&\frac{\Gamma_{22}^{\,22}}{m_{22}^2}\bJ_{[m_{22}:m_{22}]}
&\frac{\Gamma_{22}^{\,23}}{m_{22}m_{23}}\bJ_{[m_{22}:m_{23}]}
\\[4pt]
\frac{\Gamma_{21}^{\,31}}{m_{23}m_{11}}\bJ_{[m_{23}:m_{11}]}
&\frac{\Gamma_{22}^{\,32}}{m_{23}m_{12}}\bJ_{[m_{23}:m_{12}]}
&\frac{\Gamma_{22}^{\,33}}{m_{23}m_{13}}\bJ_{[m_{23}:m_{13}]}
&\frac{\Gamma_{22}^{\,31}}{m_{23}m_{21}}\bJ_{[m_{23}:m_{21}]}
&\frac{\Gamma_{22}^{\,32}}{m_{23}m_{22}}\bJ_{[m_{23}:m_{22}]}
&\frac{\Gamma_{22}^{\,33}}{m_{23}^2}\bJ_{[m_{23}:m_{23}]}
\end{bmatrix}$
},
\end{align*}
with

\[
\boldsymbol{\Gamma}=(\bP\otimes\bQ)\,\bM^{-1}\,(\bR\otimes\bS)
\]
and
\[
\Gamma_{ab}^{\,tu}
=\big[(\bP\otimes\bQ)\,\bM^{-1}\,(\bR\otimes\bS)\big]_{(a,t),(b,u)}
=\sum_{c=1}^{2}\sum_{v=1}^{3}\frac{p_{ac}\,q_{tv}\,r_{cb}\,s_{vu}}{m_{cv}}.
\]

Therefore, we can write 
\begin{align*}
    [(\bP\otimes\bQ)_{\text{cell}}\circledast\tildebJ_{\bm}]
[(\bR\otimes\bS)_{\text{cell}}\circledast\tildebJ_{\bm}]&=[(\bP\otimes\bQ)\,\bM^{-1}\,(\bR\otimes\bS)]_{\text{cell}}\circledast\tildebJ_{\bm}.
\end{align*}

\begin{align*}
14).\quad &
[(\bP\otimes\bQ)_{\text{cell}}\circledast\tildebJ_{\bm}\bI_{\bm}^f]
[(\bR\otimes\bS)_{\text{cell}}\circledast\tildebJ_{\bm}]
\\&=
\resizebox{1.1\textwidth}{!}{$
\begin{bmatrix}
\frac{p_{11}q_{11}\,f(m_{11})}{m_{11}^2}\bJ_{[m_{11}:m_{11}]}
&\frac{p_{11}q_{12}\,f(m_{12})}{m_{11}m_{12}}\bJ_{[m_{11}:m_{12}]}
&\frac{p_{11}q_{13}\,f(m_{13})}{m_{11}m_{13}}\bJ_{[m_{11}:m_{13}]}
&\frac{p_{12}q_{11}\,f(m_{21})}{m_{11}m_{21}}\bJ_{[m_{11}:m_{21}]}
&\frac{p_{12}q_{12}\,f(m_{22})}{m_{11}m_{22}}\bJ_{[m_{11}:m_{22}]}
&\frac{p_{12}q_{13}\,f(m_{23})}{m_{11}m_{23}}\bJ_{[m_{11}:m_{23}]}
\\[4pt]
\frac{p_{11}q_{21}\,f(m_{11})}{m_{12}m_{11}}\bJ_{[m_{12}:m_{11}]}
&\frac{p_{11}q_{22}\,f(m_{12})}{m_{12}^2}\bJ_{[m_{12}:m_{12}]}
&\frac{p_{11}q_{23}\,f(m_{13})}{m_{12}m_{13}}\bJ_{[m_{12}:m_{13}]}
&\frac{p_{12}q_{21}\,f(m_{21})}{m_{12}m_{21}}\bJ_{[m_{12}:m_{21}]}
&\frac{p_{12}q_{22}\,f(m_{22})}{m_{12}m_{22}}\bJ_{[m_{12}:m_{22}]}
&\frac{p_{12}q_{23}\,f(m_{23})}{m_{12}m_{23}}\bJ_{[m_{12}:m_{23}]}
\\[4pt]
\frac{p_{11}q_{31}\,f(m_{11})}{m_{13}m_{11}}\bJ_{[m_{13}:m_{11}]}
&\frac{p_{11}q_{32}\,f(m_{12})}{m_{13}m_{12}}\bJ_{[m_{13}:m_{12}]}
&\frac{p_{11}q_{33}\,f(m_{13})}{m_{13}^2}\bJ_{[m_{13}:m_{13}]}
&\frac{p_{12}q_{31}\,f(m_{21})}{m_{13}m_{21}}\bJ_{[m_{13}:m_{21}]}
&\frac{p_{12}q_{32}\,f(m_{22})}{m_{13}m_{22}}\bJ_{[m_{13}:m_{22}]}
&\frac{p_{12}q_{33}\,f(m_{23})}{m_{13}m_{23}}\bJ_{[m_{13}:m_{23}]}
\\[6pt]
\frac{p_{21}q_{11}\,f(m_{11})}{m_{21}m_{11}}\bJ_{[m_{21}:m_{11}]}
&\frac{p_{21}q_{12}\,f(m_{12})}{m_{21}m_{12}}\bJ_{[m_{21}:m_{12}]}
&\frac{p_{21}q_{13}\,f(m_{13})}{m_{21}m_{13}}\bJ_{[m_{21}:m_{13}]}
&\frac{p_{22}q_{11}\,f(m_{21})}{m_{21}^2}\bJ_{[m_{21}:m_{21}]}
&\frac{p_{22}q_{12}\,f(m_{22})}{m_{21}m_{22}}\bJ_{[m_{21}:m_{22}]}
&\frac{p_{22}q_{13}\,f(m_{23})}{m_{21}m_{23}}\bJ_{[m_{21}:m_{23}]}
\\[4pt]
\frac{p_{21}q_{21}\,f(m_{11})}{m_{22}m_{11}}\bJ_{[m_{22}:m_{11}]}
&\frac{p_{21}q_{22}\,f(m_{12})}{m_{22}m_{12}}\bJ_{[m_{22}:m_{12}]}
&\frac{p_{21}q_{23}\,f(m_{13})}{m_{22}m_{13}}\bJ_{[m_{22}:m_{13}]}
&\frac{p_{22}q_{21}\,f(m_{21})}{m_{22}m_{21}}\bJ_{[m_{22}:m_{21}]}
&\frac{p_{22}q_{22}\,f(m_{22})}{m_{22}^2}\bJ_{[m_{22}:m_{22}]}
&\frac{p_{22}q_{23}\,f(m_{23})}{m_{22}m_{23}}\bJ_{[m_{22}:m_{23}]}
\\[4pt]
\frac{p_{21}q_{31}\,f(m_{11})}{m_{23}m_{11}}\bJ_{[m_{23}:m_{11}]}
&\frac{p_{21}q_{32}\,f(m_{12})}{m_{23}m_{12}}\bJ_{[m_{23}:m_{12}]}
&\frac{p_{21}q_{33}\,f(m_{13})}{m_{23}m_{13}}\bJ_{[m_{23}:m_{13}]}
&\frac{p_{22}q_{31}\,f(m_{21})}{m_{23}m_{21}}\bJ_{[m_{23}:m_{21}]}
&\frac{p_{22}q_{32}\,f(m_{22})}{m_{23}m_{22}}\bJ_{[m_{23}:m_{22}]}
&\frac{p_{22}q_{33}\,f(m_{23})}{m_{23}^2}\bJ_{[m_{23}:m_{23}]}
\end{bmatrix}$
}\\&\quad\times
 \resizebox{1.1\textwidth}{!}{$
\begin{bmatrix}
\frac{r_{11}s_{11}}{m_{11}^2}\bJ_{[m_{11}:m_{11}]}
&\frac{r_{11}s_{12}}{m_{11}m_{12}}\bJ_{[m_{11}:m_{12}]}
&\frac{r_{11}s_{13}}{m_{11}m_{13}}\bJ_{[m_{11}:m_{13}]}
&\frac{r_{12}s_{11}}{m_{11}m_{21}}\bJ_{[m_{11}:m_{21}]}
&\frac{r_{12}s_{12}}{m_{11}m_{22}}\bJ_{[m_{11}:m_{22}]}
&\frac{r_{12}s_{13}}{m_{11}m_{23}}\bJ_{[m_{11}:m_{23}]}
\\[4pt]
\frac{r_{11}s_{21}}{m_{12}m_{11}}\bJ_{[m_{12}:m_{11}]}
&\frac{r_{11}s_{22}}{m_{12}^2}\bJ_{[m_{12}:m_{12}]}
&\frac{r_{11}s_{23}}{m_{12}m_{13}}\bJ_{[m_{12}:m_{13}]}
&\frac{r_{12}s_{21}}{m_{12}m_{21}}\bJ_{[m_{12}:m_{21}]}
&\frac{r_{12}s_{22}}{m_{12}m_{22}}\bJ_{[m_{12}:m_{22}]}
&\frac{r_{12}s_{23}}{m_{12}m_{23}}\bJ_{[m_{12}:m_{23}]}
\\[4pt]
\frac{r_{11}s_{31}}{m_{13}m_{11}}\bJ_{[m_{13}:m_{11}]}
&\frac{r_{11}s_{32}}{m_{13}m_{12}}\bJ_{[m_{13}:m_{12}]}
&\frac{r_{11}s_{33}}{m_{13}^2}\bJ_{[m_{13}:m_{13}]}
&\frac{r_{12}s_{31}}{m_{13}m_{21}}\bJ_{[m_{13}:m_{21}]}
&\frac{r_{12}s_{32}}{m_{13}m_{22}}\bJ_{[m_{13}:m_{22}]}
&\frac{r_{12}s_{33}}{m_{13}m_{23}}\bJ_{[m_{13}:m_{23}]}
\\[6pt]
\frac{r_{21}s_{11}}{m_{21}m_{11}}\bJ_{[m_{21}:m_{11}]}
&\frac{r_{21}s_{12}}{m_{21}m_{12}}\bJ_{[m_{21}:m_{12}]}
&\frac{r_{21}s_{13}}{m_{21}m_{13}}\bJ_{[m_{21}:m_{13}]}
&\frac{r_{22}s_{11}}{m_{21}^2}\bJ_{[m_{21}:m_{21}]}
&\frac{r_{22}s_{12}}{m_{21}m_{22}}\bJ_{[m_{21}:m_{22}]}
&\frac{r_{22}s_{13}}{m_{21}m_{23}}\bJ_{[m_{21}:m_{23}]}
\\[4pt]
\frac{r_{21}s_{21}}{m_{22}m_{11}}\bJ_{[m_{22}:m_{11}]}
&\frac{r_{21}s_{22}}{m_{22}m_{12}}\bJ_{[m_{22}:m_{12}]}
&\frac{r_{21}s_{23}}{m_{22}m_{13}}\bJ_{[m_{22}:m_{13}]}
&\frac{r_{22}s_{21}}{m_{22}m_{21}}\bJ_{[m_{22}:m_{21}]}
&\frac{r_{22}s_{22}}{m_{22}^2}\bJ_{[m_{22}:m_{22}]}
&\frac{r_{22}s_{23}}{m_{22}m_{23}}\bJ_{[m_{22}:m_{23}]}
\\[4pt]
\frac{r_{21}s_{31}}{m_{23}m_{11}}\bJ_{[m_{23}:m_{11}]}
&\frac{r_{21}s_{32}}{m_{23}m_{12}}\bJ_{[m_{23}:m_{12}]}
&\frac{r_{21}s_{33}}{m_{23}m_{13}}\bJ_{[m_{23}:m_{13}]}
&\frac{r_{22}s_{31}}{m_{23}m_{21}}\bJ_{[m_{23}:m_{21}]}
&\frac{r_{22}s_{32}}{m_{23}m_{22}}\bJ_{[m_{23}:m_{22}]}
&\frac{r_{22}s_{33}}{m_{23}^2}\bJ_{[m_{23}:m_{23}]}
\end{bmatrix}$
}\\&
=\resizebox{1.1\textwidth}{!}{$
\begin{bmatrix}
\frac{\Gamma_{11}^{\,11}}{m_{11}^2}\bJ_{[m_{11}:m_{11}]}
&\frac{\Gamma_{12}^{\,12}}{m_{11}m_{12}}\bJ_{[m_{11}:m_{12}]}
&\frac{\Gamma_{12}^{\,13}}{m_{11}m_{13}}\bJ_{[m_{11}:m_{13}]}
&\frac{\Gamma_{12}^{\,11}}{m_{11}m_{21}}\bJ_{[m_{11}:m_{21}]}
&\frac{\Gamma_{12}^{\,12}}{m_{11}m_{22}}\bJ_{[m_{11}:m_{22}]}
&\frac{\Gamma_{12}^{\,13}}{m_{11}m_{23}}\bJ_{[m_{11}:m_{23}]}
\\[4pt]
\frac{\Gamma_{11}^{\,21}}{m_{12}m_{11}}\bJ_{[m_{12}:m_{11}]}
&\frac{\Gamma_{11}^{\,22}}{m_{12}^2}\bJ_{[m_{12}:m_{12}]}
&\frac{\Gamma_{11}^{\,23}}{m_{12}m_{13}}\bJ_{[m_{12}:m_{13}]}
&\frac{\Gamma_{12}^{\,21}}{m_{12}m_{21}}\bJ_{[m_{12}:m_{21}]}
&\frac{\Gamma_{12}^{\,22}}{m_{12}m_{22}}\bJ_{[m_{12}:m_{22}]}
&\frac{\Gamma_{12}^{\,23}}{m_{12}m_{23}}\bJ_{[m_{12}:m_{23}]}
\\[4pt]
\frac{\Gamma_{11}^{\,31}}{m_{13}m_{11}}\bJ_{[m_{13}:m_{11}]}
&\frac{\Gamma_{11}^{\,32}}{m_{13}m_{12}}\bJ_{[m_{13}:m_{12}]}
&\frac{\Gamma_{11}^{\,33}}{m_{13}^2}\bJ_{[m_{13}:m_{13}]}
&\frac{\Gamma_{12}^{\,31}}{m_{13}m_{21}}\bJ_{[m_{13}:m_{21}]}
&\frac{\Gamma_{12}^{\,32}}{m_{13}m_{22}}\bJ_{[m_{13}:m_{22}]}
&\frac{\Gamma_{12}^{\,33}}{m_{13}m_{23}}\bJ_{[m_{13}:m_{23}]}
\\[6pt]
\frac{\Gamma_{21}^{\,11}}{m_{21}m_{11}}\bJ_{[m_{21}:m_{11}]}
&\frac{\Gamma_{22}^{\,12}}{m_{21}m_{12}}\bJ_{[m_{21}:m_{12}]}
&\frac{\Gamma_{22}^{\,13}}{m_{21}m_{13}}\bJ_{[m_{21}:m_{13}]}
&\frac{\Gamma_{22}^{\,11}}{m_{21}^2}\bJ_{[m_{21}:m_{21}]}
&\frac{\Gamma_{22}^{\,12}}{m_{21}m_{22}}\bJ_{[m_{21}:m_{22}]}
&\frac{\Gamma_{22}^{\,13}}{m_{21}m_{23}}\bJ_{[m_{21}:m_{23}]}
\\[4pt]
\frac{\Gamma_{21}^{\,21}}{m_{22}m_{11}}\bJ_{[m_{22}:m_{11}]}
&\frac{\Gamma_{22}^{\,22}}{m_{22}m_{12}}\bJ_{[m_{22}:m_{12}]}
&\frac{\Gamma_{22}^{\,23}}{m_{22}m_{13}}\bJ_{[m_{22}:m_{13}]}
&\frac{\Gamma_{22}^{\,21}}{m_{22}m_{21}}\bJ_{[m_{22}:m_{21}]}
&\frac{\Gamma_{22}^{\,22}}{m_{22}^2}\bJ_{[m_{22}:m_{22}]}
&\frac{\Gamma_{22}^{\,23}}{m_{22}m_{23}}\bJ_{[m_{22}:m_{23}]}
\\[4pt]
\frac{\Gamma_{21}^{\,31}}{m_{23}m_{11}}\bJ_{[m_{23}:m_{11}]}
&\frac{\Gamma_{22}^{\,32}}{m_{23}m_{12}}\bJ_{[m_{23}:m_{12}]}
&\frac{\Gamma_{22}^{\,33}}{m_{23}m_{13}}\bJ_{[m_{23}:m_{13}]}
&\frac{\Gamma_{22}^{\,31}}{m_{23}m_{21}}\bJ_{[m_{23}:m_{21}]}
&\frac{\Gamma_{22}^{\,32}}{m_{23}m_{22}}\bJ_{[m_{23}:m_{22}]}
&\frac{\Gamma_{22}^{\,33}}{m_{23}^2}\bJ_{[m_{23}:m_{23}]}
\end{bmatrix}$
},
\end{align*}
with
\[
f(\bm)
=\begin{bmatrix}
f(m_{11}) & &&&&\\
&f(m_{12}) &&&&\\
&&f(m_{13}) &&&\\
&&&f(m_{21}) &&\\
&&&&f(m_{22}) &\\
&&&&&f(m_{23}) \\
\end{bmatrix}
\]

\[
\boldsymbol{\Pi}=(\bP\otimes\bQ)\,f(\bm)\bM^{-1}\,(\bR\otimes\bS).
\]
and
\[
\Pi_{ab}^{\,tu}
=\big[(\bP\otimes\bQ)\,f(\bm)\bM^{-1}\,(\bR\otimes\bS)\big]_{(a,t),(b,u)}
=\sum_{c=1}^{2}\sum_{v=1}^{3}\frac{p_{ac}\,q_{tv}\,r_{cb}\,s_{vu}f(m_{cv})}{m_{cv}}.
\]
Therefore, we can write 
\begin{align*}
    [(\bP\otimes\bQ)_{\text{cell}}\circledast\tildebJ_{\bm}\bI_{\bm}^f]
[(\bR\otimes\bS)_{\text{cell}}\circledast\tildebJ_{\bm}]&=[(\bP\otimes\bQ)\,f(\bm)\bM^{-1}\,(\bR\otimes\bS)]_{\text{cell}}\circledast\tildebJ_{\bm}.
\end{align*}

\begin{align*}
15).\quad &
[(\bP\otimes\bQ)_{\text{cell}}\circledast\bI_{\bm}^f\tildebJ_{\bm}]
[(\bR\otimes\bS)_{\text{cell}}\circledast\tildebJ_{\bm}]
\\&=
\resizebox{1.1\textwidth}{!}{$
\begin{bmatrix}
\frac{f(m_{11})\,p_{11}q_{11}}{m_{11}^2}\bJ_{[m_{11}:m_{11}]}
&\frac{f(m_{11})\,p_{11}q_{12}}{m_{11}m_{12}}\bJ_{[m_{11}:m_{12}]}
&\frac{f(m_{11})\,p_{11}q_{13}}{m_{11}m_{13}}\bJ_{[m_{11}:m_{13}]}
&\frac{f(m_{11})\,p_{12}q_{11}}{m_{11}m_{21}}\bJ_{[m_{11}:m_{21}]}
&\frac{f(m_{11})\,p_{12}q_{12}}{m_{11}m_{22}}\bJ_{[m_{11}:m_{22}]}
&\frac{f(m_{11})\,p_{12}q_{13}}{m_{11}m_{23}}\bJ_{[m_{11}:m_{23}]}
\\[4pt]
\frac{f(m_{12})\,p_{11}q_{21}}{m_{12}m_{11}}\bJ_{[m_{12}:m_{11}]}
&\frac{f(m_{12})\,p_{11}q_{22}}{m_{12}^2}\bJ_{[m_{12}:m_{12}]}
&\frac{f(m_{12})\,p_{11}q_{23}}{m_{12}m_{13}}\bJ_{[m_{12}:m_{13}]}
&\frac{f(m_{12})\,p_{12}q_{21}}{m_{12}m_{21}}\bJ_{[m_{12}:m_{21}]}
&\frac{f(m_{12})\,p_{12}q_{22}}{m_{12}m_{22}}\bJ_{[m_{12}:m_{22}]}
&\frac{f(m_{12})\,p_{12}q_{23}}{m_{12}m_{23}}\bJ_{[m_{12}:m_{23}]}
\\[4pt]
\frac{f(m_{13})\,p_{11}q_{31}}{m_{13}m_{11}}\bJ_{[m_{13}:m_{11}]}
&\frac{f(m_{13})\,p_{11}q_{32}}{m_{13}m_{12}}\bJ_{[m_{13}:m_{12}]}
&\frac{f(m_{13})\,p_{11}q_{33}}{m_{13}^2}\bJ_{[m_{13}:m_{13}]}
&\frac{f(m_{13})\,p_{12}q_{31}}{m_{13}m_{21}}\bJ_{[m_{13}:m_{21}]}
&\frac{f(m_{13})\,p_{12}q_{32}}{m_{13}m_{22}}\bJ_{[m_{13}:m_{22}]}
&\frac{f(m_{13})\,p_{12}q_{33}}{m_{13}m_{23}}\bJ_{[m_{13}:m_{23}]}
\\[6pt]
\frac{f(m_{21})\,p_{21}q_{11}}{m_{21}m_{11}}\bJ_{[m_{21}:m_{11}]}
&\frac{f(m_{21})\,p_{21}q_{12}}{m_{21}m_{12}}\bJ_{[m_{21}:m_{12}]}
&\frac{f(m_{21})\,p_{21}q_{13}}{m_{21}m_{13}}\bJ_{[m_{21}:m_{13}]}
&\frac{f(m_{21})\,p_{22}q_{11}}{m_{21}^2}\bJ_{[m_{21}:m_{21}]}
&\frac{f(m_{21})\,p_{22}q_{12}}{m_{21}m_{22}}\bJ_{[m_{21}:m_{22}]}
&\frac{f(m_{21})\,p_{22}q_{13}}{m_{21}m_{23}}\bJ_{[m_{21}:m_{23}]}
\\[4pt]
\frac{f(m_{22})\,p_{21}q_{21}}{m_{22}m_{11}}\bJ_{[m_{22}:m_{11}]}
&\frac{f(m_{22})\,p_{21}q_{22}}{m_{22}m_{12}}\bJ_{[m_{22}:m_{12}]}
&\frac{f(m_{22})\,p_{21}q_{23}}{m_{22}m_{13}}\bJ_{[m_{22}:m_{13}]}
&\frac{f(m_{22})\,p_{22}q_{21}}{m_{22}m_{21}}\bJ_{[m_{22}:m_{21}]}
&\frac{f(m_{22})\,p_{22}q_{22}}{m_{22}^2}\bJ_{[m_{22}:m_{22}]}
&\frac{f(m_{22})\,p_{22}q_{23}}{m_{22}m_{23}}\bJ_{[m_{22}:m_{23}]}
\\[4pt]
\frac{f(m_{23})\,p_{21}q_{31}}{m_{23}m_{11}}\bJ_{[m_{23}:m_{11}]}
&\frac{f(m_{23})\,p_{21}q_{32}}{m_{23}m_{12}}\bJ_{[m_{23}:m_{12}]}
&\frac{f(m_{23})\,p_{21}q_{33}}{m_{23}m_{13}}\bJ_{[m_{23}:m_{13}]}
&\frac{f(m_{23})\,p_{22}q_{31}}{m_{23}m_{21}}\bJ_{[m_{23}:m_{21}]}
&\frac{f(m_{23})\,p_{22}q_{32}}{m_{23}m_{22}}\bJ_{[m_{23}:m_{22}]}
&\frac{f(m_{23})\,p_{22}q_{33}}{m_{23}^2}\bJ_{[m_{23}:m_{23}]}
\end{bmatrix}$
}
\\&\quad\times
 \resizebox{1.1\textwidth}{!}{$
\begin{bmatrix}
\frac{r_{11}s_{11}}{m_{11}^2}\bJ_{[m_{11}:m_{11}]}
&\frac{r_{11}s_{12}}{m_{11}m_{12}}\bJ_{[m_{11}:m_{12}]}
&\frac{r_{11}s_{13}}{m_{11}m_{13}}\bJ_{[m_{11}:m_{13}]}
&\frac{r_{12}s_{11}}{m_{11}m_{21}}\bJ_{[m_{11}:m_{21}]}
&\frac{r_{12}s_{12}}{m_{11}m_{22}}\bJ_{[m_{11}:m_{22}]}
&\frac{r_{12}s_{13}}{m_{11}m_{23}}\bJ_{[m_{11}:m_{23}]}
\\[4pt]
\frac{r_{11}s_{21}}{m_{12}m_{11}}\bJ_{[m_{12}:m_{11}]}
&\frac{r_{11}s_{22}}{m_{12}^2}\bJ_{[m_{12}:m_{12}]}
&\frac{r_{11}s_{23}}{m_{12}m_{13}}\bJ_{[m_{12}:m_{13}]}
&\frac{r_{12}s_{21}}{m_{12}m_{21}}\bJ_{[m_{12}:m_{21}]}
&\frac{r_{12}s_{22}}{m_{12}m_{22}}\bJ_{[m_{12}:m_{22}]}
&\frac{r_{12}s_{23}}{m_{12}m_{23}}\bJ_{[m_{12}:m_{23}]}
\\[4pt]
\frac{r_{11}s_{31}}{m_{13}m_{11}}\bJ_{[m_{13}:m_{11}]}
&\frac{r_{11}s_{32}}{m_{13}m_{12}}\bJ_{[m_{13}:m_{12}]}
&\frac{r_{11}s_{33}}{m_{13}^2}\bJ_{[m_{13}:m_{13}]}
&\frac{r_{12}s_{31}}{m_{13}m_{21}}\bJ_{[m_{13}:m_{21}]}
&\frac{r_{12}s_{32}}{m_{13}m_{22}}\bJ_{[m_{13}:m_{22}]}
&\frac{r_{12}s_{33}}{m_{13}m_{23}}\bJ_{[m_{13}:m_{23}]}
\\[6pt]
\frac{r_{21}s_{11}}{m_{21}m_{11}}\bJ_{[m_{21}:m_{11}]}
&\frac{r_{21}s_{12}}{m_{21}m_{12}}\bJ_{[m_{21}:m_{12}]}
&\frac{r_{21}s_{13}}{m_{21}m_{13}}\bJ_{[m_{21}:m_{13}]}
&\frac{r_{22}s_{11}}{m_{21}^2}\bJ_{[m_{21}:m_{21}]}
&\frac{r_{22}s_{12}}{m_{21}m_{22}}\bJ_{[m_{21}:m_{22}]}
&\frac{r_{22}s_{13}}{m_{21}m_{23}}\bJ_{[m_{21}:m_{23}]}
\\[4pt]
\frac{r_{21}s_{21}}{m_{22}m_{11}}\bJ_{[m_{22}:m_{11}]}
&\frac{r_{21}s_{22}}{m_{22}m_{12}}\bJ_{[m_{22}:m_{12}]}
&\frac{r_{21}s_{23}}{m_{22}m_{13}}\bJ_{[m_{22}:m_{13}]}
&\frac{r_{22}s_{21}}{m_{22}m_{21}}\bJ_{[m_{22}:m_{21}]}
&\frac{r_{22}s_{22}}{m_{22}^2}\bJ_{[m_{22}:m_{22}]}
&\frac{r_{22}s_{23}}{m_{22}m_{23}}\bJ_{[m_{22}:m_{23}]}
\\[4pt]
\frac{r_{21}s_{31}}{m_{23}m_{11}}\bJ_{[m_{23}:m_{11}]}
&\frac{r_{21}s_{32}}{m_{23}m_{12}}\bJ_{[m_{23}:m_{12}]}
&\frac{r_{21}s_{33}}{m_{23}m_{13}}\bJ_{[m_{23}:m_{13}]}
&\frac{r_{22}s_{31}}{m_{23}m_{21}}\bJ_{[m_{23}:m_{21}]}
&\frac{r_{22}s_{32}}{m_{23}m_{22}}\bJ_{[m_{23}:m_{22}]}
&\frac{r_{22}s_{33}}{m_{23}^2}\bJ_{[m_{23}:m_{23}]}
\end{bmatrix}$
}\\&
=
\resizebox{1.1\textwidth}{!}{$
\begin{bmatrix}
\frac{\Gamma_{11}^{\,11}f(m_{11})}{m_{11}^2}\bJ_{[m_{11}:m_{11}]}
&\frac{\Gamma_{12}^{\,12}f(m_{11})}{m_{11}m_{12}}\bJ_{[m_{11}:m_{12}]}
&\frac{\Gamma_{12}^{\,13}f(m_{11})}{m_{11}m_{13}}\bJ_{[m_{11}:m_{13}]}
&\frac{\Gamma_{12}^{\,11}f(m_{11})}{m_{11}m_{21}}\bJ_{[m_{11}:m_{21}]}
&\frac{\Gamma_{12}^{\,12}f(m_{11})}{m_{11}m_{22}}\bJ_{[m_{11}:m_{22}]}
&\frac{\Gamma_{12}^{\,13}f(m_{11})}{m_{11}m_{23}}\bJ_{[m_{11}:m_{23}]}
\\[4pt]
\frac{\Gamma_{11}^{\,21}f(m_{12})}{m_{12}m_{11}}\bJ_{[m_{12}:m_{11}]}
&\frac{\Gamma_{11}^{\,22}f(m_{12})}{m_{12}^2}\bJ_{[m_{12}:m_{12}]}
&\frac{\Gamma_{11}^{\,23}f(m_{12})}{m_{12}m_{13}}\bJ_{[m_{12}:m_{13}]}
&\frac{\Gamma_{12}^{\,21}f(m_{12})}{m_{12}m_{21}}\bJ_{[m_{12}:m_{21}]}
&\frac{\Gamma_{12}^{\,22}f(m_{12})}{m_{12}m_{22}}\bJ_{[m_{12}:m_{22}]}
&\frac{\Gamma_{12}^{\,23}f(m_{12})}{m_{12}m_{23}}\bJ_{[m_{12}:m_{23}]}
\\[4pt]
\frac{\Gamma_{11}^{\,31}f(m_{13})}{m_{13}m_{11}}\bJ_{[m_{13}:m_{11}]}
&\frac{\Gamma_{11}^{\,32}f(m_{13})}{m_{13}m_{12}}\bJ_{[m_{13}:m_{12}]}
&\frac{\Gamma_{11}^{\,33}f(m_{13})}{m_{13}^2}\bJ_{[m_{13}:m_{13}]}
&\frac{\Gamma_{12}^{\,31}f(m_{13})}{m_{13}m_{21}}\bJ_{[m_{13}:m_{21}]}
&\frac{\Gamma_{12}^{\,32}f(m_{13})}{m_{13}m_{22}}\bJ_{[m_{13}:m_{22}]}
&\frac{\Gamma_{12}^{\,33}f(m_{13})}{m_{13}m_{23}}\bJ_{[m_{13}:m_{23}]}
\\[6pt]
\frac{\Gamma_{21}^{\,11}f(m_{21})}{m_{21}m_{11}}\bJ_{[m_{21}:m_{11}]}
&\frac{\Gamma_{22}^{\,12}f(m_{21})}{m_{21}m_{12}}\bJ_{[m_{21}:m_{12}]}
&\frac{\Gamma_{22}^{\,13}f(m_{21})}{m_{21}m_{13}}\bJ_{[m_{21}:m_{13}]}
&\frac{\Gamma_{22}^{\,11}f(m_{21})}{m_{21}^2}\bJ_{[m_{21}:m_{21}]}
&\frac{\Gamma_{22}^{\,12}f(m_{21})}{m_{21}m_{22}}\bJ_{[m_{21}:m_{22}]}
&\frac{\Gamma_{22}^{\,13}f(m_{21})}{m_{21}m_{23}}\bJ_{[m_{21}:m_{23}]}
\\[4pt]
\frac{\Gamma_{21}^{\,21}f(m_{22})}{m_{22}m_{11}}\bJ_{[m_{22}:m_{11}]}
&\frac{\Gamma_{22}^{\,22}f(m_{22})}{m_{22}m_{12}}\bJ_{[m_{22}:m_{12}]}
&\frac{\Gamma_{22}^{\,23}f(m_{22})}{m_{22}m_{13}}\bJ_{[m_{22}:m_{13}]}
&\frac{\Gamma_{22}^{\,21}f(m_{22})}{m_{22}m_{21}}\bJ_{[m_{22}:m_{21}]}
&\frac{\Gamma_{22}^{\,22}f(m_{22})}{m_{22}^2}\bJ_{[m_{22}:m_{22}]}
&\frac{\Gamma_{22}^{\,23}f(m_{22})}{m_{22}m_{23}}\bJ_{[m_{22}:m_{23}]}
\\[4pt]
\frac{\Gamma_{21}^{\,31}f(m_{23})}{m_{23}m_{11}}\bJ_{[m_{23}:m_{11}]}
&\frac{\Gamma_{22}^{\,32}f(m_{23})}{m_{23}m_{12}}\bJ_{[m_{23}:m_{12}]}
&\frac{\Gamma_{22}^{\,33}f(m_{23})}{m_{23}m_{13}}\bJ_{[m_{23}:m_{13}]}
&\frac{\Gamma_{22}^{\,31}f(m_{23})}{m_{23}m_{21}}\bJ_{[m_{23}:m_{21}]}
&\frac{\Gamma_{22}^{\,32}f(m_{23})}{m_{23}m_{22}}\bJ_{[m_{23}:m_{22}]}
&\frac{\Gamma_{22}^{\,33}f(m_{23})}{m_{23}^2}\bJ_{[m_{23}:m_{23}]}
\end{bmatrix}$
},
\end{align*}
with
\[
\boldsymbol{\Gamma}=(\bP\otimes\bQ)\,\bM^{-1}\,(\bR\otimes\bS)
\]
and
\[
\Gamma_{ab}^{\,tu}
=\big[(\bP\otimes\bQ)\,\bM^{-1}\,(\bR\otimes\bS)\big]_{(a,t),(b,u)}
=\sum_{c=1}^{2}\sum_{v=1}^{3}\frac{p_{ac}\,q_{tv}\,r_{cb}\,s_{vu}}{m_{cv}}.
\]
Therefore, we can write 
\begin{align*}
    [(\bP\otimes\bQ)_{\text{cell}}\circledast \bI_{\bm}^f\tildebJ_{\bm}]
[(\bR\otimes\bS)_{\text{cell}}\circledast\tildebJ_{\bm}]&=[(\bP\otimes\bQ)\,\bM^{-1}\,(\bR\otimes\bS)]_{\text{cell}}\circledast \bI_{\bm}^f\tildebJ_{\bm}.
\end{align*}

\end{proof}
\subsection*{Proof of Lemma 3}

\begin{proof}
Set $\bD= \bC-\bA$, so that $\bD \ge 0$ entrywise. Then
\[
\bC \bB - \bA \bB = (\bC-\bA)\bB = \bD \bB.
\]
For any $i,j$,
\[
(\bD \bB)_{ij} = \sum_{k=1}^n \bD_{ik} \bB_{kj}.
\]
Since $\bD_{ik} \ge 0$ and $\bB_{kj} \ge 0$ for all $k,j$, each term in the sum is nonnegative, hence $(\bD \bB)_{ij}\ge 0$. This proves $\bA \bB \le \bC \bB$ entrywise.

If for a fixed row $i$ there exists some $k$ with $\bD_{ik}>0$, then for any column $j$, if there is at least one such $k$ with $\bB_{kj}>0$, we obtain
\[
(\bD \bB)_{ij} \ge \bD_{ik} \bB_{kj} > 0,
\]
so the $(i,j)$ entry of $\bC \bB-\bA \bB$ is strictly positive. Conversely, if in row $i$ all $\bD_{ik}=0$, or if for a given $j$ all $k$ with $\bD_{ik}>0$ also satisfy $\bB_{kj}=0$, then $(\bD \bB)_{ij}=0$ and the corresponding entries of $\bA \bB$ and $\bC \bB$ are equal. This yields the stated condition for strict inequality.
\end{proof}

\subsection*{Proof of Lemma 4}
\begin{proof}
As $g,h\to\infty$, we have 
\begin{align*}
&    (\barbJ{g}\otimes\bI_h)_{\text{cell}}\circledast\tildebJ_{\bm}=\bO_{[n:n]}\left(\frac{1}{gm_L^2}\right),
\qquad    (\bI_g\otimes\barbJ{h})_{\text{cell}}\circledast\tildebJ_{\bm}=\bO_{[n:n]}\left(\frac{1}{hm_L^2}\right),
\\&    (\barbJ{g}\otimes\barbJ{h})_{\text{cell}}\circledast\tildebJ_{\bm}=\bO_{[n:n]}\left(\frac{1}{ghm_L^2}\right),
\end{align*}
where $\bO_{[n:n]}(a)$ denotes an $n \times n$ matrix with all entries of order $O(a)$.
For $l=1$, we have
\begin{align*}
\big[(\barbJ{g}\otimes\bI_h)_{\text{cell}}\circledast\tildebJ_{\bm}
		+(\bI_g\otimes\barbJ{h})_{\text{cell}}\circledast\tildebJ_{\bm}
		+(\barbJ{g}\otimes\barbJ{h})_{\text{cell}}\circledast\tildebJ_{\bm}\big]
=\bO_{[n:n]}\left(\frac{1}{gm_L^2} +\frac{1}{hm_L^2}+ \frac{1}{ghm_L^2}\right).   
\end{align*}
	Following Lemma 1 and $\barbJ{a}\barbJ{a}=\barbJ{a}$,  we have
	\begin{align*}
		&\left[(\barbJ{g}\otimes\bI_h)_{\text{cell}}\circledast\tildebJ_{\bm}\right]
		\left[(\barbJ{g}\otimes\bI_h)_{\text{cell}}\circledast\tildebJ_{\bm}\right]
		\le m_L^{-1}\left[(\barbJ{g}\otimes\bI_h)_{\text{cell}}\circledast\tildebJ_{\bm}\right],
		\\&
		\left[(\barbJ{g}\otimes\bI_h)_{\text{cell}}\circledast\tildebJ_{\bm}\right]  \left[(\bI_g\otimes\barbJ{h})_{\text{cell}}\circledast\tildebJ_{\bm}\right]\le m_L^{-1}\left[(\barbJ{g}\otimes\barbJ{h})_{\text{cell}}\circledast\tildebJ_{\bm}\right],
		\\&
		\left[(\barbJ{g}\otimes\bI_h)_{\text{cell}}\circledast\tildebJ_{\bm}\right]  \left[(\barbJ{g}\otimes\barbJ{h})_{\text{cell}}\circledast\tildebJ_{\bm}\right]\le m_L^{-1}\left[(\barbJ{g}\otimes\barbJ{h})_{\text{cell}}\circledast\tildebJ_{\bm}\right],
		\\&
		\left[(\bI_g\otimes\barbJ{h})_{\text{cell}}\circledast\tildebJ_{\bm}\right]
		\left[(\bI_g\otimes\barbJ{h})_{\text{cell}}\circledast\tildebJ_{\bm}\right]\le m_L^{-1}\left[(\bI_{g}\otimes\barbJ{h})_{\text{cell}}\circledast\tildebJ_{\bm}\right],
		\\&
		\left[(\bI_g\otimes\barbJ{h})_{\text{cell}}\circledast\tildebJ_{\bm}\right]  \left[(\barbJ{g}\otimes\barbJ{h})_{\text{cell}}\circledast\tildebJ_{\bm}\right]\le m_L^{-1}\left[(\barbJ{g}\otimes\barbJ{h})_{\text{cell}}\circledast\tildebJ_{\bm}\right],
		\\&
		\left[(\barbJ{g}\otimes\barbJ{h})_{\text{cell}}\circledast\tildebJ_{\bm}\right] \left[(\barbJ{g}\otimes\barbJ{h})_{\text{cell}}\circledast\tildebJ_{\bm}\right]\le m_L^{-1}\left[(\barbJ{g}\otimes\barbJ{h})_{\text{cell}}\circledast\tildebJ_{\bm}\right].
	\end{align*}
	For $l=2$, following Lemma 3, we have 
	\begin{align*}
	&	\big[(\barbJ{g}\otimes\bI_h)_{\text{cell}}\circledast\tildebJ_{\bm}
		+(\bI_g\otimes\barbJ{h})_{\text{cell}}\circledast\tildebJ_{\bm}
		+(\barbJ{g}\otimes\barbJ{h})_{\text{cell}}\circledast\tildebJ_{\bm}\big]^2
		\\&
		=
		\left[(\barbJ{g}\otimes\bI_h)_{\text{cell}}\circledast\tildebJ_{\bm}\right]
		\left[(\barbJ{g}\otimes\bI_h)_{\text{cell}}\circledast\tildebJ_{\bm}\right]
		+
		\left[(\barbJ{g}\otimes\bI_h)_{\text{cell}}\circledast\tildebJ_{\bm}\right]  \left[(\bI_g\otimes\barbJ{h})_{\text{cell}}\circledast\tildebJ_{\bm}\right]
		\\&\quad
		+
			\left[(\barbJ{g}\otimes\bI_h)_{\text{cell}}\circledast\tildebJ_{\bm}\right]  \left[(\barbJ{g}\otimes\barbJ{h})_{\text{cell}}\circledast\tildebJ_{\bm}\right]
		+
 \left[(\bI_g\otimes\barbJ{h})_{\text{cell}}\circledast\tildebJ_{\bm}\right]
 \left[(\barbJ{g}\otimes\bI_h)_{\text{cell}}\circledast\tildebJ_{\bm}\right] 
			\\&\quad
	+
		\left[(\bI_g\otimes\barbJ{h})_{\text{cell}}\circledast\tildebJ_{\bm}\right]
	\left[(\bI_g\otimes\barbJ{h})_{\text{cell}}\circledast\tildebJ_{\bm}\right]
	+
		\left[(\bI_g\otimes\barbJ{h})_{\text{cell}}\circledast\tildebJ_{\bm}\right]  \left[(\barbJ{g}\otimes\barbJ{h})_{\text{cell}}\circledast\tildebJ_{\bm}\right]
				\\&\quad
		+
\left[(\barbJ{g}\otimes\barbJ{h})_{\text{cell}}\circledast\tildebJ_{\bm}\right]
			\left[(\barbJ{g}\otimes\bI_h)_{\text{cell}}\circledast\tildebJ_{\bm}\right]  
			+
\left[(\bI_g\otimes\barbJ{h})_{\text{cell}}\circledast\tildebJ_{\bm}\right]
	\left[(\bI_g\otimes\barbJ{h})_{\text{cell}}\circledast\tildebJ_{\bm}\right]
		\\&\quad
+		
			\left[(\barbJ{g}\otimes\barbJ{h})_{\text{cell}}\circledast\tildebJ_{\bm}\right] \left[(\barbJ{g}\otimes\barbJ{h})_{\text{cell}}\circledast\tildebJ_{\bm}\right]			
\\&
\le
\frac{1}{m_L}\left[(\barbJ{g}\otimes\bI_h)_{\text{cell}}\circledast\tildebJ_{\bm}\right]
+\frac{1}{m_L}\left[(\bI_{g}\otimes\barbJ{h})_{\text{cell}}\circledast\tildebJ_{\bm}\right]
+ \frac{7}{m_L}\left[(\barbJ{g}\otimes\barbJ{h})_{\text{cell}}\circledast\tildebJ_{\bm}\right].
		\end{align*}
For $l=3$,  following Lemma 3, we have 
\begin{align*}
		&	\big[(\barbJ{g}\otimes\bI_h)_{\text{cell}}\circledast\tildebJ_{\bm}
	+(\bI_g\otimes\barbJ{h})_{\text{cell}}\circledast\tildebJ_{\bm}
	+(\barbJ{g}\otimes\barbJ{h})_{\text{cell}}\circledast\tildebJ_{\bm}\big]^3
\\&
\le 
\left(
\frac{1}{m_L}\left[(\barbJ{g}\otimes\bI_h)_{\text{cell}}\circledast\tildebJ_{\bm}\right]
+\frac{1}{m_L}\left[(\bI_{g}\otimes\barbJ{h})_{\text{cell}}\circledast\tildebJ_{\bm}\right]
+ \frac{7}{m_L}\left[(\barbJ{g}\otimes\barbJ{h})_{\text{cell}}\circledast\tildebJ_{\bm}\right]
\right)\times
\\&\quad
\big[(\barbJ{g}\otimes\bI_h)_{\text{cell}}\circledast\tildebJ_{\bm}
+(\bI_g\otimes\barbJ{h})_{\text{cell}}\circledast\tildebJ_{\bm}
+(\barbJ{g}\otimes\barbJ{h})_{\text{cell}}\circledast\tildebJ_{\bm}\big]
\\&
\le 
 \frac{1}{m_L^2}\left[(\barbJ{g}\otimes\bI_h)_{\text{cell}}\circledast\tildebJ_{\bm}\right]
+\frac{1}{m_L^2}\left[(\bI_{g}\otimes\barbJ{h})_{\text{cell}}\circledast\tildebJ_{\bm}\right]
+ \frac{25}{m_L^2}\left[(\barbJ{g}\otimes\barbJ{h})_{\text{cell}}\circledast\tildebJ_{\bm}\right].
\end{align*}
Similarly,  following Lemma 3, we have 
\begin{align*}
		&	\big[(\barbJ{g}\otimes\bI_h)_{\text{cell}}\circledast\tildebJ_{\bm}
	+(\bI_g\otimes\barbJ{h})_{\text{cell}}\circledast\tildebJ_{\bm}
	+(\barbJ{g}\otimes\barbJ{h})_{\text{cell}}\circledast\tildebJ_{\bm}\big]^l
	\\&
	\le 
\frac{1}{m_L^{l-1}}\left[(\barbJ{g}\otimes\bI_h)_{\text{cell}}\circledast\tildebJ_{\bm}\right]
	+\frac{1}{m_L^{l-1}}\left[(\bI_{g}\otimes\barbJ{h})_{\text{cell}}\circledast\tildebJ_{\bm}\right]
	+ \frac{3^{l-1}+4\sum_{i=2}^l 3^{i-2}}{m_L^{l-1}}\left[(\barbJ{g}\otimes\barbJ{h})_{\text{cell}}\circledast\tildebJ_{\bm}\right].
\end{align*} 
As $g,h\to\infty$, we have
\begin{align*}
&\frac{1}{m_L^{l-1}}\left[(\barbJ{g}\otimes\bI_h)_{\text{cell}}\circledast\tildebJ_{\bm}\right]=
		\bO_{[n:n]}\left(\frac{1}{gm_L^{l+1}}\right),
\quad
\frac{1}{m_L^{l-1}}\left[(\bI_{g}\otimes\barbJ{h})_{\text{cell}}\circledast\tildebJ_{\bm}\right]
=
\bO_{[n:n]}\left(\frac{1}{hm_L^{l+1}}\right),
\end{align*}

	\begin{align*}
	 \frac{3^{l-1}+4\sum_{i=2}^l 3^{i-2}}{m_L^{l-1}}\left[(\barbJ{g}\otimes\barbJ{h})_{\text{cell}}\circledast\tildebJ_{\bm}\right]&=
 \frac{3^l-2}{m_L^{l-1}}
\left[(\barbJ{g}\otimes\barbJ{h})_{\text{cell}}\circledast\tildebJ_{\bm}\right]
=\bO_{[n:n]}\left(\frac{3^{l}-2}{ghm_L^{l+1}}\right).
	\end{align*}
Therefore, we have 
\begin{align*}
		&	\big[(\barbJ{g}\otimes\bI_h)_{\text{cell}}\circledast\tildebJ_{\bm}
	+(\bI_g\otimes\barbJ{h})_{\text{cell}}\circledast\tildebJ_{\bm}
	+(\barbJ{g}\otimes\barbJ{h})_{\text{cell}}\circledast\tildebJ_{\bm}\big]^k
   \\& =\bO_{[n:n]}\left(\frac{1}{gm_L^{l+1}}+\frac{1}{hm_L^{l+1}}+\frac{3^{l}-2}{ghm_L^{l+1}}\right)
   \\&
=  \bO_{[n:n]}\left(\frac{1}{\min(g,h) m_L^{l+1}}+\frac{3^{l}-2}{ghm_L^{l+1}}\right) .
\end{align*}

\end{proof}

\subsection*{Proof of Lemma 5}

\begin{proof}
 
Lower bound: for $g> 1$ we have $\log(g) \ge 0$, hence 
\begin{align*}
d_2^{\sqrt{\log(g)}}\ge 1, \qquad\text{then}\qquad   \frac{ d_2^{\sqrt{\log (g)}}}{g} \ge \frac{1}{g}.
\end{align*}
Since $g^{-{\varepsilon}}\le 1$ for all $g >1$ and $\varepsilon\in(0,1)$, it follows that 
\begin{align*}
   \frac{ d_2^{\sqrt{\log (g)}}}{g} \ge \frac{1}{g^{1+\varepsilon}} \qquad\text{for all $g> 1$}.
\end{align*}
 
Upper bound: fix $\varepsilon\in(0,1)$ and define $G_\varepsilon=\exp[\{\log(d_2)\}^2/\varepsilon^2]$. If $g\ge G_\varepsilon$, then 
\begin{align*}
\sqrt{\log(g)}\ge \frac{\log (d_2) }{\varepsilon},\quad\text{equivalently}\quad
\frac{\log (d_2)}{\sqrt{\log(g)}}\le \varepsilon,\quad\text{so}\quad
\log(d_2) \sqrt{\log(g)}\le \varepsilon\log(g).
\end{align*}
 Exponentiating gives
 \begin{align*}
  d_2^{\sqrt{\log(g)}}=\exp\{(\log(d_2))\sqrt{\log(g)}\}\le \exp\{\varepsilon\log(g)\}=g^{\varepsilon} .
 \end{align*}
Dividing by $g$ yields
\begin{align*}
     \frac{d_2^{\sqrt{\log (g)}}}{g}\le \frac{1}{g^{1-\varepsilon}} \quad
\text{for all $g\ge G_\varepsilon$.}
\end{align*}
\end{proof}

\subsection*{Proof of Lemma 6}

\begin{proof}
We use the properties of the Khatri--Rao product from Lemma~\ref{lma1} to show that all elements of the matrices 
$\mathcal{R}$, $(-\bU_2^{-1}\mathcal{R})^l$ and $-\mathcal{R}(-\bU_2^{-1}\mathcal{R})^l$ converge to zero as $g,h\to\infty$.

\medskip
\noindent
1) For $\mathcal{R}$, we write
\begin{align*}
\mathcal{R}
=&\, m_U\Bigl(\frac{1}{\lambda_5}-\frac{1}{\lambda_7}\Bigr)
(\barbJ{g}\otimes\bI_h)_{\text{cell}}\circledast\tildebJ_{\bm}
+m_U\Bigl(\frac{1}{\lambda_3}-\frac{1}{\lambda_7}\Bigr)
(\bI_g\otimes\barbJ{h})_{\text{cell}}\circledast\tildebJ_{\bm}
\\
&\quad +m_U\Bigl(\frac{1}{\lambda_1}-\frac{1}{\lambda_3}
-\frac{1}{\lambda_5}+\frac{1}{\lambda_7}\Bigr)
(\barbJ{g}\otimes\barbJ{h})_{\text{cell}}\circledast\tildebJ_{\bm}.
\end{align*}
As $g,h\to\infty$, we have
\begin{align*}
&m_U\Bigl(\frac{1}{\lambda_3}-\frac{1}{\lambda_7}\Bigr)
=\frac{-hm_U^2\siga^2}{(\sige^2+m_U\siggama^2+hm_U\siga^2)(\sige^2+m_U\siggama^2)}
\to \frac{-m_U}{\sige^2+m_U\siggama^2}
=O(1),
\\
&m_U\Bigl(\frac{1}{\lambda_5}-\frac{1}{\lambda_7}\Bigr)
=\frac{-gm_U^2\sigb^2}{(\sige^2+m_U\siggama^2+gm_U\sigb^2)(\sige^2+m_U\siggama^2)}
\to \frac{-m_U}{\sige^2+m_U\siggama^2}
=O(1),
\\
&m_U\Bigl(\frac{1}{\lambda_1}-\frac{1}{\lambda_3}
-\frac{1}{\lambda_5}+\frac{1}{\lambda_7}\Bigr)
=\frac{gm_U\sigb^2(\lambda_1\lambda_3-\lambda_5\lambda_7)}
{\lambda_1\lambda_3\lambda_5\lambda_7}
\to \frac{1}{\sige^2+m_U\siggama^2}
=O(1).
\end{align*}
Following Lemma 4, as $g,h\to\infty$, we obtain
\begin{align*}
\mathcal{R}
=\bO_{[n:n]}\left(\frac{1}{gm_L^2}
+\frac{1}{hm_L^2}
+\frac{1}{ghm_L^2}\right)
=\bO_{[n:n]}\left(\frac{1}{\min(g,h)m_L^2}\right).
\end{align*}

\medskip
\noindent
2) For $-\bU_1\bU_2^{-1}\mathcal{R}$ and $-\bU_2^{-1}\mathcal{R}\bU_1$, we write
\begin{align*}
\bU_2^{-1}
&=\left(\diag\left[(a_{ij}+b_{ij})\bI_{m_{ij}}+c_{ij}\bJ_{m_{ij}}\right]\right)^{-1}
\\
&=\diag\left[\{(a_{ij}+b_{ij})\bI_{m_{ij}}+c_{ij}\bJ_{m_{ij}}\}^{-1}\right]
\\
&=\diag\left[\frac{1}{a_{ij}+b_{ij}}\left(\bI_{m_{ij}}
-\frac{c_{ij}}{a_{ij}+b_{ij}+m_{ij}c_{ij}}\bJ_{m_{ij}}\right)\right].
\end{align*}
Then,
\begin{align*}
\bU_1\bU_2^{-1}
&=\diag\left[b_{ij}\bI_{m_{ij}}+c_{ij}\bJ_{m_{ij}}\right]
\diag\left[\frac{1}{a_{ij}+b_{ij}}\left(\bI_{m_{ij}}
-\frac{c_{ij}}{a_{ij}+b_{ij}+m_{ij}c_{ij}}\bJ_{m_{ij}}\right)\right]
\\
&=\diag\left[
\frac{1}{a_{ij}+b_{ij}}\bI_{m_{ij}}
+\frac{a_{ij}c_{ij}}{(a_{ij}+b_{ij})(a_{ij}+b_{ij}+m_{ij}c_{ij}}\bJ_{m_{ij})}
\right].
\end{align*}
Since $\bU_1\bU_2^{-1}$ is block diagonal with finite elements and each block has finite dimension $m_{ij}\times m_{ij}$, as $g,h\to\infty$, we obtain
\begin{align*}
\bU_1\bU_2^{-1}\mathcal{R}
=\bO_{[n:n]}\left(\frac{1}{\min(g,h)m_L^2}\right)
\quad\text{and}\quad
\bU_2^{-1}\mathcal{R}\bU_1
=\bO_{[n:n]}\left(\frac{1}{\min(g,h)m_L^2}\right).
\end{align*}

\medskip
\noindent
3) For $-\mathcal{R}\bU_2^{-1}\mathcal{R}$, we first calculate
\begin{align*}
&-\bU_2^{-1}\mathcal{R}
\\
&=\diag\left[\frac{1}{a_{ij}+b_{ij}}\left(\bI_{m_{ij}}
-\frac{c_{ij}}{a_{ij}+b_{ij}+m_{ij}c_{ij}}\bJ_{m_{ij}}\right)\right](-\mathcal{R})   
\\
&=\diag\left[\frac{1}{a_{ij}+b_{ij}} \bI_{m_{ij}}\right](-\mathcal{R})
+\diag\left[\frac{-c_{ij}}{(a_{ij}+b_{ij})(a_{ij}+b_{ij}+m_{ij}c_{ij})}\bJ_{m_{ij}}\right](-\mathcal{R}).
\end{align*}
For $i=1,\ldots,g$ and $j=1,\ldots,h$, we have
\begin{align*}
&0\le \frac{1}{a_{ij}+b_{ij}}
=\frac{m_{ij}(m_U-m_{ij})\sige^2}{m_U^2}
\le \frac{\sige^2}{4},
\\
&0\le
\frac{-c_{ij}}{(a_{ij}+b_{ij})(a_{ij}+b_{ij}+m_{ij}c_{ij})}
=\frac{(m_U-m_{ij})^2\sige^2\siggama^2}
{m_U^2(\sige^2+m_{ij}\siggama^2)}
\le \frac{(m_U-m_L)^2\sige^2\siggama^2}
{m_U^2(\sige^2+m_L\siggama^2)}.
\end{align*}
Since 
$\diag\left[{1}/{(a_{ij}+b_{ij})}\bI_{m_{ij}}\right]$ and 
$\diag\left[{-c_{ij}}/{((a_{ij}+b_{ij})(a_{ij}+b_{ij}+m_{ij}c_{ij}))}\bJ_{m_{ij}}\right]$ 
are block-diagonal matrices with nonnegative entries, 
and $-\mathcal{R}$ is symmetric with nonnegative entries, 
it follows from Lemma 3 that
\begin{align*}
-\bU_2^{-1}\mathcal{R}
&\le \frac{\sige^2}{4}\bI_n(-\mathcal{R})
+\frac{(m_U-m_L)^2\sige^2\siggama^2}{m_U^2(\sige^2+m_L\siggama^2)}
\big[(\bI_g\otimes\bI_h)_{\text{cell}}\circledast\tildebJ_{\bm}\big](-\mathcal{R})
\\
&=\frac{\sige^2}{4}\bigg[-m_U\Bigl(\frac{1}{\lambda_5}-\frac{1}{\lambda_7}\Bigr)
(\barbJ{g}\otimes\bI_h)_{\text{cell}}\circledast\tildebJ_{\bm}
-m_U\Bigl(\frac{1}{\lambda_3}-\frac{1}{\lambda_7}\Bigr)
(\bI_g\otimes\barbJ{h})_{\text{cell}}\circledast\tildebJ_{\bm}
\\
&\qquad -m_U\Bigl(\frac{1}{\lambda_1}-\frac{1}{\lambda_3}
-\frac{1}{\lambda_5}+\frac{1}{\lambda_7}\Bigr)
(\barbJ{g}\otimes\barbJ{h})_{\text{cell}}\circledast\tildebJ_{\bm}\bigg]
\\
&\quad +\frac{(m_U-m_L)^2\sige^2\siggama^2}{m_U^2(\sige^2+m_U\siggama^2)}
\big[(\bI_g\otimes\bI_h)_{\text{cell}}\circledast\tildebJ_{\bm}\big]
\bigg[-m_U\Bigl(\frac{1}{\lambda_5}-\frac{1}{\lambda_7}\Bigr)
(\barbJ{g}\otimes\bI_h)_{\text{cell}}\circledast\tildebJ_{\bm}
\\
&\qquad -m_U\Bigl(\frac{1}{\lambda_3}-\frac{1}{\lambda_7}\Bigr)
(\bI_g\otimes\barbJ{h})_{\text{cell}}\circledast\tildebJ_{\bm}
-m_U\Bigl(\frac{1}{\lambda_1}-\frac{1}{\lambda_3}
-\frac{1}{\lambda_5}+\frac{1}{\lambda_7}\Bigr)
(\barbJ{g}\otimes\barbJ{h})_{\text{cell}}\circledast\tildebJ_{\bm}\bigg].
\end{align*}
Let
\begin{align*}
d_1
=\frac{m_U\sige^2}{4(\sige^2+m_U\siggama^2)}
\max\left[
\frac{hm_U\siga^2}{\sige^2+m_U\siggama^2+hm_U\siga^2},
\frac{gm_U\siga^2}{\sige^2+m_U\siggama^2+gm_U\sigb^2}
\right],
\end{align*}
then we have
\begin{align*}
&-\frac{\sige^2}{4}m_U\Bigl(\frac{1}{\lambda_3}-\frac{1}{\lambda_7}\Bigr)
=\frac{hm_U^2\sige^2\siga^2}
{4(\sige^2+m_U\siggama^2)(\sige^2+m_U\siggama^2+hm_U\siga^2)}
\le d_1,
\\
&-\frac{\sige^2}{4}m_U\Bigl(\frac{1}{\lambda_5}-\frac{1}{\lambda_7}\Bigr)
=\frac{gm_U^2\sige^2\sigb^2}
{4(\sige^2+m_U\siggama^2)(\sige^2+m_U\siggama^2+gm_U\sigb^2)}
\le d_1,
\\
&-\frac{\sige^2}{4}m_U\Bigl(\frac{1}{\lambda_1}-\frac{1}{\lambda_3}
-\frac{1}{\lambda_5}+\frac{1}{\lambda_7}\Bigr)
\\
&=\frac{gm_U^2\sige^2\sigb^2}
{4(\sige^2+m_U\siggama^2+gm_U\sigb^2)(\sige^2+m_U\siggama^2)}
-\frac{gm_U^2\sige^2\sigb^2}
{4(\sige^2+m_U\siggama^2+hm_U\siga^2+gm_U\sigb^2)
(\sige^2+m_U\siggama^2+hm_U\siga^2)}
\\&
\le d_1,
\end{align*}
and
\begin{align*}
&-\frac{(m_U-m_L)^2\sige^2\siggama^2}{m_U^2(\sige^2+m_U\siggama^2)}
m_U\Bigl(\frac{1}{\lambda_3}-\frac{1}{\lambda_7}\Bigr)
=\frac{hm_U^2(m_U-m_L)^2\sige^2\siggama^2\siga^2}
{m_U^2(\sige^2+m_U\siggama^2)^2(\sige^2+m_U\siggama^2+hm_U\siga^2)}
\le d_1,
\\
&-\frac{(m_U-m_L)^2\sige^2\siggama^2}{m_U^2(\sige^2+m_U\siggama^2)}
m_U\Bigl(\frac{1}{\lambda_5}-\frac{1}{\lambda_7}\Bigr)
=\frac{gm_U^2(m_U-m_L)^2\sige^2\siggama^2\sigb^2}
{m_U^2(\sige^2+m_U\siggama^2)^2(\sige^2+m_U\siggama^2+gm_U\sigb^2)}
\le d_1,
\\
&-\frac{(m_U-m_L)^2\sige^2\siggama^2}{m_U^2(\sige^2+m_U\siggama^2)}
m_U\Bigl(\frac{1}{\lambda_1}-\frac{1}{\lambda_3}
-\frac{1}{\lambda_5}+\frac{1}{\lambda_7}\Bigr)
\\
&=\frac{gm_U^2(m_U-m_L)^2\sige^2\siggama^2\sigb^2}
{m_U^2(\sige^2+m_U\siggama^2)^2(\sige^2+m_U\siggama^2+gm_U\sigb^2)}
\\&\qquad
-\frac{gm_U^2(m_U-m_L)^2\sige^2\siggama^2\sigb^2}
{m_U^2(\sige^2+m_U\siggama^2)(\sige^2+m_U\siggama^2+hm_U\siga^2)
(\sige^2+m_U\siggama^2+hm_U\siga^2+gm_U\sigb^2)}
\\&\le d_1.
\end{align*}
Applying Lemma 3 again, we have
\begin{align*}
-\bU_2^{-1}\mathcal{R}
&\le d_1\big[(\barbJ{g}\otimes\bI_h)_{\text{cell}}\circledast\tildebJ_{\bm}
+(\bI_g\otimes\barbJ{h})_{\text{cell}}\circledast\tildebJ_{\bm}
+(\barbJ{g}\otimes\barbJ{h})_{\text{cell}}\circledast\tildebJ_{\bm}\big]
\\
&\quad +d_1\big[(\bI_g\otimes\bI_h)_{\text{cell}}\circledast\tildebJ_{\bm}\big]
\big[(\barbJ{g}\otimes\bI_h)_{\text{cell}}\circledast\tildebJ_{\bm}
+(\bI_g\otimes\barbJ{h})_{\text{cell}}\circledast\tildebJ_{\bm}
+(\barbJ{g}\otimes\barbJ{h})_{\text{cell}}\circledast\tildebJ_{\bm}\big].
\end{align*}
Following Lemma 1 and using $\barbJ{a}\barbJ{a}=\barbJ{a}$, we obtain
\begin{align*}
&\big[(\bI_g\otimes\bI_h)_{\text{cell}}\circledast\tildebJ_{\bm}\big]
\big[(\bI_g\otimes\barbJ{h})_{\text{cell}}\circledast\tildebJ_{\bm}\big]
\le m_L^{-1}\big[(\bI_g\otimes\barbJ{h})_{\text{cell}}\circledast\tildebJ_{\bm}\big],
\\
&\big[(\bI_g\otimes\bI_h)_{\text{cell}}\circledast\tildebJ_{\bm}\big]
\big[(\barbJ{g}\otimes\bI_h)_{\text{cell}}\circledast\tildebJ_{\bm}\big]
\le m_L^{-1}\big[(\barbJ{g}\otimes\bI_h)_{\text{cell}}\circledast\tildebJ_{\bm}\big],
\\
&\big[(\bI_g\otimes\bI_h)_{\text{cell}}\circledast\tildebJ_{\bm}\big]
\big[(\barbJ{g}\otimes\barbJ{h})_{\text{cell}}\circledast\tildebJ_{\bm}\big]
\le m_L^{-1}\big[(\barbJ{g}\otimes\barbJ{h})_{\text{cell}}\circledast\tildebJ_{\bm}\big].
\end{align*}
Hence,
\begin{equation}\label{eq: U2R}
-\bU_2^{-1}\mathcal{R}
\le \frac{d_1(m_L+1)}{m_L}
\big[(\barbJ{g}\otimes\bI_h)_{\text{cell}}\circledast\tildebJ_{\bm}
+(\bI_g\otimes\barbJ{h})_{\text{cell}}\circledast\tildebJ_{\bm}
+(\barbJ{g}\otimes\barbJ{h})_{\text{cell}}\circledast\tildebJ_{\bm}\big].
\end{equation}
Since  
\begin{equation}\label{Positive R cons}
    \begin{split}
&0<-m_U\Bigl(\frac{1}{\lambda_3}-\frac{1}{\lambda_7}\Bigr)<1, \quad
0<-m_U\Bigl(\frac{1}{\lambda_5}-\frac{1}{\lambda_7}\Bigr)<1, \\&
0<-m_U\Bigl(\frac{1}{\lambda_1}-\frac{1}{\lambda_3}
-\frac{1}{\lambda_5}+\frac{1}{\lambda_7}\Bigr)<1,
    \end{split}
\end{equation}
we have
\begin{align*}
\mathcal{R}\bU_2^{-1}\mathcal{R}
\le \frac{d_1(m_L+1)}{m_L}
\big[(\barbJ{g}\otimes\bI_h)_{\text{cell}}\circledast\tildebJ_{\bm}
+(\bI_g\otimes\barbJ{h})_{\text{cell}}\circledast\tildebJ_{\bm}
+(\barbJ{g}\otimes\barbJ{h})_{\text{cell}}\circledast\tildebJ_{\bm}\big]^2.
\end{align*}
Therefore, by Lemma 4, as $g,h\to\infty$, we obtain
\begin{align*}
\mathcal{R}\bU_2^{-1}\mathcal{R}
=\bO_{[n:n]}\left(\frac{1}{gm_L^3}
+\frac{1}{hm_L^3}
+\frac{7}{ghm_L^3}\right)
=\bO_{[n:n]}\left(\frac{1}{\min(g,h)m_L^3}\right).
\end{align*}

\medskip
\noindent
4) For $\sum_{l=1}^{s(g,h)}\bU_1(-\bU_2^{-1}\mathcal{R})^{l+1}$ and $\sum_{l=1}^{s(g,h)}\bU_1(-\bU_2^{-1}\mathcal{R})^{l}\bU_2^{-1}\bU_1$, we first calculate $(-\bU_2^{-1}\mathcal{R})^{l}$. From \eqref{eq: U2R} and following Lemma 3 and 4, we have
\begin{equation} \label{K terms: UR}
    \begin{split}
&(-\bU_2^{-1}\mathcal{R})^{l}
\le \frac{d_1^l(m_L+1)^l}{m_L^{l}}
\big[(\barbJ{g}\otimes\bI_h)_{\text{cell}}\circledast\tildebJ_{\bm}
+(\bI_g\otimes\barbJ{h})_{\text{cell}}\circledast\tildebJ_{\bm}
+(\barbJ{g}\otimes\barbJ{h})_{\text{cell}}\circledast\tildebJ_{\bm}\big]^{l}.
    \end{split}
\end{equation}
Following Lemma 4, we obtain
\begin{align*}
(-\bU_2^{-1}\mathcal{R})^l
&=
\frac{d_1^l(m_L+1)^l}{m_L^l}\bO_{[n:n]}\left(\frac{1}{\min(g,h) m_L^{l+1}}+\frac{3^{l}-2}{ghm_L^{l+1}}\right)
\\
&=\bO_{[n:n]}\left(\frac{d_1^l(m_L+1)^l}{\min(g,h)m_L^{2l+1}}+\frac{(3^{l}-2)d_1^l(m_L+1)^l}{ghm_L^{2l+1}}\right).
\end{align*}
Now we calculate
\begin{align*}
\sum_{l=1}^{s(g,h)} (-\bU_2^{-1}\mathcal{R})^l
&=\bO_{[n:n]}\left(\sum_{l=1}^{s(g,h)}\frac{d_1^l(m_L+1)^l}{\min(g,h)m_L^{2l+1}}+\sum_{l=1}^{s(g,h)}\frac{(3^{l}-2)d_1^l(m_L+1)^l}{ghm_L^{2l+1}}\right).
\end{align*}
We denote $d_2={d_1(m_L+1)}/{m_L^{2}}<\sige^2/(2m_L\siggama^2)<\infty$, then we have
\begin{align*}
\sum_{l=1}^{s(g,h)}\frac{d_1^l(m_L+1)^l}{\min(g,h)m_L^{2l+1}}
&=\frac{1}{\min(g,h)m_L}\sum_{l=1}^{s(g,h)}
\left[\frac{d_1(m_L+1)}{m_L^{2}}\right]^l
\\
&=\frac{1}{\min(g,h)m_L}\sum_{l=1}^{s(g,h)} d_2^l
\\
&=\frac{1}{\min(g,h)m_L}\frac{d_2(1-d_2^{s(g,h)})}{1-d_2}
\\
&=\frac{d_2}{\min(g,h)m_L(1-d_2)}
-
\frac{d_2}{m_L(1-d_2)}\frac{d_2^{s(g,h)}}{\min(g,h)}.
\end{align*}
Since $d_2/[m_L(1-d_2)]<\sige^2/(2m_L^2\siggama^2)$, as $g,h\to\infty$, we have
\begin{align*}
\frac{d_2}{\min(g,h)m_L(1-d_2)} =O\left(\frac{1}{\min(g,h)m_L^2}\right).
\end{align*}
If $d_2\le 1$ (for example $\sige^2/(2m_L\siggama^2)\le 1$), as $g,h\to\infty$, we have
\begin{align*}
\frac{d_2}{m_L(1-d_2)}\frac{d_2^{s(g,h)}}{\min(g,h)}
=O\left(\frac{1}{\min(g,h)m_L^2}\right).
\end{align*}
If $d_2>1$, following Lemma 5, as $g,h\to\infty$, we have
\begin{align*}
\frac{d_2}{m_L(1-d_2)}\frac{d_2^{s(g,h)}}{\min(g,h)}
=O\left(\frac{1}{\min(g^{1-\varepsilon},h^{1-\varepsilon})m_L^2}\right),
\qquad \text{where $\varepsilon\in (0,1)$.}
\end{align*}
Combining the above equations, we have
\begin{equation}\label{sum kterm1}
\begin{split}
\sum_{l=1}^{s(g,h)}\frac{d_1^l(m_L+1)^l}{\min(g,h)m_L^{2l+1}}
&=O\left(\frac{1}{\min(g,h)m_L^2}\right)
+O\left(\frac{1}{\min(g^{1-\varepsilon},h^{1-\varepsilon})m_L^2}\right)
\\
&=O\left(\frac{1}{\min(g^{1-\varepsilon},h^{1-\varepsilon})m_L^2}\right),
\end{split}
   \end{equation}
where $\varepsilon\in [0,1)$. Now we consider
\begin{align*}
\sum_{l=1}^{s(g,h)}\frac{(3^{l}-2)d_1^l(m_L+1)^l}{ghm_L^{2l+1}}
&=\frac{1}{ghm_L}\sum_{l=1}^{s(g,h)}\left[\frac{3d_1(m_L+1)}{m_L^2 }\right]^l
-\frac{2}{ghm_L}\sum_{l=1}^{s(g,h)}\left[\frac{d_1(m_L+1)}{m_L^2 }\right]^l
\\
&=\frac{1}{ghm_L}\sum_{l=1}^{s(g,h)}(3d_2)^l
-\frac{2}{ghm_L}\sum_{l=1}^{s(g,h)}d_2^l
\\
&=\frac{1}{ghm_L}\frac{3d_2[1-(3d_2)^{s(g,h)}]}{1-3d_2}
-
\frac{2}{ghm_L}\frac{d_2(1-d_2^{s(g,h)})}{1-d_2}.
\end{align*}
Similarly, following Lemma 5, as $g,h\to\infty$, we have
\begin{equation}\label{sum kterm2}
\begin{split}
\sum_{l=1}^{s(g,h)}\frac{(3^{l}-2)d_1^l(m_L+1)^l}{ghm_L^{2l+1}}
&=O\left(\frac{1}{ghm_L^2}\right)
+O\left(\frac{1}{\min(g^{1-\varepsilon}h,gh^{1-\varepsilon})m_L^2}\right)
\\
&=O\left(\frac{1}{\min(g^{1-\varepsilon}h,gh^{1-\varepsilon})m_L^2}\right),
\end{split}
\end{equation}
where $\varepsilon\in [0,1)$. Therefore, as $g,h\to\infty$, we have
\begin{align*}
\sum_{l=1}^{s(g,h)} (-\bU_2^{-1}\mathcal{R})^l
&=\bO_{[n:n]}\left(
\frac{1}{\min(g^{1-\varepsilon},h^{1-\varepsilon})m_L^2}
+\frac{1}{\min(g^{1-\varepsilon}h,gh^{1-\varepsilon})m_L^2}
\right)
\\
&=\bO_{[n:n]}\left(
\frac{1}{\min(g^{1-\varepsilon},h^{1-\varepsilon})m_L^2}
\right),
\end{align*}
where $\varepsilon\in [0,1)$. Further, as $g,h\to\infty$, we can obtain
\begin{align*}
\sum_{l=1}^{s(g,h)} \bU_1(-\bU_2^{-1}\mathcal{R})^{l+1}
&=\bO_{[n:n]}\left(
\frac{1}{\min(g^{1-\varepsilon},h^{1-\varepsilon})m_L^2}
\right),
\\
\sum_{l=1}^{s(g,h)} \bU_1(-\bU_2^{-1}\mathcal{R})^{l}\bU_2^{-1}\bU_1
&=\bO_{[n:n]}\left(
\frac{1}{\min(g^{1-\varepsilon},h^{1-\varepsilon})m_L^2}
\right),
\end{align*}
because $\bU_1$ and $\bU_2^{-1}\bU_1$ are block diagonal with finite elements and each block has finite dimension $m_{ij}\times m_{ij}$.

\medskip
\noindent
5) For
$\sum_{l=1}^{s(g,h)}-\mathcal{R} (-\bU_2^{-1}\mathcal{R})^{l+1}$ and 
$\sum_{l=1}^{s(g,h)}-\mathcal{R} (-\bU_2^{-1}\mathcal{R})^l\bU_2^{-1}\bU_1$, we first calculate
$-\mathcal{R} (-\bU_2^{-1}\mathcal{R})^l$.
Following  \eqref{Positive R cons}--\eqref{K terms: UR} and Lemma 3, we have
\begin{align*}
&-\mathcal{R}(-\bU_2^{-1}\mathcal{R})^{l}
\le \frac{d_1^l(m_L+1)^l}{m_L^{l}}
\big[(\barbJ{g}\otimes\bI_h)_{\text{cell}}\circledast\tildebJ_{\bm}
+(\bI_g\otimes\barbJ{h})_{\text{cell}}\circledast\tildebJ_{\bm}
+(\barbJ{g}\otimes\barbJ{h})_{\text{cell}}\circledast\tildebJ_{\bm}\big]^{l+1}.
\end{align*}
Therefore, by Lemma 4, as $g,h\to\infty$, we have
\begin{align*}
-\mathcal{R}(-\bU_2^{-1}\mathcal{R})^l
&=
\frac{d_1^l(m_L+1)^l}{m_L^l}\bO_{[n:n]}\left(\frac{1}{\min(g,h) m_L^{l+2}}+\frac{3^{l+1}-2}{ghm_L^{l+2}}\right)
\\
&=\bO_{[n:n]}\left(\frac{d_1^l(m_L+1)^l}{\min(g,h)m_L^{2l+2}}+\frac{(3^{l+1}-2)d_1^l(m_L+1)^l}{ghm_L^{2l+2}}\right).
\end{align*}

Now we calculate
\begin{align*}
\sum_{l=1}^{s(g,h)} -\mathcal{R}(-\bU_2^{-1}\mathcal{R})^l
&=\bO_{[n:n]}\left(\sum_{l=1}^{s(g,h)}\frac{d_1^l(m_L+2)^l}{\min(g,h)m_L^{2l+2}}+\sum_{l=1}^{s(g,h)}\frac{(3^{l+1}-2)d_1^l(m_L+1)^l}{ghm_L^{2l+2}}\right).
\end{align*}
Following \eqref{sum kterm1} and \eqref{sum kterm2}, as $g,h\to\infty$, we have
\begin{align*}
\sum_{l=1}^{s(g,h)} -\mathcal{R}(-\bU_2^{-1}\mathcal{R})^l
&=\bO_{[n:n]}\left(
\frac{1}{\min(g^{1-\varepsilon},h^{1-\varepsilon})m_L^3}
\right),
\end{align*}
where $\varepsilon\in [0,1)$. Further, as $g,h\to\infty$, we can obtain
\begin{align*}
\sum_{l=1}^{s(g,h)} -\mathcal{R}(-\bU_2^{-1}\mathcal{R})^l\bU_2^{-1}\bU_1
&=\bO_{[n:n]}\left(
\frac{1}{\min(g^{1-\varepsilon},h^{1-\varepsilon})m_L^3}
\right),
\end{align*}
because  $\bU_2^{-1}\bU_1$ is block diagonal with finite elements and each block has finite dimension $m_{ij}\times m_{ij}$.
\end{proof}

\newpage
\section{Results for Crossed Random Effect Models without Interaction}

The two-way crossed random effect with no interaction model is
\begin{equation}\label{two-way cross model no interaction}
	y_{ijk}=\mu(\bx_{ijk})+\alpha_i+\beta_j+e_{ijk},
\end{equation}
for $i=1,\ldots,g$, $j=1,\ldots,h$, $k=1,\ldots,m_{ij}$,
where $\mu(\bx_{ijk})$ is the conditional mean of the response given the covariate $\bx_{ijk}$ (also called the regression function), $\alpha_i$ is the random effect due to the $i$th row, $\beta_j$ is the random effect  due to the $j$th column, and $e_{ijk}$ is the error term.  The variables $\alpha_i$, $\beta_j$, and $e_{ijk}$ are assumed mutually independent with zero means and variances $\sigma_\alpha^2$, $\sigma_\beta^2$,  and $\sigma_e^2$, respectively.

The variance-covariance matrix for the model~(\ref{two-way cross model no interaction}) can be written in a similar form with $\gamma$, $\siggama^2$ and $\bZ_3$ omitted, given by 
\begin{align*}
\bV^\star=\sige^2\bI_n+\bD^\star, \quad\text{where}\quad  \bD^\star=\sigma_\alpha^2(\bI_g\otimes\bJ_h)_{\text{cell}}\circledast\bJ_{\bm}+\sigma_\beta^2(\bJ_g\otimes\bI_h)_{\text{cell}}\circledast\bJ_{\bm}.
\end{align*}
The corresponding modified covariance matrix is
\begin{equation} \label{eq:Vtildestar}
\check{\bV}^{\star} = \frac{\sigma_e^2}{m_U} \bI_{\bm}^M + \bD^{\star}.
\end{equation}
We introduce the star into the notation for some parameters and  variance-covariance matrix in the model (\ref{two-way cross model no interaction}) because this  reduces possible confusion with model (\ref{two-way cross model}). We have the following results.

\begin{theorem} \label{theom4}
 Let $\bC$ be the $n \times n$ orthogonal matrix with $\bC^\top = [\bC_1, \ldots, \bC_8]$, where each block $\bC_i$ is specified in Table~\ref{Matrix C:component}. Then $\bC^\top \bC = \bI_n$.
     	 For the model without interaction in (\ref{two-way cross model no interaction}), 
         we have
	\begin{displaymath}
		\begin{aligned}
			\bC  m_U\barbI_{\bm} \check{\bV}^{\star} \barbI_{\bm}\bC^\top &=
			\diag\bigl[
			\lambda_1^{\star},\, \lambda_0\bI_{[m_{gh}-1:m_{gh}-1]},\, \lambda_3^{\star}\bI_{h-1},\, \lambda_0\bI_{[\sum_{j=1}^{h-1}(m_{gj}-1):\sum_{j=1}^{h-1}(m_{gj}-1)]},\, \lambda_5^{\star}\bI_{g-1}, \\
			&\quad
			\lambda_0\bI_{[\sum_{i=1}^{g-1}(m_{ih}-1):\sum_{i=1}^{g-1}(m_{ih}-1)]},\, \lambda_0\bI_{(g-1)(h-1)},\, \lambda_0\bI_{[\sum_{i=1}^{g-1}\sum_{j=1}^{h-1}(m_{ij}-1):\sum_{i=1}^{g-1}\sum_{j=1}^{h-1}(m_{ij}-1)]}
			\bigr] \\
			&= \bLambda_\lambda^{\star},
		\end{aligned}
	\end{displaymath}
	where the four distinct characteristic roots of $ m_U\barbI_{\bm} \check{\bV}^{\star} \barbI_{\bm}$ are 
    $\lambda_0=\sige^2$
	$\lambda_1^{\star} = \sige^2+ hm_U\sigma_\alpha^2 + gm_U\sigma_\beta^2$, 
	$\lambda_3^{\star} = \sige^2 + hm_U\sigma_\alpha^2$,and
	$\lambda_5^{\star} = \sige^2 + gm_U\sigma_\beta^2$.
\end{theorem}

\begin{corollary}\label{corollary2}
Let $\barbJ{a}^{0} = \bI_a$ and $\barbJ{a}^{1} = \barbJ{a}$. 
Consider the inverse of the modified covariance for the model without interaction \eqref{two-way cross model no interaction}, denoted by $\check{\bV}^{-1}$. Then, we have
\begin{align}
\check{\bV}^{\star-1}
    	=\frac{1}{\lambda_0}m_U\tildebI_{\bm}+m_U\left[\sum_{\bii=\bzero_{[2:1]}}^{\bone_{[2:1]}}\delta_{i_1i_2}^{\star} (\barbJ{g}^{i_1}\otimes \barbJ{h}^{i_2})\right]_{\text{cell}}\circledast\tildebJ_{\bm}, 
\end{align}
where $m_U\tildebI_{\bm}=m_U\bI_{1/\bm}$, $\delta_{00}^{\star}=0$, $ \delta_{01}^{\star}={1}/{\lambda_5^{\star}}$,$\delta_{10}^{\star}={1}/{\lambda_3^{\star}}$ and $	\delta_{11}^{\star}={1}/{\lambda_1^{\star}}-{1}/{\lambda_3^{\star}}-{1}/{\lambda_5^{\star}}$.
Here, the exponents $i_1$ and $i_2$ are also used as subscripts of $\delta$ for notational convenience and readability. They are treated as components of the binary vector $\bii$, with the subscripts written in reverse natural order to allow interpretation of $\bii$ as binary numbers. The summation $\sum_{\bii = \bzero}^{\bone}$ denotes the sum over all four binary vectors from $00$ to $11$.
\end{corollary}

\begin{theorem} 
Suppose $g,h\to\infty$ and all variance components are bounded. Then
\begin{align}
\bV^{\star-1}
=\frac{1}{\sige^2}\bI_n
+\bO_{[n:n]}\!\left(
\frac{1}{\min(g^{1-\varepsilon},h^{1-\varepsilon})m_L^2}
\right),
\end{align}
where $0\le\varepsilon<1$. When $\sige^2/(2m_L^2\siggama^2)\le1$ or $m_{ij}=m$, we have $\varepsilon=0$.
\end{theorem}
\begin{theorem}
Let $\Delta = (m_U - m_L)/m_U$.Suppose all variance components are bounded and $ 0<\Delta< 1/2$. Then the $r$th order approximation to the inverse of the covariance matrix $\bV^{\star}$ is
\begin{equation}
   \bV^{\star-1} = \left( \sum_{l=0}^{r} (-\sige^2)^{l} (\check{\bV}^{\star-1} \bI_{\bm}^{\Delta})^l \right) \check{\bV}^{\star-1} + \bO_{[n:n]}\left( \frac{\Delta^{r+1}}{(1 - \Delta)^{r+1}} \right), 
\end{equation}
where $\bI_{\bm}^{\Delta} = \diag((1 - m_{11}/m_U)\bI_{m_{11}}, \ldots, (1 - m_{gh}/m_U)\bI_{m_{gh}})$, and $\bO_{[n:n]}(a)$ denotes an $n \times n$ matrix with all entries of order $O(a)$.
\end{theorem}

\begin{theorem}
Define $\bE_{ijk} = \sigma_e^2 \left(1 - {m_{ij}}/{m_U} \right) \bI_{n(ijk)}$, where $\bI_{n(ijk)}$ denotes the $n \times n$ diagonal matrix with a one at the diagonal position corresponding to the lexicographic index of $(i,j,k)$, and zeros elsewhere.  
For each $(a,b,c)$ with $a \in \{1,\ldots,g\}$, $b \in \{1,\ldots,h\}$, and $c \in \{1,\ldots,m_{ab}\}$, 
define $\mathcal{L}_{ab} = \{(i,j):\ i < a \text{ or } (i = a,\ j < b)\}$ and
$\bW_{abc+1}^{\star}= \check{\bV}^{\star} 
+ \sum_{(i,j) \in \mathcal{L}_{ab}} \sum_{k=1}^{m_{ij}} \bE_{ijk}
+ \sum_{k=1}^{c} \bE_{abk},
$
with $\bW_{111}^{\star} = \check{\bV}^{\star}$. Assume that each matrix $\bW_{abc}^{\star}$ is nonsingular. Then
\begin{align*}
\bW_{111}^{\star-1} = \check{\bV}^{\star-1} \text{ and } \quad \bW_{abc+1}^{\star-1} = \bW_{abc}^{\star-1} - \kappa_{abc}^{\star} \bW_{abc}^{\star-1} \bE_{abc} \bW_{abc}^{\star-1},
\end{align*}
where $\kappa_{abc}^{\star} =1/[1+\tr(\bW_{abc}^{\star-1}\bE_{abc})]$.
In particular, the inverse of $\bV^{\star}$ is  
\begin{align*}
\bV^{\star-1} = \bW_{gh\,m_{gh}}^{\star-1} - \kappa_{ghm_{gh}} \bW_{gh\,m_{gh}}^{-\star1} \bE_{gh\,m_{gh}} \bW_{gh\,m_{gh}}^{\star-1}.
\end{align*}
\end{theorem}

\newpage
\section{Full Simulation Results}

\begin{figure}[!b]
    	\includegraphics[width=0.45\linewidth]{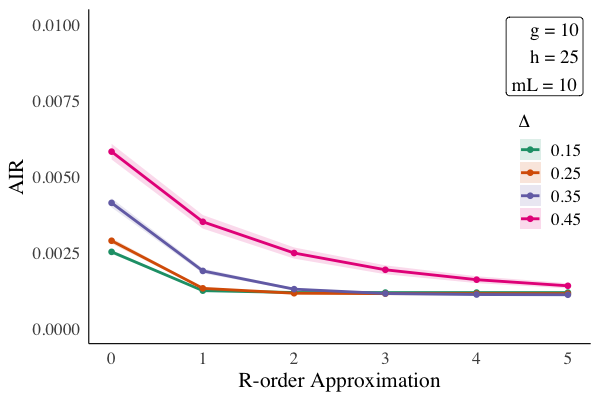}
	\includegraphics[width=0.45\linewidth]{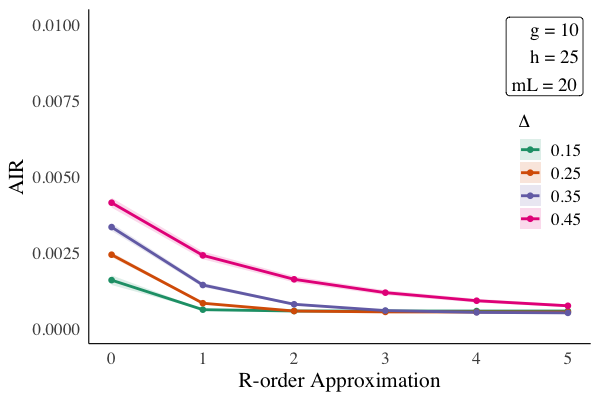}
	\includegraphics[width=0.45\linewidth]{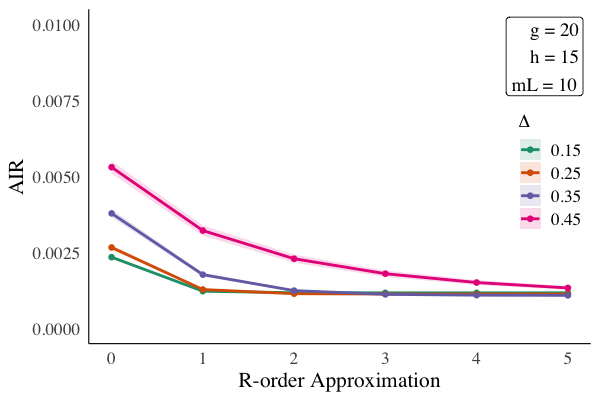}
	\includegraphics[width=0.45\linewidth]{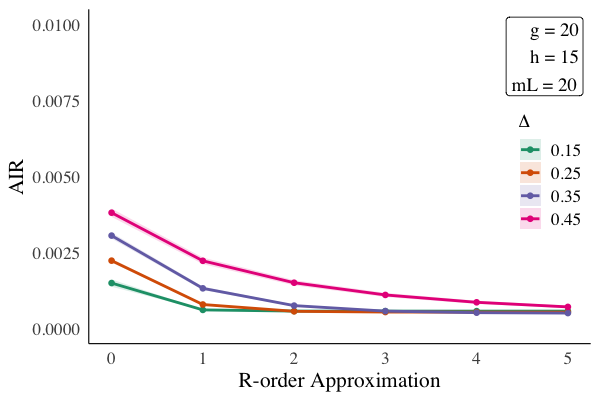}
	\includegraphics[width=0.45\linewidth]{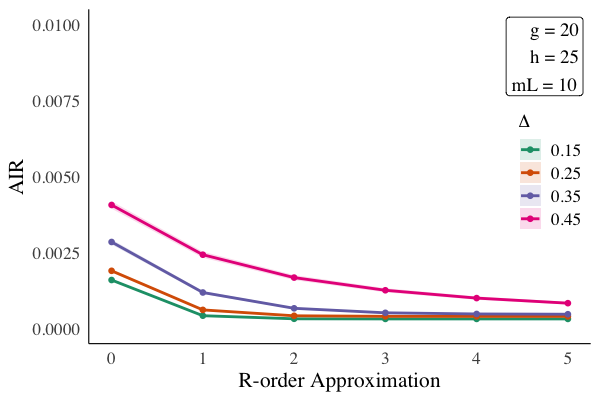}
	\includegraphics[width=0.45\linewidth]{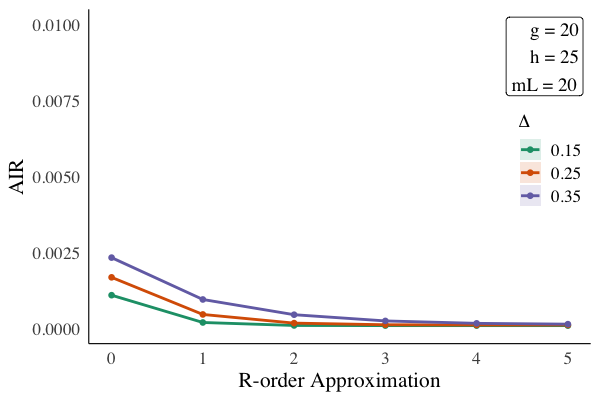}
	\caption{Non-asymptotic Setting: Average Inversion Residual with shaded areas representing one standard deviation}
\end{figure}

\end{document}